\documentclass[a4paper]{article}

\usepackage{amssymb}
\usepackage{graphics}
\usepackage{tabularx}
\usepackage{color}
\usepackage{natbib}
\usepackage{rotating}
\usepackage{authblk}
\usepackage[english]{babel}
\usepackage[utf8x]{inputenc}
\usepackage[T1]{fontenc}

\usepackage[a4paper,top=3cm,bottom=2cm,left=3cm,right=3cm,marginparwidth=1.75cm]{geometry}
\usepackage{amsmath}
\usepackage{amsmath}
\usepackage{graphicx}
\usepackage[colorinlistoftodos]{todonotes}
\usepackage[colorlinks=true, allcolors=blue]{hyperref}
\usepackage{fancyhdr}
\usepackage{float}
\usepackage{xcolor}

\usepackage{lipsum}

\title{The \textit{Herschel}\thanks{{\it Herschel} is an ESA space observatory with science instruments provided by European-led Principal Investigator consortia and with important participation from NASA.}/PACS Point Source Catalogue Explanatory Supplement}
\author[1]{G. Marton}
\author[2]{L. Calzoletti}
\author[2]{A. M. Perez Garcia}
\author[1]{C. Kiss}
\author[3]{R. Paladini}
\author[2]{B. Altieri}
\author[2]{M. S\'anchez Portal}
\author[2]{M. Kidger}
\author[ ]{\\ the Herschel Point Source Catalogue Working Group}

\affil[1]{Konkoly Observatory, Research Centre for Astronomy and Earth Sciences, Hungarian Academy of Sciences, H-1121 Budapest}
\affil[2]{European Space Astronomy Centre (ESAC)/ESA, PO Box 78, 28690 Villanueva de la Ca\~nada, Madrid, Spain}
\affil[3]{Infrared Processing Analysis Center, California Institute of Technology, 770 South Wilson Ave., Pasadena, CA 91125, USA}

\begin{document}

\newcommand{\lele}[3]{{#1}\,$\le$\,{#2}\,$\le$\,{#3}}
\newcommand{\tauav}[1]{$\tau_{#1}$/A$_{\rm V}$}
\newcommand{\iav}[1]{I$_{#1}$/A$_{\rm V}$}
\newcommand{\arcsec}{$\prime\prime$}
\newcommand{\degr}{$^{\circ}$}
\newcommand{\mjysr}{MJy\,sr$^{-1}$}
\newcommand{\dr}{2012\,DR$_{30}$}
\newcommand{\angstrom}{\mbox{\normalfont\AA}}
\newcommand\sun{\hbox{$\odot$}}
\def\lesssim{\mathrel{\hbox{\rlap{\hbox{\lower3pt\hbox{$\sim$}}}\hbox{\raise2pt\hbox{$<$}}}}}
\def\degr{\hbox{$^\circ$}}
\def\arcmin{\hbox{$^\prime$}}
\def\arcsec{\hbox{$^{\prime\prime}$}}
\def\utw{\smash{\rlap{\lower5pt\hbox{$\sim$}}}}
\def\udtw{\smash{\rlap{\lower6pt\hbox{$\approx$}}}}
\def\fd{\hbox{${,}\!\!^{\rm d}$}}
\def\fh{\hbox{${,}\!\!^{\rm h}$}}
\def\fm{\hbox{${,}\!\!^{\rm m}$}}
\def\fs{\hbox{${,}\!\!^{\rm s}$}}
\def\fdg{\hbox{${,}\!\!^\circ$}}
\def\farcm{\hbox{${,}\mkern-4mu^\prime$}}
\def\farcs{\hbox{${,}\!\!\arcsec$}}
\def\fp{\hbox{${,}\!\!^{\scriptscriptstyle\rm p}$}}
\newcommand{\tiunit}{$\mathrm{J\,m^{-2}\,s^{-1/2}\,K^{-1}}$}
\newcommand{\chired}{$\chi^2_{\rm r}$}

\maketitle
\begin{figure}[H]
\centering
\includegraphics[width=1.0\textwidth]{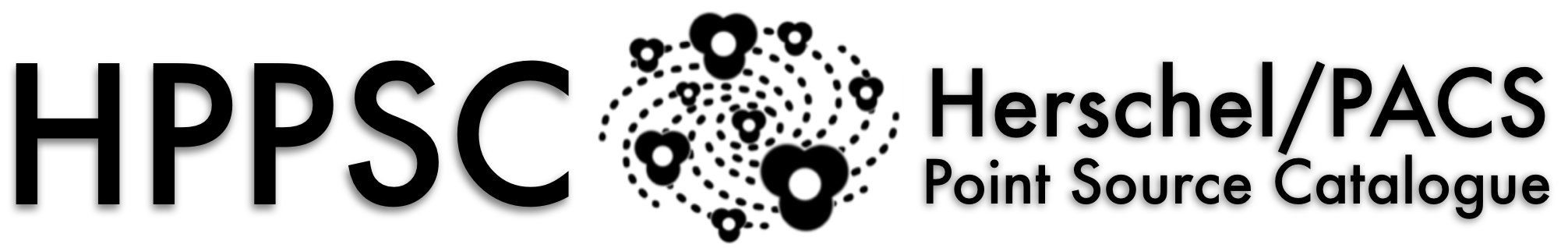}
\end{figure}

\begin{abstract}
The \textit{Herschel} Space Observatory was the fourth cornerstone mission in the European Space Agency (ESA) science programme. It had excellent broad band imaging capabilities in the far–infrared (FIR) and sub–millimetre part of the electromagnetic spectrum. Although the spacecraft finished observing in 2013, it left a large legacy dataset that is far from having been fully explored and still has a great potential for new scientific discoveries. The PACS and SPIRE photometric cameras observed about 8\% of the sky in six different wavebands. This document describes the \textit{Herschel}/PACS Point Source Catalogue (HPPSC), a FIR catalogue based on the broad-band photometric observations of the PACS instrument with filters centred at 70, 100 and 160 $\mu$m.
  
We analysed all combined, Level 2.5/Level 3 \textit{Herschel}/PACS photometric observations including 682 Parallel Mode, 12932 nominal mode and 1644 SSO maps. 
The PACS photometer maps that we produced were generated by applying the JScanam task of the \textit{Herschel} Interactive Processing Environment (HIPE) v13.0.0. Sources were identified with the HIPE implementation of SUSSEXtractor, and the flux densities obtained by aperture photometry. 
We found a total of 108\,319 point sources that are considered to be reliable in the 70\,$\mu$m maps, 131\,322 at 100\,$\mu$m and 251\,392 point sources in the 160\,$\mu$m maps. In addition, our quality control algorithm identified 546\,587 candidate sources that were found to be extended and 7\,185\,160 features which did not pass the signal-to-noise and other criteria to be considered reliable sources. These sources were included in the Extended Source List and Rejected Source List of the HPPSC, respectively. 
The calculated completeness and photometric accuracy values are based on simulations, where artificial sources were injected into the observational timeline with well controlled flux density values. The actual completeness is a complex function of the source flux, photometric band and the background complexity. 


\end{abstract}

\clearpage

\tableofcontents

\clearpage

\subsection*{List of Acronyms}
\begin{tabular}{ll}
AOR & Astronomical Observation Request\\
ESA & European Space Agency \\
FIR & far-infrared \\
FWHM & Full Width at Half Maximum \\
GH & GOODS-\textit{Herschel}\\
Hi-GAL & \textit{Herschel} infrared Galactic Plane Survey\\
HIPE & \textit{Herschel} Interactive Processing Environment\\
HPESL & \textit{Herschel}/PACS  Extended Source  List\\
HPF & High-Pass Filtering\\
HPPSC & 	\textit{Herschel}/PACS Point Source Catalogue\\
HPRSL & \textit{Herschel}/PACS  Rejected Source  List\\
HSA & \textit{Herschel} Science Archive\\
N$_S$ &  Structure noise \\
NASA & National Aeronautics and Space Administration\\
PACS & Photodetector Array Camera and Spectrometer\\
PEP & PACS Evolutionary Probe\\
PSF & Point Spread Function \\
ROI & Region Of Interest\\
S/N$_S$ & Structure noise based Signal-to-Noise Ratio\\
S/N$_R$ & Background RMS based Signal-to-Noise Ratio\\
smm & sub-millimetre \\
SPG & Standard Product Generation\\
SPIRE & Spectral and Photometric Imaging Receiver\\
SSO & Solar System Object \\
STRN & Structure noise \\
TOD & Time Ordered Data\\
\end{tabular}

\clearpage

\subsection*{Document Change History}

\begin{center}
\begin{tabular}{|l|l|}
\hline
& \\
Date  & Changed\\
& \\
\hline
& \\
12--April--2017 & First version \\
&\\
\hline
&\\
24--April--2017 & Language editing \\
& \\
\hline
&\\
05--May--2017 & First public version \\
& \\
\hline
\end{tabular}
\end{center}

\clearpage

\section{General Information}
\subsection{Purpose and Structure of the Document}
The \textit{Herschel}/PACS Point Source Catalogue (HPPSC) Explanatory Supplement is part of the first public version of the Catalogue. This Explanatory Supplement is intended to be a complete and self-contained description of the products generated for and the methods of generation used in the HPPSC. The first part is an introduction and overview of the mission, the PACS instrument and practical issues related to the work. The second part gives a detailed description of the procedure that was used in order to generate the HPPSC, including the products, the methods of detection and the extraction pipeline, as well as the quality assessment. The third part gives details of the resulting products, including the Catalogue itself and additional lists. The fourth part details the additional material, including the \textit{Herschel}/PACS Extended Source List (HPESL), the \textit{Herschel}/PACS Rejected Source List (HPRSL), and the observation table. The observation table contains information on the map that was used for the extraction of a given Catalogue object. The final chapter describes the validation procedures, including internal validation by using simulations and external validation by comparing our results with already published catalogues. 

This document includes information that are essential in order to understand our products. However, more detail might be necessary for some users. Additional background information is found in the \href{http://herschel.esac.esa.int/Docs/PACS/pdf/pacs_om.pdf}{PACS Observer's Manual}. A deeper understanding of the map products can be obtained from the \href{http://herschel.esac.esa.int/twiki/pub/Public/PacsCalibrationWeb/pacs_mapmaking_report_ex_sum_v3.pdf}{PACS Map-Making Report}. A detailed analysis of the PACS photometer Point Spread function is available \href{http://herschel.esac.esa.int/twiki/pub/Public/PacsCalibrationWeb/bolopsf_22.pdf}{here}. The Point Source Flux Calibration is described in \href{http://herschel.esac.esa.int/twiki/pub/Public/PacsCalibrationWeb/Balog_ExpAstr_2013.pdf}{Balog et al., 2014} and confirmed with asteroids in \href{http://herschel.esac.esa.int/twiki/pub/Public/PacsCalibrationWeb/Mueller_ExpAstr_2013.pdf}{M\"uller et al., 2014}. Documents of the colour corrections can be found in \href{http://herschel.esac.esa.int/twiki/pub/Public/PacsCalibrationWeb/cc_report_v1.pdf}{M\"uller et al., 2011} and \href{http://herschel.esac.esa.int/twiki/pub/Public/PacsCalibrationWeb/PICC-CR-TN-044.pdf}{here}.

\subsection{Purpose of the Catalogue}
The \textit{Herschel} Space Observatory ceased to take science observations on 29 April, 2013. It left a large legacy dataset that is far from having been fully explored and still has potential for new scientific discoveries. Some source catalogues have already been produced by individual observing programmes, which are available via the  \href{https://www.cosmos.esa.int/web/herschel/user-provided-data-products}{\textit{Herschel} User Provided Data Products}. However, there are still many observations that have not yet been analysed for their full science content.

To maximise the science return of the PACS data sets, we built a homogeneous \textit{Herschel}/PACS Point Source Catalogue (HPPSC) from all the scan map observations (see Section~\ref{mapRep}). Our source extraction enables a systematic and unbiased comparison of sensitivity across different Herschel fields that single programs will generally not be able to provide. The extracted point sources contain individual YSOs and unresolved YSO clusters in our Galaxy, as well as dusty extragalactic objects of the local and distant Universe. This rich dataset provides a pool of diverse celestial targets for scientists to study in more detail the early phases of star formation, galaxy properties, both locally and at large distances and galaxy evolution through time. In addition to enabling the possibility of statistical analysis of stellar and galaxy clusters, it also provides an excellent target list for follow-up observations with current state--of--art and future facilities such as ALMA and JWST.

While we did our best to deliver a catalogue that can be used for scientific analysis as it is, caveats are still present. We encourage all users of the HPPSC to read the Cautionary Notes below, and to do their science with these notes in mind.

\subsection{Cautionary Notes}

\subsubsection{Catalogue and additional material}
The primary goal of our work was to create a reliable Point Source Catalogue that provides information that can be easily and safely used for scientific study. Therefore, we provide detailed information of this work, including data description, calibration and quality assessment. However, \textit{Herschel} did observe many other objects that could be potentially interesting for the users. Our additional source lists, the \textit{Herschel}/PACS Extended Source List (HPESL) and the Rejected Source List (HPRSL) contain information that were extracted with the pipeline designed for the point sources, but that are too extended to be considered as point sources, and/or fail to pass one or more of our quality criteria (Figure~\ref{workflow3}). \textbf{These additional lists can be used at the user's own risk. We encourage strongly users to check the entries found in these lists and revise them using their own methods.}

\subsubsection{Photometry}

Our photometric accuracy is within 1\% in relative to the photometry of the PACS calibrator stars: see subsection~\ref{balog}. However, these values are based on measurements that were specifically designed for flux calibration, with multiple visits, 20$\arcsec$s$^{-1}$ scan speed and isolated sources on flat background. Their Spectral Energy Distribution is well known, therefore colour corrections were also possible. These conditions apply only for a very small number of sources observed by \textit{Herschel}. Therefore, colour corrections were not made for the published values: it is up to the user to calculate them. Also, it is important to note, that the values listed in the Catalogue are \textbf{in-band} flux density values.

\subsubsection{Coverage \& Completeness}\label{CoveCompl}

The \textit{Herschel} Space Observatory, as its name indicates, was operated as an observatory, meaning that instead of an all-sky coverage it was pointed towards carefully selected targets/fields. The summed area of the observations, including overlaps, is $\sim$7\% of the sky across the three PACS bands. However, not all observations could be used for Catalogue generation, as we required that the observations were processed at least up to Level 2.5 (see Section~\ref{mapRep}): 126 Parallel Mode observations and 4452 Scan map observations do not satisfy this requirement.
Generally, observations reduced up to a lower processing level (Level 2) are spurious observations in which the sky fields are recovered by higher level products (Level 2.5 or Level 3). Nevertheless there are some large surveys the scanning strategies prevented the generation of standard Level 2.5. The most important of these are the North/South Galactic Poles fields (about 490 sqdg), belonging to \href{http://www.h-atlas.org}{The Herschel Thousand Degree Survey (H-ATLAS)} Key Program and the \href{https://irsa.ipac.caltech.edu/data/Herschel/HERITAGE/docs/aj_148_6_124.pdf}{HERschel Inventory of the Agents of Galaxy Evolution (HERITAGE)} Key Program: this last consists of the LMC (8$\degr \times 8.5\degr$), the SMC (5$\degr \times 5\degr$), and a 4$\degr \times 4\degr$ region of the SMC Tail at two different epochs. So, the final estimated sky coverage of our Catalogue is $\sim$4.5\%.

We made a considerable effort to understand how the completeness of our detections depends on the flux level of the sources and on the complexity of the environment. More details are found in Section~\ref{completeness}. It is important to note that, especially in star forming regions, sources tend to be located in groups or clusters. In many cases the cluster members are so close together that they bias the S/N estimation, causing false rejection of sources despite our best effort to preserve them.

\subsubsection{Homogeneity}

\textit{Herschel} carried out a multitude of observing programmes with different scientific goals. Due to the observing strategies and parameters of these programmes, the sky coverage is very arbitrary. One of the main goals with the \textit{Herschel} Point Source Catalogues was to carry out a homogeneous source extraction, so the results can be compared, taking into account scan speed, sampling rate, repetition factor. We used the same pipeline for all maps, which was of course tailored for the different observing bands. Similarly, the input PSFs for the source detection algorithm (SUSSEXtractor) were chosen as a function of the observing mode and scan speed.

\subsubsection{Cross--matching}

The HPPSC is a blind catalogue, i.e., extraction of sources is not based on prior detections at different wavelengths. Extraction of sources in each band was carried out separately. Band merging and cross-identifications are out of the scope of this work.

\subsubsection{Reliability}
The unprecedented resolution of \textit{Herschel} in the FIR and at sub-mm wavelengths allowed us to map the dusty ISM with a detail never before seen. The separation of the background fluctuation and the semi-extended, or point sources is mathematically challenging and triggered the development of many algorithms tailored for star forming regions observed by \textit{Herschel}. These algorithms are very sophisticated and complex tools that are not meant for general use. We needed to select an algorithm that is both fast and easy to use for all kinds of fields, even if that comes at the cost of reliability, or completeness. The \href{http://iopscience.iop.org/article/10.1086/515393/pdf}{SUSSEXtractor} is mainly sensitive to point sources, but has also identified brightness fluctuations of the background as sources. We refer to the \href{ftp://cdsarc.u-strasbg.fr/pub/cats/II/125/psc.txt}{IRAS Explanatory Supplement} and state that the  sky  at  60-200  $\mu$m  \textbf{is  dominated by filaments termed "infrared
 cirrus"  which,  although  concentrated  in  the  Galactic plane, can be found
 almost all the way up to the Galactic poles.  The primary, deleterious effects
 of the cirrus are that it can generate well-confirmed point and small extended
 sources  that  are  actually  pieces  of  degree-- or arcminute--sized  structures rather than
 isolated,  discrete  objects} and  that it  can  corrupt measurements of true point sources.

Despite our conservative approach to filtering out spurious sources, based on various quality criteria, such objects remained occasionally  in our catalogue. If the user is not confident that an astronomical object is present at the desired sky position, eyeballing of the map is encouraged. All the data (signal--to--noise, FWHM, flux ratio, etc., see Section~\ref{sub:consolidation}) that we used in the process of source table cleaning and quality assessment, are included in the Point Source Catalogue files and provided for the users. Scientists are encouraged to create their own criteria, or improve ours if necessary, so the resulting source list is more suited for their scientific needs.

\subsubsection{Tracked maps of Solar system targets}

As it is detailed below, maps of Solar system objects have typically been observed in tracked mode, i.e. taking into account the apparent movement of the target. These maps were reprocessed in the 'rest' sky frame for the HPPSC. In the present version of the catalogue no identification of the moving Solar system targets is performed, but the users are encouraged to check against the possible solar system target contamination using publicly available solar system object databases (e.g. Herschel Catalogue of Solar System Observations of Romero, Kidger \& Rengel, in prep.). The affected OBSIDs are marked by an 'SSOMAP' flag in the Observation Table. 

\subsection{How to access the HPPSC data products}

The HPPSC data products are available in three different ways:
\begin{itemize}
\item Separate, compressed csv tables via the \href{https://www.cosmos.esa.int/web/herschel/pacs-point-source-catalogue}{ESA COSMOS website}
\item \href{http://sky.esa.int/}{ESA Sky}
\item \href{http://irsa.ipac.caltech.edu/Missions/herschel.html}{Catalog Search Tool for Herschel data} available from NASA's Infrared Science Archive (IRSA)
\end{itemize}

\subsection{Referencing the HPPSC}
The relevant rules of the \href{https://www.cosmos.esa.int/web/herschel/publishing-rules-guidelines}{\textit{Herschel} Publishing Rules and Guidelines} apply for scientific results based on the HPPSC. It has been agreed that in all papers using \textit{Herschel} data a mandatory footnote on the first page should be included. The footnote serves multiple purposes, including providing credit and simplifying publication tracking. The footnote is: \textbf{"Herschel is an ESA space observatory with science instruments provided by European-led Principal Investigator consortia and with important participation from NASA."} The standard references to \textit{Herschel} is Pilbratt, G.L., Riedinger, J.R., Passvogel, T. et al. 2010, A\&A, 518, L1. The standard reference for the PACS instrument is Poglitsch, A., Waelkens, C., Geis, N. et al. 2010, A\&A, 518, L2. 

For the time being the HPPSC can be cited by means of the the arXiv publication of the Explanatory Supplement.


\newpage

\section{Introduction}
\subsection{Mission overview}
The \textit{Herschel} Space Observatory \href{http://www.aanda.org/articles/aa/pdf/2010/10/aa14759-10.pdf}{(Pilbratt et al. 2010)} was the fourth cornerstone mission in the European Space Agency (ESA) science programme. It had a primary mirror of 3.5\,m in diameter that allowed an unprecedented spatial resolution and sensitivity at far–infrared (FIR) and sub–millimetre (smm) wavelengths.

\textit{Herschel} operated successfully from June 2009 to 29 April 2013 when it ran out of the liquid helium coolant required to maintain the operational temperatures for the instruments' detectors. The three instruments on-board covered the FIR and smm spectral range from 55 to 671 $\mu$m. The  Photodetector Array Camera and Spectrometer \href{http://www.aanda.org/articles/aa/pdf/2010/10/aa14535-10.pdf}{(PACS, Poglitsch et al. 2010)} and the Spectral and Photometric Imaging REceiver \href{http://www.aanda.org/articles/aa/pdf/2010/10/aa14519-10.pdf}{(SPIRE, Griffin et al. 2010)} were able to make both spectroscopic and photometric observations, while the Heterodyne Instrument for the Far Infrared \href{http://www.aanda.org/articles/aa/pdf/2010/10/aa14698-10.pdf}{(HIFI, de Graauw et al. 2010)} was a purely spectroscopic instrument. Over 35\,000 observations were made during the more than 25\,000 hour long mission. A large legacy dataset was obtained that is far from having been fully analysed and which still has a great potential for new scientific discoveries.

The observing time was allocated to both Guaranteed and Open Time Programmes and some source catalogues have already been produced by these observing programs, however there are many observations that remained unexplored. To maximise the scientific return of the \textit{Herschel} photometric observations, the \textit{Herschel}/SPIRE Point Source Catalogue (\href{https://www.cosmos.esa.int/web/herschel/spire-point-source-catalogue}{HSPSC}) and the \textit{Herschel}/PACS Point Source Catalogue (HPPSC) were generated, the latter described in the present Explanatory Supplement. 

Our homogeneous source extraction enables a systematic and unbiased comparison of sensitivity across the different \textit{Herschel} fields that single programs will generally not be able to provide. The extracted point sources include mainly individual YSOs and unresolved YSO clusters of our Galaxy, as well as the dusty extragalactic objects of the local and distant Universe. Such a huge dataset can help scientists better to understand the early phases of star and galaxy formation in addition to the possibility of carrying out statistical analysis of stellar and galaxy clusters to unravel astrophysical evolution laws through time. It will also provide an excellent target list for future proposals.

\subsection{The PACS Photometer}
\begin{figure}[H]
\begin{center}
\includegraphics[width=0.80\textwidth]{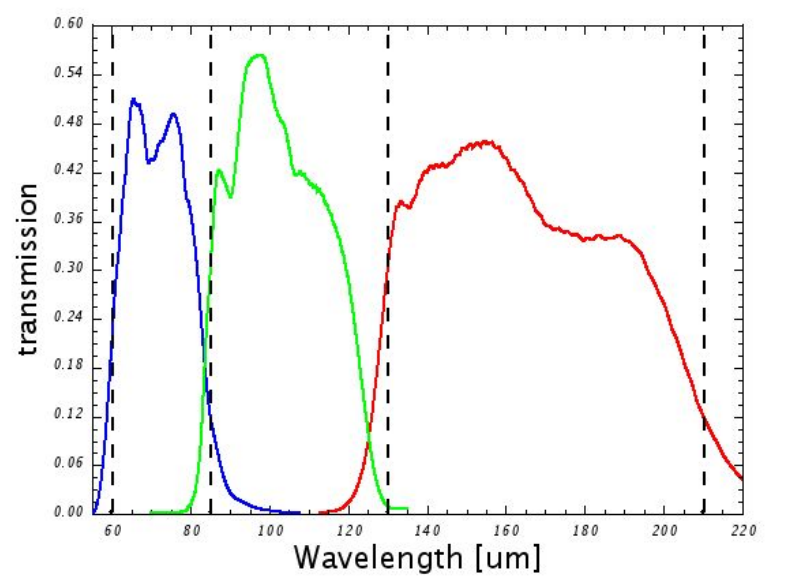}
\caption{Filter transmissions of the PACS filters. The graph represents the overall transmission of the filters combined with the dichroic and the detector relative response in each of the three bands of the photometer. The dashed vertical lines mark the original intended (design values) of the band edges.} \label{transmission}
\end{center}
\end{figure}

The PACS photometer was a dual-band instrument composed of two filled silicon bolometer arrays: the Blue camera consisted of 32$\times$64 pixels and the Red camera of 16$\times$32 pixels. The Blue camera could be used with two filters centered at 70 $\mu$m and 100 $\mu$m (blue and green band), while the Red camera acquired observations at 160 $\mu$m nominal wavelength (red band). The filter transmission curves are shown in Figure~\ref{transmission}. The two cameras allowed two-band simultaneous observations: one of the two bands from the Blue camera plus the Red camera, resulting in pairs of observations taken at 70/160 or 100/160\,$\mu$m. Both cameras had a field of view of $\sim1.75^\prime \times 3.5^\prime$, with close to Nyquist beam sampling in each band.

HPPSC data were taken from 430 unique observing programs, including major Guaranteed Time Key Programs and small, individual Open Time Programs. The key parameters that characterise the PACS observations are the following:
\begin{itemize}
\item {Observing mode: the observing mode could either be the nominal Scan Map or Parallel Mode. In the latter the observations were acquired together with the SPIRE instrument, with a simultaneous five-band photometry of the same field of view. In both modes, the bolometer read-out frequency was 40\,Hz, but due to data-rate limitations, an on-board averaging of 4 frames was performed, except for the blue and green bands in Parallel Mode where the averaging was increased to 8 frames. We emphasize that the HPPSC does not contain any extraction from chop-nod observations, as in the chop-nod mode the same source appears multiple times on the processed images, as descibed in \href{http://adsabs.harvard.edu/abs/2013ExA....36..631N}{Nielbock et al. (2013)}.}    

\item {Scan speed: the standard satellite scanning speed was 20$\arcsec$s$^{-1}$. The highest speed of 60$\arcsec$s$^{-1}$ was dedicated to large Galactic surveys and Parallel Mode maps. We also included 47 calibration observations where the scan speed was 10$\arcsec$s$^{-1}$.}

\item {Coverage/ depth of the observations:
depending on the science goals of the observations different scanning strategies were applied. The shallowest maps are those acquired in Parallel Mode observations, due to the high scan speed and the application of a single repetition. The deepest maps are the ones where low or medium scan speed was selected and the number of repetitions was high. These are typically dedicated observations of point sources. As mentioned above, in many cases Level 2.5 observations were combined into Level 3 images.}
\end{itemize}

 The shape of the PACS Point Spread Function (PSF) is dominated by a tri--lobate pattern in all three bands. A detailed description can be found in \href{http://herschel.esac.esa.int/twiki/pub/Public/PacsCalibrationWeb/bolopsf_22.pdf}{Lutz (2015)}. The PSF can be approximated by 2-dimensional Gaussian fits, as given in  Table~\ref{tab:psffwhm} below. For low and medium scan speeds the typical FWHMs are 5.5$^{\prime\prime}$, 6.7$^{\prime\prime}$ and 11$^{\prime\prime}$ in the 70, 100 and 160\,$\mu$m bands, respectively. These beams also define the achievable spatial resolution and set the limit on the separability of nearby sources.

\begin{table}[H]
\begin{tabular}{lccrcrcr} \hline
Mode&AMA&\multicolumn{2}{c}{blue 70$\mu$m}&\multicolumn{2}{c}{green 100$\mu$m}&\multicolumn{2}{c}{red 160$\mu$m}\\
    &          &FWHM   &PA         &FWHM   &PA        &FWHM   &PA\\ 
    &$^{\circ}$&$\arcsec$&$^{\circ}$&$^{\prime\prime}$&$^{\circ}$&$^{\prime\prime}$&$^{\circ}$\\\hline
Prime 10~$\arcsec$s$^{-1}$&$+$63&
5.20$\times$ 5.56&     &6.54$\times$ 6.78&     &10.38$\times$11.97&  6.1\\
Prime 20~$\arcsec$s$^{-1}$&$+$63&
5.41$\times$ 5.72&     &6.66$\times$ 6.89&     &10.55$\times$12.08&  9.1\\
Prime 60~$\arcsec$s$^{-1}$&$+$63&
5.70$\times$ 9.05& 61.7&6.84$\times$ 9.81& 61.8&11.39$\times$13.37& 41.2\\
Parallel 20~$\arcsec$s$^{-1}$&$+$42&
5.44$\times$ 6.51& 30.8&6.62$\times$ 7.44& 31.1&10.29$\times$12.20&  8.5\\
Parallel 20~$\arcsec$s$^{-1}$&$-$42&
5.31$\times$ 6.68&-26.5&6.53$\times$ 7.56&-27.0&10.37$\times$12.27& -3.4\\
Parallel 60~$\arcsec$s$^{-1}$&$+$42&
5.85$\times$12.58& 43.7&6.99$\times$13.15& 43.9&10.90$\times$14.09& 27.7\\
Parallel 60~$\arcsec$s$^{-1}$&$-$42&
5.69$\times$12.74&-36.9&6.87$\times$13.41&-37.1&11.01$\times$14.53&-23.7\\ \hline
Parallel 20~$\arcsec$s$^{-1}$&$+$42,$-$42&
5.74$\times$6.26&  0.4& 6.98$\times$ 7.42&$-$2.9&10.46$\times$12.27&3.1\\  
Parallel 60~$\arcsec$s$^{-1}$&$+$42,$-$42&
8.80$\times$9.60&$-$4.4& 9.73$\times$10.69&$-$3.8&11.51$\times$13.65&5.3\\ \hline 
\end{tabular}
\caption{FWHM of the PACS PSF for several important cases. 2-dimensional Gaussian fits were used to derive the FWHM for the small and large axis. For noticeably non-round PSF cores, the position angle east of the spacecraft Z direction is noted. The array to map angle (AMA) of the scan is also specified. The maps used to derive the FWHM have been created by photProject with map pixel size 1$^{\arcsec}$\ and pixfrac=1.
Entries above the line refer to single direction scans, showing the in-scan elongation for fast scan and for Parallel Mode. Entries below the line refer to coadded Parallel Mode crossed scans, where a cross-like PSF emerges from co-adding the two elongated PSFs.}
\label{tab:psffwhm}
\end{table}
\subsection{The PACS Standard Products}
\begin{figure}[H]
\begin{center}
\includegraphics[width=0.90\textwidth]{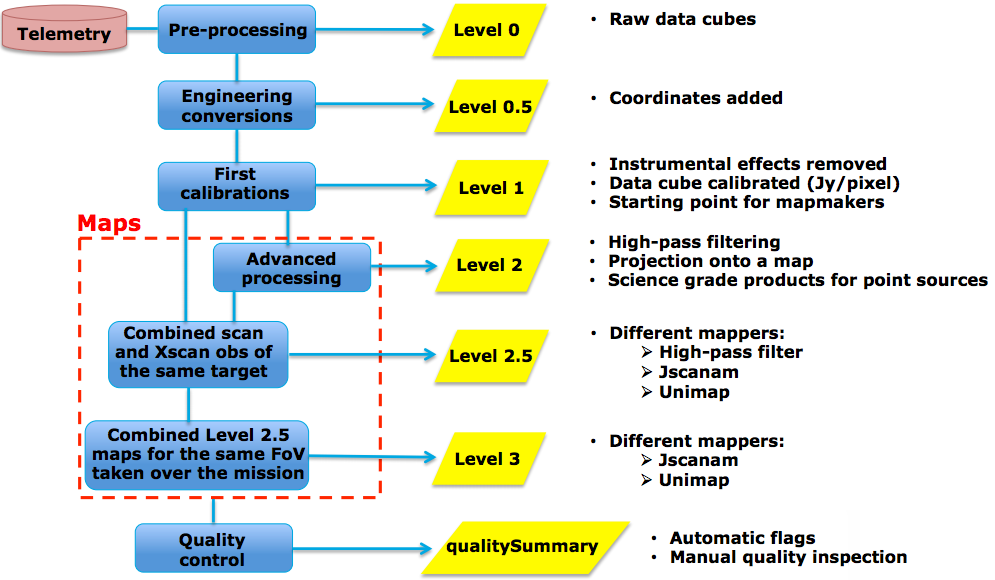}
\caption{SPG chain from Level 0 to Level 3. The raw data are stored in Level 0 cubes. After numerous levels of noise removal and processing, the Level 2.5 and Level 3 maps are the final products that are used for the HPPSC generation.} \label{products}
\end{center}
\end{figure}

PACS observations were processed automatically by SPG (Standard Product Generator) and placed in the \textit{Herschel} Science Archive (HSA). Several products belonging to different processing Levels are generated by  SPG chain (see Figure~\ref{products}); they are available in the HSA for users to download (see the PACS Observing Manual for a detailed description of the data product levels). Starting from the raw telemetry, Level 0 frames are generated and processed up to Level 1. In these steps, instrumental effects (1/f noise, common mode drift, cross-talk) are removed and frames are calibrated in Jy/pixel.
Level 1 frames are the starting point for the mapmakers:
\begin{itemize}
\item Level 2 maps are generated by using the High-Pass Filtering (HPF) pipeline. Large scale structures (noise and extended emission) are removed by means of a sliding median
filter on individual bolometer timeline. These maps are not suited to recovering extended emission. 
\item Level 2.5 maps are generated by combining scan and cross-scan observations of the same sky field (acquired in the same observing mode), by using three different mappers: HPF, \href{http://adsabs.harvard.edu/abs/2015MNRAS.447.1471P}{Unimap} and JScanam. Level 2.5 HPF maps are the averages of the corresponding Level 2 HPF maps. The Unimap mapper exploits the Generalised Least Square approach with the pixel noise compensation for removing 1/f noise, while the JScanam mappers is a Java implementation of \href{http://adsabs.harvard.edu/abs/2013PASP..125.1126R}{Scanamorphos de-striper method} that does not rely on any noise model nor on filtering and exploits the redundancy to derive the drifts from the data. Unimap and JScanam maps are science ready, not absolutely calibrated and are reliable for recovering both extended emission and point-like sources. 
\item Level 3 maps are the average of Unimap and JScanam Level 2.5 maps of any overlapping areas. Additional criteria are that: the scan speed is uniform for all the observations; the observations combined are from the same observing program; they contain at least 180 seconds worth of data; there is at least one scan and cross-scan observation in all combined observations and that for the blue channel data only, the filter is the same.  For the red channel, all observations are used without regard to the filter used in the blue channel.

\end{itemize}

\newpage

\section{Catalogue generation -- data \& methods}

\subsection{Map repository}\label{mapRep}

The HPPSC makes use of JScanam SPG 13.0.0 maps because they are considered to be science ready for catalogue purposes. The JScanam pipeline is  stable and has been validated from SPG12 on. The updates produced up to the legacy version of the software (SPG14.2.1) do not introduce changes in the JScanam maps. That is also the reason why the JScanam Level 2.5 maps are selected as the standard maps in the HSA.
Although in JScanam maps (as for all PACS maps) the zero-level of the background surface brightness is not known, it is not a concern when one performs photometry of point or compact sources, either as aperture or as synthetic photometry, because a certain background level is always subtracted in the source flux extraction procedure.
The absolute calibration for point sources is very accurate, with a relative accuracy $<$5\% in all PACS bands \href{http://adsabs.harvard.edu/abs/2014ExA....37..129B}{Balog et al. (2014)}).

We considered the most reliable and deepest maps available in the HSA for a specific field for our repository. When available, the JScanam Level 3 map is taken for each observed field, otherwise we use the related Level 2.5 maps. In some cases, HPF Level 2 maps are the highest level products available. These are not included in the repository, because of the low reliability of photometric analysis of Level 2 maps and the impossibility of generating structure noise (N$_S$) maps (see Section~\ref{strnmaps}) of high-pass filtered maps, given that high-pass filtering removes the large scale structures from the maps. 

In some cases, large surveys made use of scanning strategies that prevented the generation of standard Level 2.5 products so these sky regions are not included in the HPPSC, nor, as mentioned earlier, are observations acquired with the chp/nod observing mode included in the catalogue, limiting the coverage of our Catalog to 4.5\% of the sky (see Section~\ref{CoveCompl}).

The HPPSC map repository includes Level 2.5 observations of Solar System Objects (SSOs), re-processed in the fixed sky frame and not in the co-moving frame of the target, as it is performed in the standard processing. The rational is to exclude SSOs from the catalogue, while including all the serendipitous detections of the SSO fields.\\
Pointing issues affecting the pointing were observed in SPG13 products during several ODs, introducing offsets for one of two reasons:
\begin{itemize}
\item Switch from the main star-tracker (STR1) to the backup unit (STR2) either due to the autonomous Failure Detection, Isolation and Recovery (FDIR) or commanded from ground for testing and calibrating the backup unit. This produces an offset of $\sim$15\arcsec{} between the desired (nominal) and actually commanded coordinates.
\item Incorrect upload of the Spacecraft Velocity Vector (SVV) to the star-tracker (STR) resulting in a incorrect light aberration correction in the coordinates of the STR guide stars. This produces a variable error in the range 0-20\arcsec{} between the desired (nominal) and actually commanded coordinates.
\end{itemize}

The problem has been fixed in SPG14. The PACS photometric observations affected by these anomalies (151 individual maps, 1\% of the repository) were updated with SPG14 products in our repository. The number of maps used for creating the HPPSC is detailed in Table~\ref{mapstats1}.

\begin{table}[H]
\centering
\small\addtolength{\tabcolsep}{-3pt}
\begin{tabular}{|l|cc|cc|cc|cc|cc|cc|}
\hline
	 &\multicolumn{6}{c}{Parallel Mode} &\multicolumn{6}{|c|}{Scan map}\\
\hline
SPG & \multicolumn{2}{c|}{13.0.0}& \multicolumn{2}{c|}{14.0.1}& \multicolumn{2}{c|}{14.2.0}& \multicolumn{2}{c|}{13.0.0}& \multicolumn{2}{c|}{14.0.1}& \multicolumn{2}{c|}{14.2.0}\\
\hline
Level & 2.5 & 3 & 2.5 & 3& 2.5 & 3& 2.5 & 3& 2.5 & 3& 2.5 & 3\\
\hline
blue&	270&	4&	7&	0&	0&	0&	3471&	86&		25&	0&	13&	1\\
green&	33&		15&	2&	0&	0&	0&	3320&	63&		18&	0&	10&	2\\
red&	292&	25&	9&	0&	0&	0&	3924&	1558&	33&	5&	19&	7\\
\hline
\end{tabular}
\caption{Number of maps in our repository in the different bands at different levels and different SPG versions.}
\label{mapstats1}
\end{table}

\begin{table}[H]
\centering
\small\addtolength{\tabcolsep}{-3pt}
\begin{tabular}{|l|ccc|ccc|ccc|}
\hline
	 &\multicolumn{3}{c|}{blue} &\multicolumn{3}{|c|}{green} & \multicolumn{3}{|c|}{red} \\
\hline
Scan speed [$\arcsec$s$^{-1}$]     &10&20&60&10&20&60&10&20&60 \\
\hline
Parallel Mode & 0 & 51 & 275 & 0 & 26 & 24 & 0 & 51 & 275 \\
Scan map	  & 14&3987& 17  & 14&3782& 10 & 24&6314&  23 \\
\hline
\end{tabular}
\caption{Number of maps in our repository in the different bands and obtained at different scan speeds.}
\label{mapstats}
\end{table}

\subsection{Structure noise maps}\label{strnmaps}

The confusion noise present in the FIR and smm photometric observations is a major limiting factor in sensitivity and photometric accuracy. Confusion noise comes either from the extragalactic background or from the cirrus emission of the Galaxy. Both are mainly due to the presence of the cold dust. It is essential to characterise the properties of the catalogue well. Completeness, flux boosting, and flux error at different sensitivity levels (depending on the depth of the observations) can be evaluated by creating simulations. We injected sources with well controlled flux into observational timelines and reprocessed the new data in the same way as the real data was. As expected we found that in extragalactic observations we are more complete than in the Galactic star forming regions. Also, the photometric accuracy is highly dependent on the complexity of the celestial environment. As our goal was to create a homogeneous catalogue, we had to find a way to treat all environments in the same way. 

\begin{figure}[H]
\begin{center}
\includegraphics[width=0.40\textwidth]{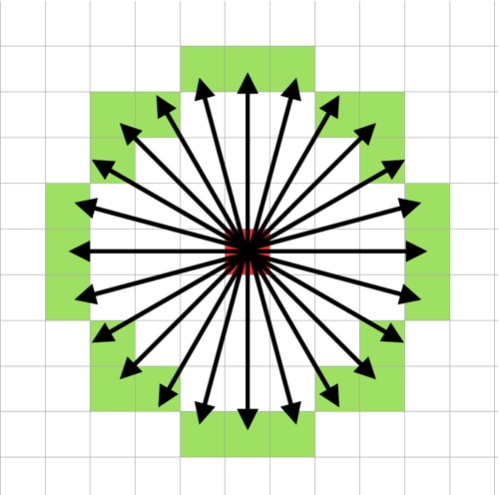}
\caption{The circular configuration used for N$_S$ calculation. The target pixel is shown with red square. The neighbouring pixels at the predefined angular distance are shown with green squares.} \label{strnconfig}
\end{center}
\end{figure}

The structure noise (N$_S$) measures the pixel fluctuation in the map (\href{http://adsabs.harvard.edu/abs/2005A\%26A...430..343K}{Kiss et al. 2005}). It can be translated into the fluctuation power of the neighbouring areas, and gives a local information instead of a general (regional average) number. For each pixel of the map the N$_S$ is calculated as the standard deviation of all flux differences between the pixel and all the surrounding pixels at a fixed distance (see Figure~\ref{strnconfig}). It includes the spatial noise of the celestial environment as well as the noise contribution from the instrument.

The N$_S$ maps were generated with an IDL script. The code reads the same maps that were used to generate the catalogue. For each pixel of the map (target pixel, shown as a red square in Figure~\ref{strnconfig}) we are looking for neighbouring pixels in 24 directions (green squares) at $\sim$30$\arcsec$ angular distance. If fewer than 24 pixels are found, the unique pixels are selected. The next step is to calculate the absolute difference between the unique neighbouring pixels and the target pixel. When the number of unique data points is greater than three, the standard deviation of these differences is stored as a pixel value in place of the target pixel. The pixel value is calculated according to :
\begin{equation}
\sigma_{strn}=\sqrt{\frac{1}{24}\sum_{i=1}^{24} (d_i-\mu)^2},
\end{equation}

where $d_i$ are the differences of each of the fluxes of the 24 (or fewer) pixels and that of the central pixel, and $\mu$ is the average of all $d_i$. The resulting N$_S$ map is then stored as a standard FITS file with the header of the original map.

\begin{figure}[H]
\begin{center}
\includegraphics[width=0.90\textwidth]{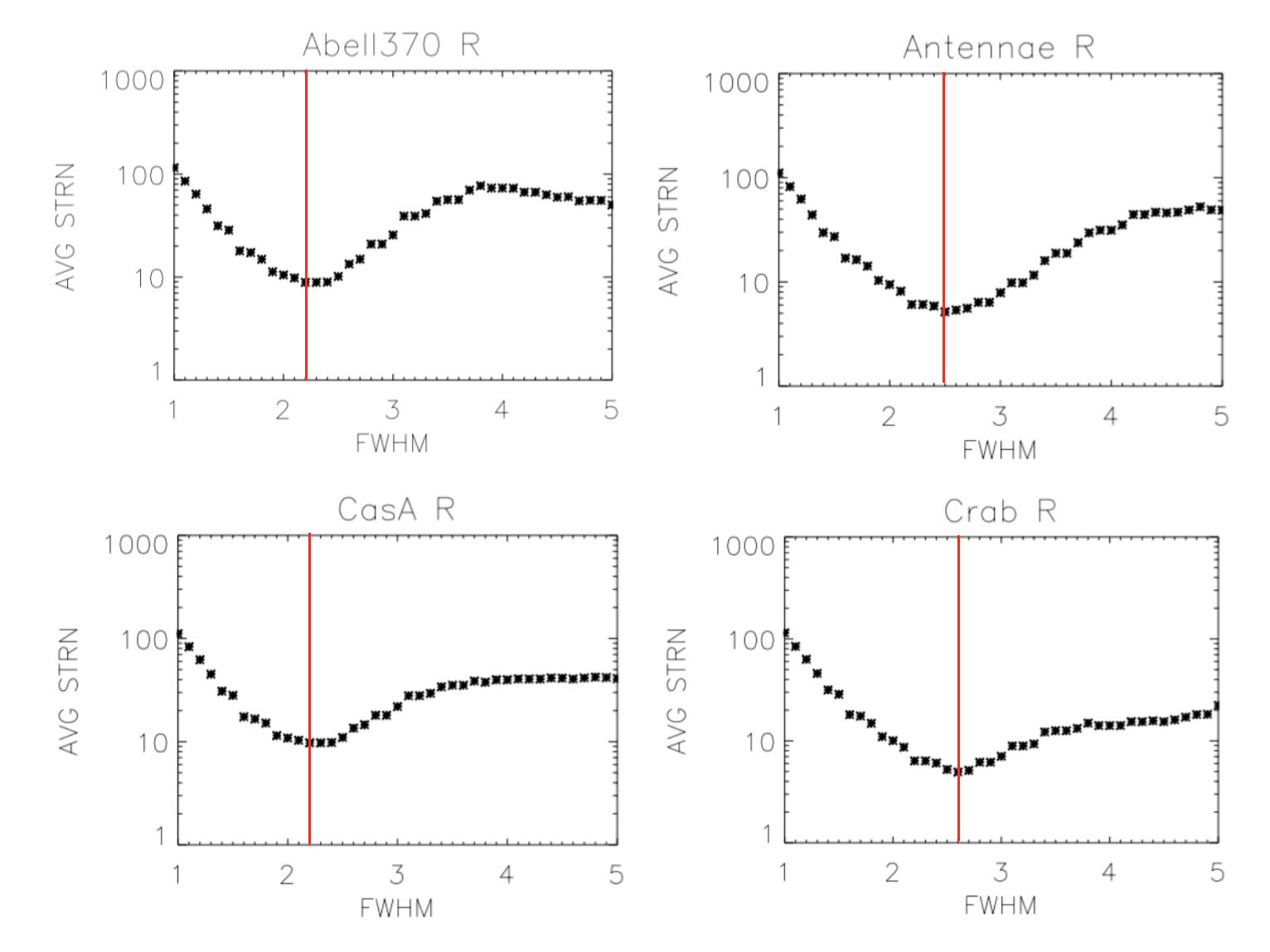}
\caption{Examples showing the average N$_S$ value of the injected sources in red Scan map simulations as a function of the angular distance between the target pixel and the neighbouring pixels. The minimum of the curves (indicated by red vertical lines) represent the angular separation where the source flux has the smallest impact on the N$_S$.}\label{strntestfig}
\end{center}
\end{figure}

The distance between the target pixel and the neighbouring pixels corresponds to a spatial frequency, that had to be optimised. Choosing too small distance would mean that the N$_S$ value includes the flux from the PSF wings causing a scaling with the source flux. To minimize this effect we created N$_S$ maps of simulations where we injected sources with 20 Jy flux and increased the angular distance between the target pixel and the neighbouring pixels. On each N$_S$ map  the N$_S$ value at the position of our artificial sources was calculated and checked as a function of the angular separation (see Figure~\ref{strntestfig}). We found that at an angular distance of $\sim$30$\arcsec$ the N$_S$ value decouples from the source flux and has a minimum value before the large scale structures start to dominate the fluctuation power. We decided to use this angular separation to create our N$_S$ maps in all bands and to attach a N$_S$ value to each of our detections. The average values where the minimum values were found for the different modes and bands are listed in Table~\ref{strnsep}.

\begin{table}[H]
\centering
\begin{tabular}{|l|c|c|c|}
\hline
	 &blue [$\arcsec$]&green [$\arcsec$]&red [$\arcsec$]\\
\hline
Parallel Mode &32$\pm$3&31$\pm$3&31$\pm$7\\
Scan map	  &28$\pm$3&26$\pm$6&27$\pm$5\\
\hline
\end{tabular}
\caption{The smallest angular distance where the source flux decouples from the N$_S$ value.}
\label{strnsep}
\end{table}

The pixel values have the same units as the input map. To avoid so-called NaN-Donuts, single NaNs in the input map are interpolated before the calculation. 

The N$_S$ value for a given source is calculated as follows: we place an aperture at the extracted position of each source, the diameter of which equals to the aperture size used for the photometry, i.e. 6$\arcsec$ for the blue camera and 12$\arcsec$ for the red camera. The total N$_S$ inside the aperture is then converted into units of MJy\,$sr^{-1}$. Eventually, the resulting N$_S$ value is attached to each source in a separate database column.

\subsection{Pipeline description}\label{pipeline}
We created our pipeline using \href{https://www.cosmos.esa.int/web/herschel/hipe-download}{HIPE} release 13.0.0. Our goal was to do as many tasks as possible within the environment to be consistent all the way from observational data processing to photometry.

\begin{figure}[H]
\centering
\includegraphics[width=0.9\textwidth]{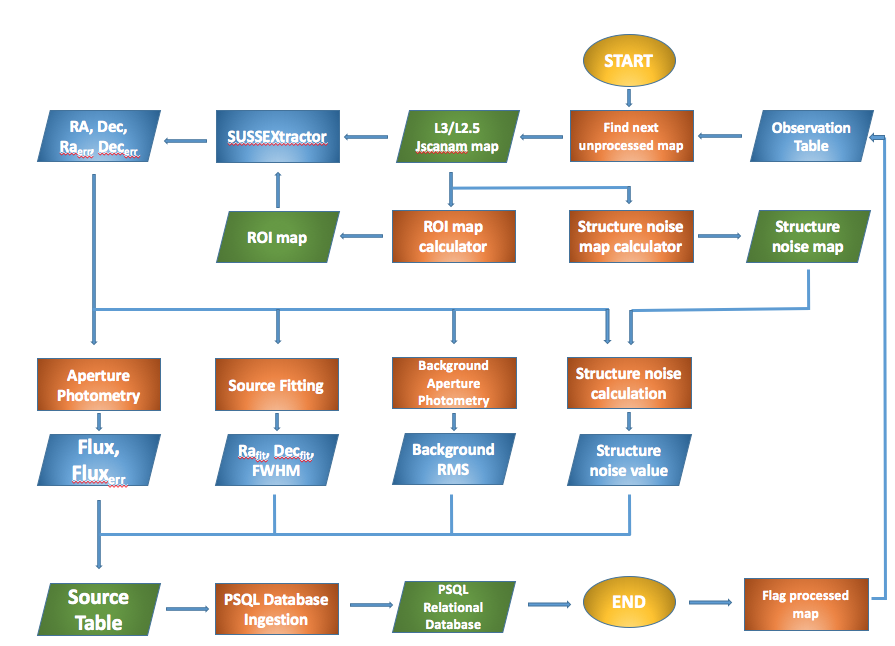}
\caption{The HPPSC pipeline. The workflow shows the steps that were performed on each map from the START to the END. This workflow was repeated for all maps.}\label{workflow1}
\end{figure}

The steps of the workflow are shown in Figure~\ref{workflow1}. Before executing the actual pipeline, the first step is to find the next unprocessed map. This step thus required a list of observations (Observation Table, see Section~\ref{ot}) that was used to check if a map was already processed. If the flag in the Observation Table allowed, the pipeline started by reading the corresponding L3/L2.5 map. 

The N$_S$ maps for the given map were created in IDL, as described in Section~\ref{strnmaps}, and at this stage of the process they were read into HIPE. 

The detection algorithm is applied on specific Region of Interest (ROI) of the maps, in order to identify the map regions with an appropriate coverage values. Slewing and turnaround regions are rejected and on the scan regions where scan and cross-scan coverage is guaranteed at constant speed, a threshold equal to 0.6 times the normalised coverage value is considered for the definition of the ROI. This threshold is not used on Parallel observations, because generally their scanning strategy does not exploit the homogeneous coverage on the field and the threshold approach generates holes in the ROI maps. For a Parallel map, the ROI is defined as the intersections of all the scanning regions of the observations that compose the map.

The JScanam map was then used to locate sources. This was done by the HIPE implementation of SUSSEXtractor, called sourceExtractorSussextracorTask (see Section~\ref{sourcedetection}).  The provided sky coordinates are the J2000.0 Right Ascension (RA) and Declination (Dec) in units of degrees. The positions are used by four different processes:
\begin{enumerate}
\item Aperture photometry by annularSkyAperturePhotometry() task
\item sourceFitter() task
\item background aperture photometry for the background RMS calculation
\item aperture photometry on the N$_S$ map to obtain the N$_S$ value
\end{enumerate}

All the resulting data are stored in a separate \textbf{comma separated value} (csv) table. Instead of handling all the individual csv tables, we created a relational PostgreSQL database, that allowed us to handle our date in a more efficient way.

\subsubsection{Source detection}\label{sourcedetection}

Source detection in the presence of the cirrus emission is a non-trivial task. There are several tools specifically designed for identifying sources on top of filamentary structures. Such algorithms are \href{http://www.herschel.fr/Phocea/file.php?class=page&file=6/aa18797-12_Men_shchikov_etal2012.pdf}{\textit{getsources}} or \href{http://www.aanda.org/articles/aa/pdf/2011/06/aa14752-10.pdf}{CuTEx}. For extragalactic/cosmological observations \href{http://adsabs.harvard.edu/abs/2000A\%26AS..147..335D}{Starfinder} is often used (e.g. \href{http://www.mpe.mpg.de/resources/PEP/DR1_tarballs/readme_PEP_global.pdf}{PEP}). These are very sophisticated tools that can be tuned for environments with various degrees of complexity. Their fine tuning and also their performance is, however, not sufficient for our  purposes. An extensive test for source detection was carried out for the \href{https://www.cosmos.esa.int/web/herschel/spire-point-source-catalogue}{\textit{Herschel}/SPIRE Point Source Catalogue}. Our findings are detailed in "Testing Point Source Extractors"  section of the \href{http://archives.esac.esa.int/hsa/legacy/HPDP/SPIRE/SPIRE-P/SPSC/SPIREPointSourceCatalogueExplanatorySupplementFull20170203.pdf}{SPSC Explanatory Supplement}.

Following our findings for the SPSC, the sourceExtractorSussextractor() task of HIPE (hereafter simply SUSSEXtractor), which is an implementation of the SUSSEXtractor algorithm \href{http://iopscience.iop.org/article/10.1086/515393/pdf}{Savage \& Oliver 2007} was used by our pipeline. This task requires an error map for which we used the "stDev" map of the JScanam products. The detection threshold was set to 3. The algorithm can use a custom PRF, hence we used the PACS PSFs as PRFs according to the wavelength and to the observation mode. Despite using the proper PRFs, the algorithm also requires the FWHM of the sources that we are looking for. Our FWHM values in the blue, green and red bands were set to 5.5$\arcsec$, 7.0$\arcsec$ and 11.5$\arcsec$. The most important information collected with the SUSSEXtractor was: RA, Dec, image X and Y pixel coordinates, RA error and Dec error.

\subsubsection{Flux extraction\label{sub:flux}}
Unlike for SPIRE sources, photometry of PACS sources can at present only obtained by aperture photometry. Therefore flux extraction was carried out using HIPE's annularSkyAperturePhotometry() task. This task performs aperture photometry of the target, enclosed by a circular aperture. The sky is estimated from a concentric annular aperture, with configurable inner and outer radii. In our case the inner and outer radii were set to 25$\arcsec$ and 35$\arcsec$, respectively. The task is able to perform five possible sky estimation. We used the default \textbf{daophot} estimation, which is an implementation of the IDL \href{https://idlastro.gsfc.nasa.gov/ftp/pro/idlphot/mmm.pro}{mmm.pro} routine. 

The flux errors provided by the task are calculated in the same way as in the IDL \href{https://idlastro.gsfc.nasa.gov/ftp/pro/idlphot/aper.pro}{aper.pro} routine. 
The error for the target flux including the background is defined as the square-root of the (absolute value of the) total flux in the target aperture (including the background). The error for the sky is defined as the standard deviation of the sky value. Note that this standard deviation is calculated only with the sky intensities that are used to calculate the sky intensity (as there are sky estimation algorithms that reject sky values). The squared error for the target flux is thus:
\begin{itemize}
\item the absolute value of the target flux from which the sky contribution has been subtracted
\item the standard deviation of the sky value * the surface of the target aperture
\item the squared standard deviation on the sky value
\end{itemize}

The annularSkyAperturePhotometry() task was used for three different purposes. 
\begin{itemize}
\item First of all, this task was responsible for the accurate source flux density measurement of the detected sources. The input coordinates were  provided by SUSSEXtractor. We used 11 different aperture sizes to measure the flux. The aperture radii in the red band were 9,10,11,12,13,14,15,16,17,18 \& 22 arcseconds and 4,5,6,7,8,9,10,11,12,13 \& 18 arcseconds in the blue and green bands. The inner and outer edge of the sky annulus was set in 25$\arcsec$ and 35$\arcsec$ in all cases. After applying the aperture correction, we were able to set quality criteria based on the flux values measured in the different apertures. 
\item The N$_S$ was also obtained with aperture photometry from the N$_S$ maps. We placed 6$\arcsec$ and 12$\arcsec$ apertures (depending on the camera) at the position of the source, on the N$_S$ maps. The N$_S$ inside the aperture was converted then to units of MJy\,sr$^{-1}$.
\item It was also used to measure the background RMS
\end{itemize}

\subsubsection{2D Gaussian fit\label{sub:2Dfit}}

The HIPE sourceFitting task is a task to fit a two-dimensional Gaussian to a source in a specified rectangular region on an image. It can use either an elongated or a circular Gaussian, and can fit a tilted background depending on whether or not there is a slope in the background. For the circular fit the model is a Gauss2DModel + PolySurfaceModel, the initial parameters are a) the maximum value in the cropped region, which  provides an initial value for the centre of the Gaussian, and its peak value, and b) the background, which is estimated via the Daophot algorithm (as for aperture photometry) on the whole rectangle. In the elongated case the initial parameter a) is the fit that would be made if the source were circular, which yields an initial value for the centre and the peak value of the Gaussian and for the background, while b) the sigma's and the rotation angle of the Gaussian are estimated from profile plotting. For all straight lines through the estimated centre with an angle between 0 and 180 degrees w.r.t. the x-axis (in steps of 15 degrees), a Gaussian combined with a constant function is fitted (deriving the initial parameters from the circular 2D-fit) and the largest sigma is taken to be the initial value for the width of the Gaussian in the x-direction and the corresponding angle as the initial value for the rotation angle of the Gaussian. The sigma in the y-direction is estimated from the fit of the line perpendicular to the previous one. The fitter used in all cases is the LevenbergMarquardtFitter.

The sourceFitting is applied on the maps at the positions provided by the SUSSEXtractor task. The assumption is that the objects are always slightly elongated, since the PSF is also not circular, thus the pipeline first tries to fit an elongated Gaussian an a rectangle in a square of  11$\times$11 pixels. If the fit did not converge in 10$^4$ iterations, then a circular fit was applied instead. The task provides a set of output parameters suitable for quality assessment. From all parameters derived of 2D Gaussian fitting (see Section~\ref{sub:flux}), we use the Gaussian peak (RA, DEC) and the  $\sigma_X$ and $\sigma_Y$  to set  the  requirements that HPPSC sources  have to be met (more details in Section~\ref{sub:consolidation}). The $\sigma$ values of the fitted Gaussians were converted to FWHM values, by multiplying them by $2\sqrt{2ln2}$.

\subsection{Quality assessment\label{sub:qa}}

\begin{figure}[H]
\centering
\includegraphics[width=0.9\textwidth]{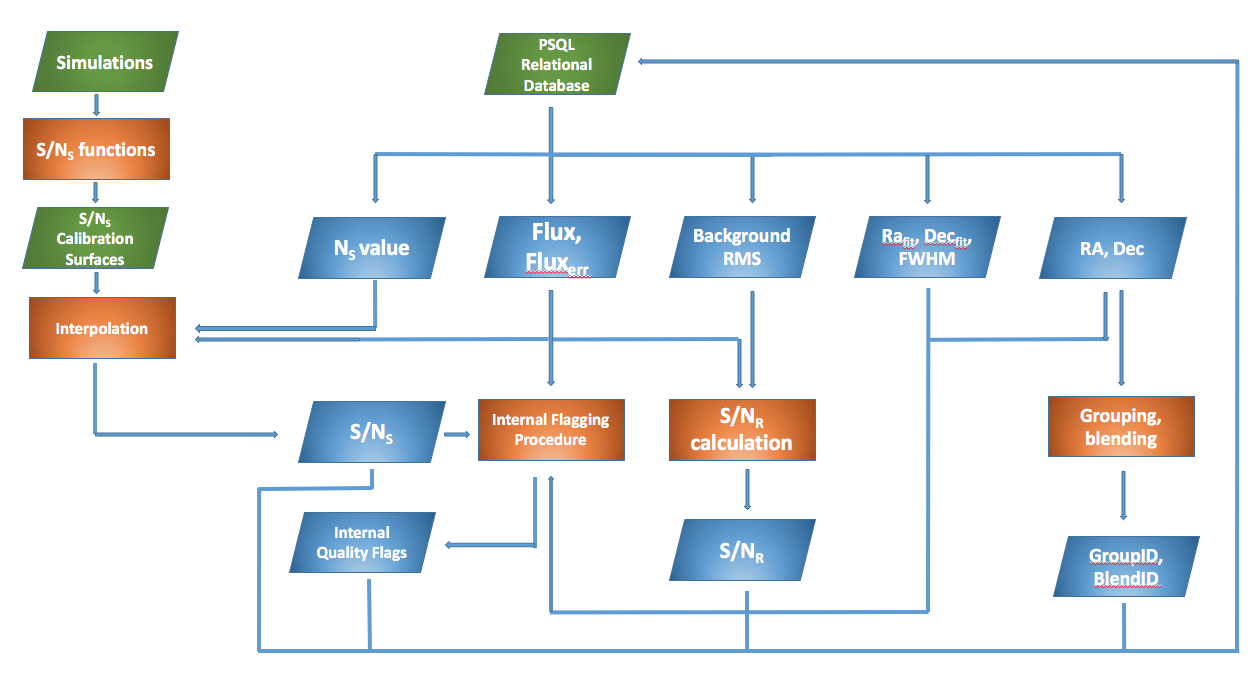}
\caption{Workflow diagram for quality assessment.}\label{workflow2}
\end{figure}

The pipeline collected a large amount of data that were essential to analyse further the detections and to create a process that is able to qualify the listed sources. First of all, the flux error provided by the annularSkyAperturePhotometry task was not intended to calculate the noise for bolometers in the presence of a strongly fluctuating background. Therefore we had to find methods that are able to give us a proper estimate of the noise. The workflow for quality assessment is presented in Figure~\ref{workflow2}. Our process used internal flags to indicate if an entry from the database matches our quality criteria, which we detail below.

\subsubsection{Structure noise based S/N}\label{sub:snr}

As it was described in the \href{http://archives.esac.esa.int/hsa/legacy/HPDP/SPIRE/SPIRE-P/SPSC/SPIREPointSourceCatalogueExplanatorySupplementFull20170203.pdf}{SPSC Explanatory Supplement}, a method was developed to give a proper statistical error of the photometry. As described in the Section~\ref{simulations} below, we injected sources into fields with various complexity. The same pipeline that we used to detect our sources and to collect photometry from real observations was used to detect our simulated sources and to measure their flux. The N$_S$ values were collected also in the same way. This procedure allowed us to compare the input flux to the measured flux as a function of the N$_S$. 

The signal to noise value depends on the source flux and on the noise, in our case the N$_S$ that is the combination of the instrument noise and the sky confusion noise. The theoretical flux ($F_{in}$) for each injected source is known and we have also the N$_S$ values for each. For a given $F_{in}$ level the $N_S$ was binned. The bin size was 25 MJy\,$sr^{-1}$. In each bin we calculated the ratio of the measured flux ($F_{out}$) and $F_{in}$. The standard deviation of the $F_{out}$/$F_{in}$ ratio ($\sigma_{F}$) is the photometric uncertainty, that we were looking for. By dividing $F_{in}$ by $\sigma_{F}$ we obtain the S/N. This S/N value is just for any given $N_S$ values. To calculate it for any $N_S$ levels, we fitted the following function to the data:
\begin{equation}
S/N=a\times N_S^b
\end{equation}

\begin{figure}[H]
\centering
\includegraphics[width=0.43\textwidth]{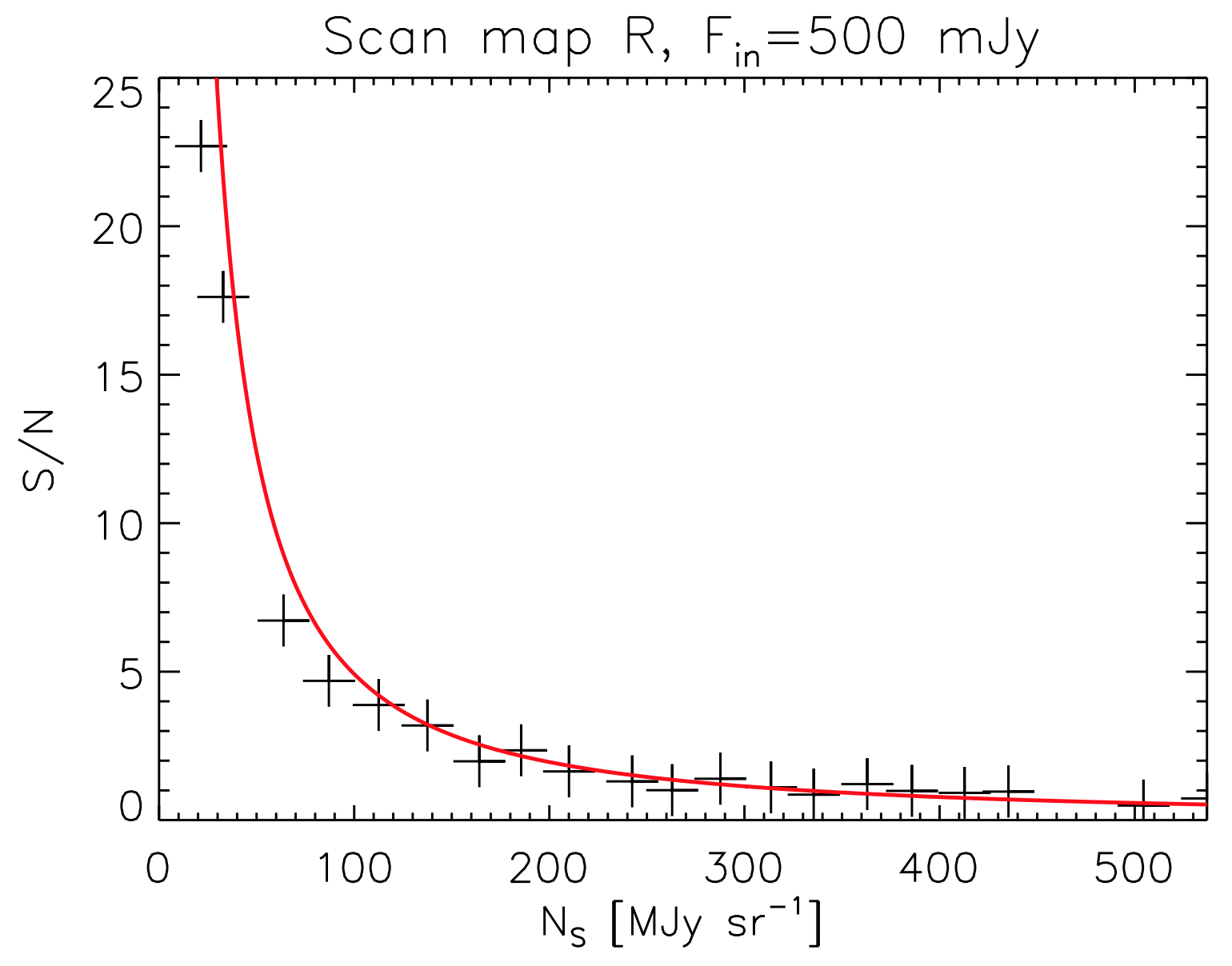}
\includegraphics[width=0.46\textwidth]{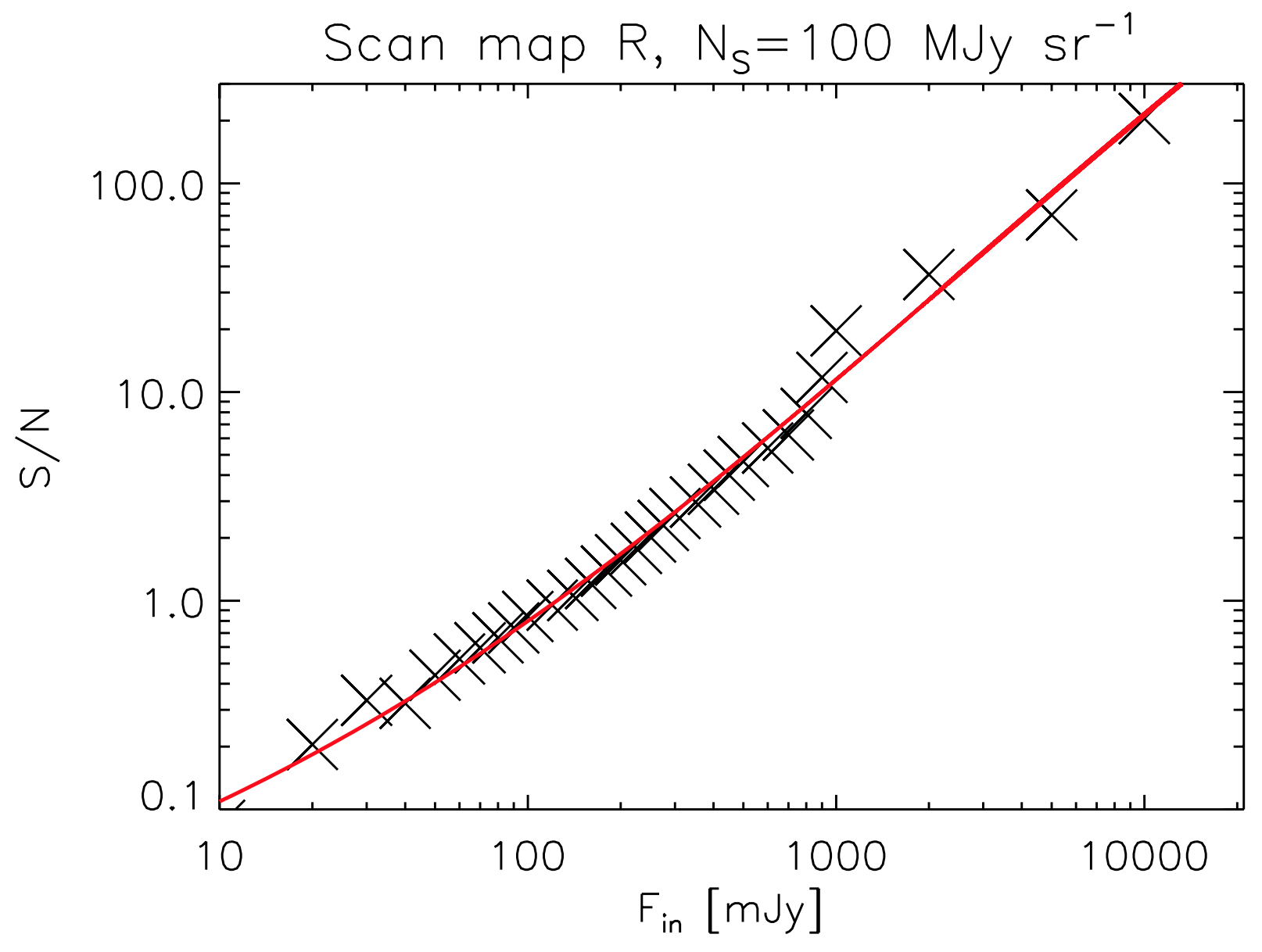}
\caption{Left: The photometric uncertainty ($\sigma_{F}$) as a function of the $N_S$ for sources with 500 mJy $F_{in}$ (black plus signs). Right: The photometric uncertainty ($\sigma_{F}$) as a function of the $F_{in}$ for 100 MJy\,$sr^{-1}$ $N_{S}$ (black plus signs). }\label{strnfin}
\end{figure}

The left panel on Figure~\ref{strnfin} shows the function fitted to the data points, allowing us to calculate the S/N value for any $N_S$ value. The data points were weighted by the number of elements in each bin. 

To have a grid sufficient sampled that can be used for interpolation, we created a one with steps of 1 MJy\,$sr^{-1}$ in the $N_S$ direction and 1 mJy in the $F_{in}$ direction. The above function was used to calculate the S/N values for all $N_S$ values on the grid, but only for the 32 input fluxlevels. To fill up the grid, we had to find a function that could be used to calculate the S/N for all flux levels at give $N_S$ level. The following equation was used:

\begin{equation}
S/N=a + \frac{b}{c + exp\frac{(x-d)}{e}}
\end{equation}

\begin{figure}[H]
\centering
\includegraphics[width=0.9\textwidth]{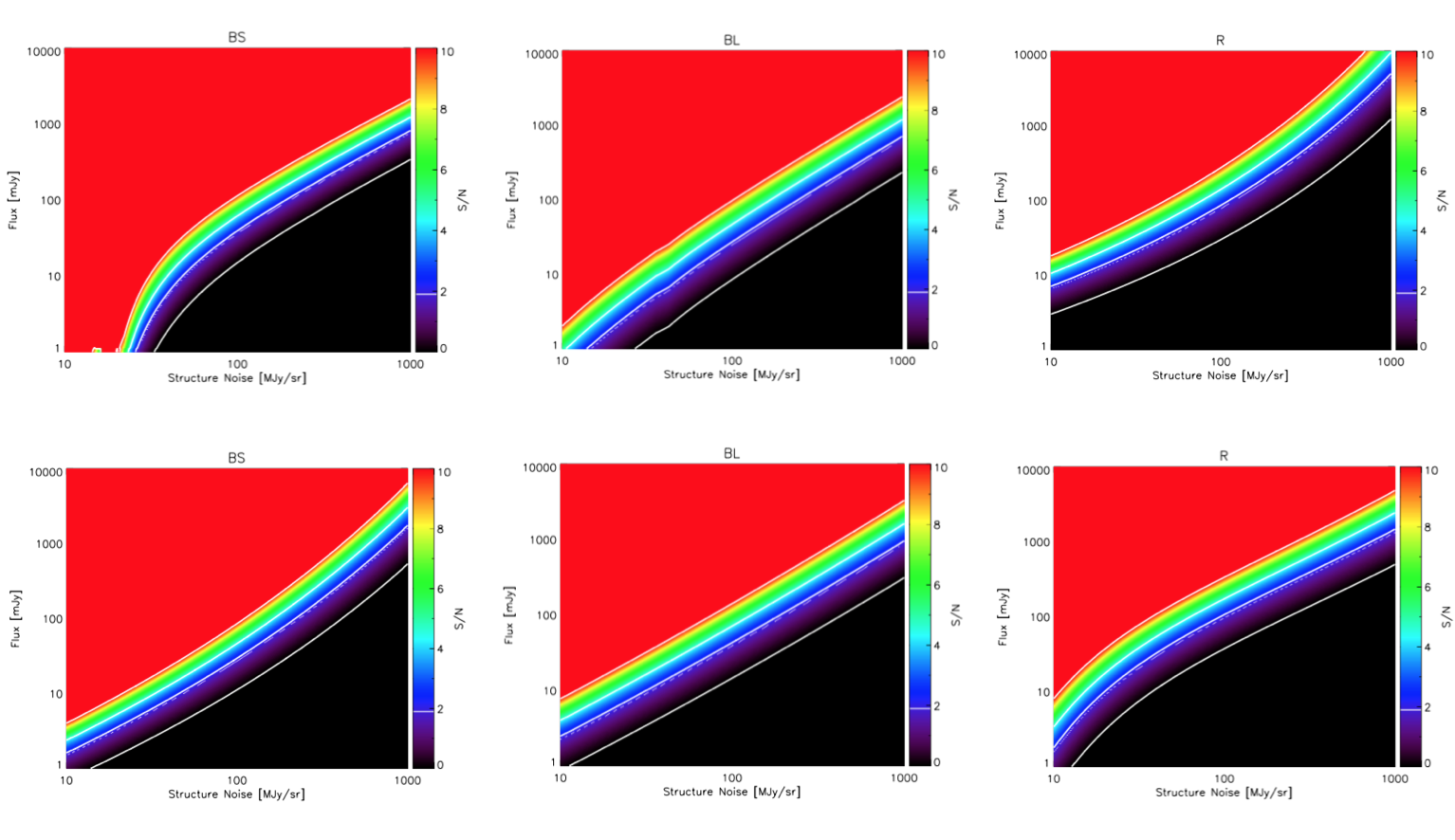}
\caption{The signal to noise ratio (S/N) surfaces for the different bands and observing modes. The upper row shows the S/N as a function of N$_S$ and flux in the 70$\mu m$, 100$\mu m$ and 160$\mu m$ bands (from left to right) for  Parallel Mode observations. Similarly, the bottom row shows the SNR surfaces for scan map observations. Solid white lines indicate S/N levels 1, 3, 5 and 10.}\label{snrsurfaces}
\end{figure}

The right panel of Figure~\ref{strnfin} shows the fitted function. After repeating the fit for all desired N$_S$ levels, the result is an array that can be used to calculate the S/N for the catalogue sources. Figure~\ref{snrsurfaces} shows the resulting arrays for each band and for the two observing modes. The S/N value for the objects in the catalogue was calculated by interpolation/extrapolation using the corresponding S/N array.

\begin{figure}[H]
\centering
\includegraphics[width=0.9\textwidth]{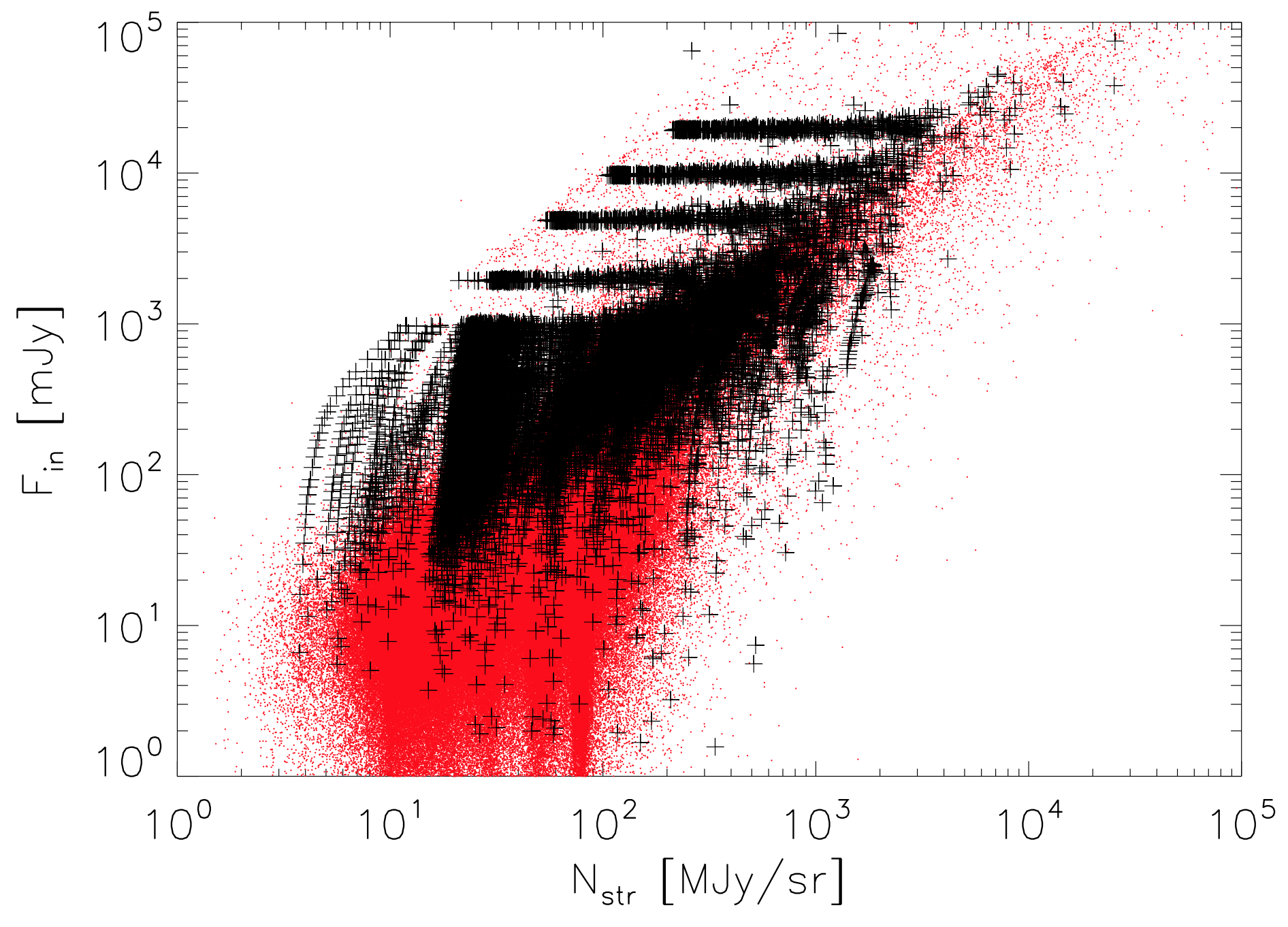}
\caption{The simulated (black plus sign) and Catalogue (red dots) sources on the N$_s$--Flux diagram.}\label{strncov}
\end{figure}

Although we created a large number of simulated maps and injected several thousand sources at each flux levels, our simulations cover only a part of the Flux--$N_{S}$ parameter space. Figure~\ref{strncov} shows where the recovered simulated sources are located on the plane. Our experience shows that the calculated S/N is reliable only between 0 and 2000 MJy\,$sr^{-1}$ $N_{S}$ and between 0 and 20 Jy source flux. Although these ranges cover $\sim$98\% of our sources, we used additional methods to estimate a proper S/N for each source, including the brightest ones that are important quality indicators for such a catalogue.

\subsubsection{Background RMS based S/N}\label{sub:stn}

A different way to estimate the properties of the noise around the source in an observation is to place various apertures around the source and to measure the flux in these apertures. The HIPE pipeline "scanmap Pointsources Photproject" follows this idea. The script creates high-pass filtered maps and then calculates the photometry at any given input coordinates. 

In order to speed up the up the procedure we implemented the photometric part as an IDL script. The core of the algorithm is to place apertures on a grid of 7$\times$7 points with an interval of 5 pixels, centred on the source. The flux was measured in each aperture and corrected for the EEF. An initial mean and standard deviation were calculated. Then, those apertures, in which the flux was higher than the mean plus twice the standard deviation, were excluded. The remaining apertures we term the blank apertures. The RMS of the background was calculated as the standard deviation of the flux measured in the blank apertures. 

\begin{figure}[H]
\centering
\includegraphics[width=0.9\textwidth]{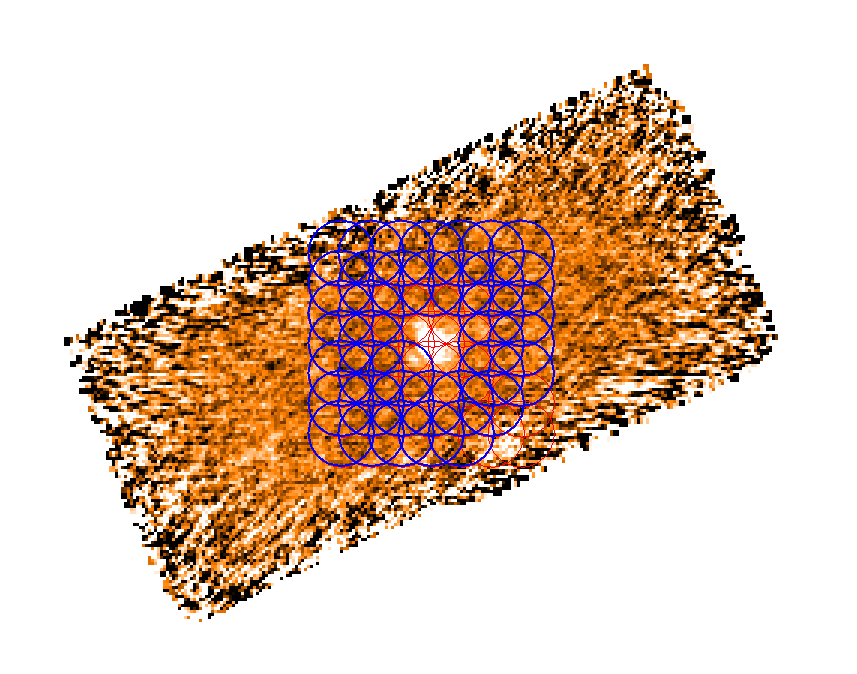}
\caption{A red band image of $\gamma$ Dra. The background RMS is calculated as the standard deviation of the flux measured in the blank apertures around the source (blue circles). }\label{blankapertures}
\end{figure}

\subsubsection{Fitted positions}

\begin{figure}[H]
\centering
\includegraphics[width=0.45\textwidth]{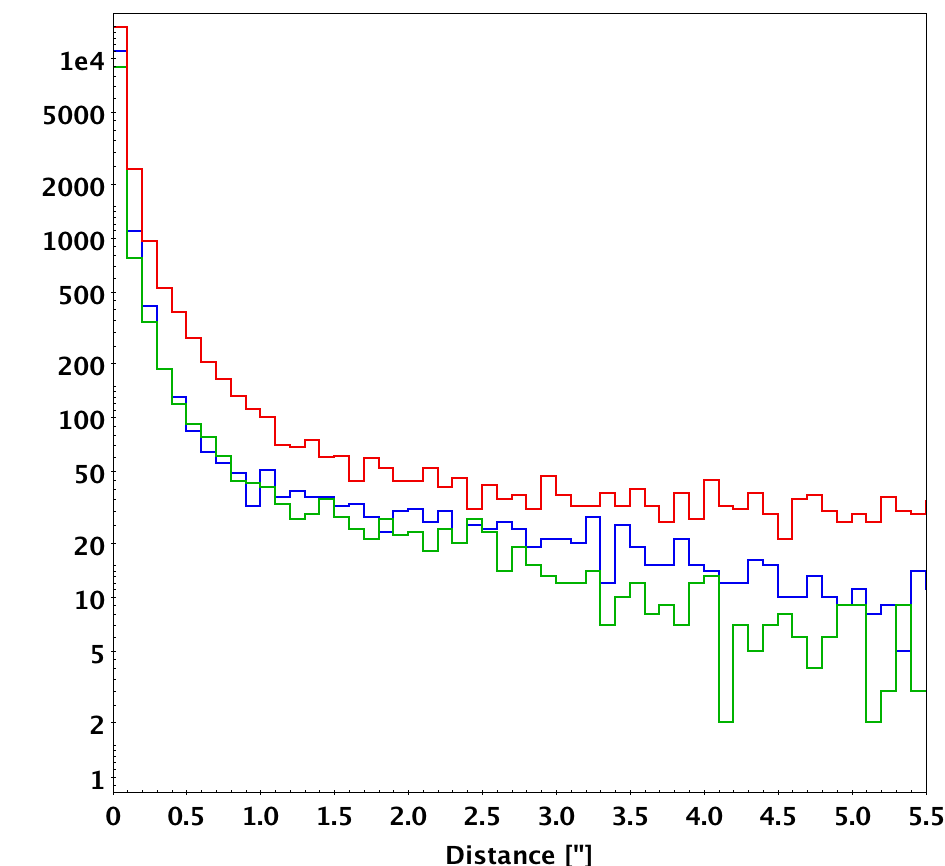}
\includegraphics[width=0.45\textwidth]{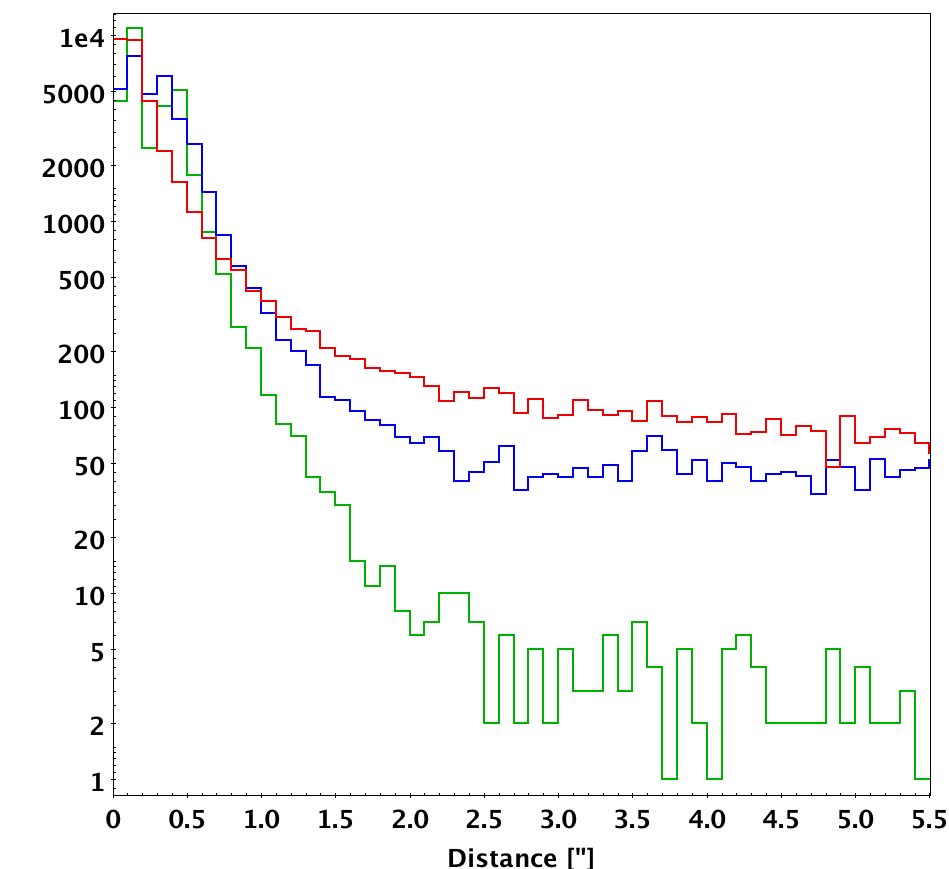}
\caption{Position uncertainty based on simulations for the scan map (left) and Parallel Mode (right). The uncertainty is measured as the angular distance between the SUSSEXtractor position and the 2D Gaussian fit source position. The blue, green and red bands are indicated with the corresponding colours. }\label{fitpos}
\end{figure}

A good quality indicator for our detections is the angular distance between the sky position provided by SUSSEXtractor and the position of the 2D Gaussian fit performed by the sourceFitting() task. We found that the distance between the two positions is very small for the reliable extractions. Figure~\ref{fitpos} shows the distribution of the separation for scan map observations (left panel) and for Parallel Mode observations (right panel).  

\subsubsection{Fluxes}
\begin{figure}[H]
\centering
\includegraphics[width=0.9\textwidth]{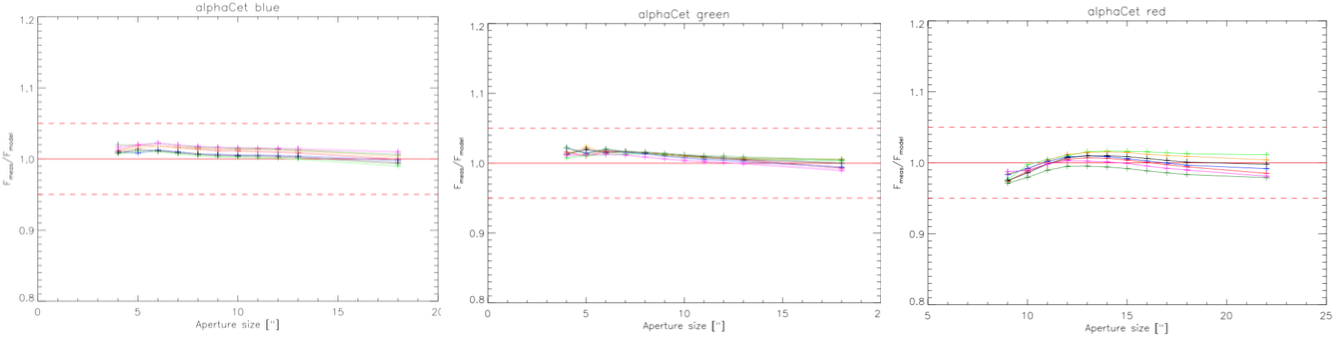}
\caption{Measured flux over the model flux for the calibration star HD18884 for different aperture size, plotted along the horizontal axis in the blue (left), green (middle) and red (right) bands. Observations made at different epochs are indicated with different colours. }\label{alphacet}
\end{figure}

As described in Section~\ref{sub:flux} the flux of our sources was measured with 11 apertures of different sizes, and the aperture correction applied to each. As shown for the calibration star HD18884 ($\alpha$Cet) in Figure~\ref{alphacet}, the flux measured in the different apertures is stable after applying the aperture correction, as only a few per cent difference can be seen. It is valid to assume that all sources having a PSF-like shape and flux distribution have a similar behaviour.

\subsubsection{FWHM}

\begin{figure}[H]
\centering
\includegraphics[width=0.9\textwidth]{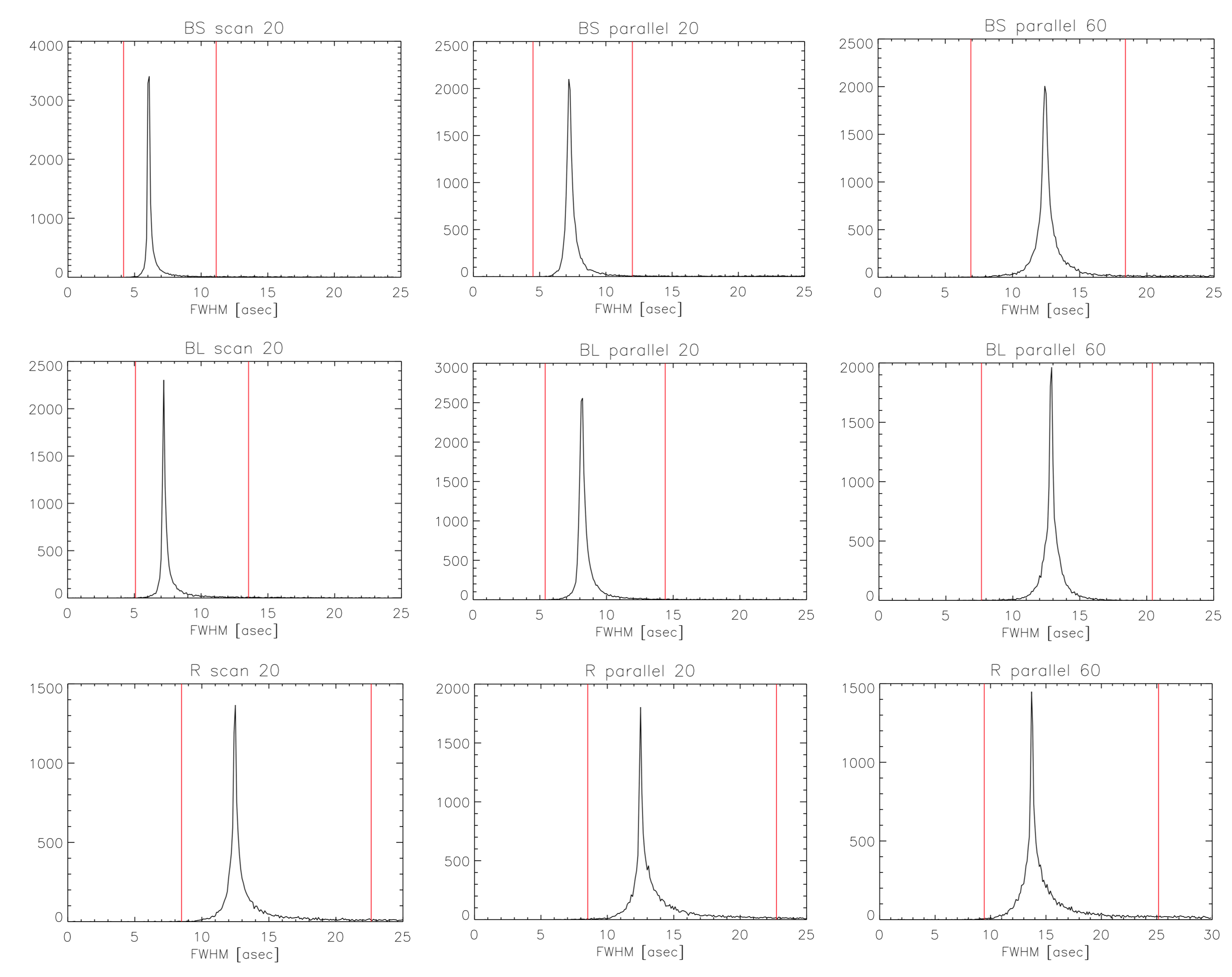}
\caption{Distribution of the FWHM values of our simulated sources in the different bands and at different scan speeds and observing modes. 0.75 and 2.0 times the PSF FWHM values are marked with red solid vertical lines.}\label{fwhmsim}
\end{figure}

The sourceFitting() task of HIPE  fitted a 2D Gaussian to our detections. If an ellipsoidal fit was possible, the fit resulted in two FWHM values, measured along the major and the minor axis. If only a circular Gaussian could be fitted then only one FWHM value was added to our database. These FWHM values were compared to the PSF FWHM values. Figure~\ref{fwhmsim} shows the FWHM value distribution for our artificial sources in different band and at different scan speeds in scan map and  Parallel Mode observations. Our simulations showed that both the shape of the  PSF and the measured FWHM can be affected by the background emission, especially at faint flux levels. 

\subsection{Object consolidation}\label{sub:consolidation}

In order to increase the reliability of our products, and to consolidate the final catalogue, we went through a process of objects consolidation. The first part  was the cleaning of our source tables. We analysed the parameters derived from the 2D Gaussian fit (centroid, FWHM, elongation), the signal-to-noise ratios of the sources, and their flux distribution.  The second part was to identify the multiple detections of the same astronomical object and to decide which entry should appears in the HPPSC.

\begin{figure}[H]
\centering
\includegraphics[width=0.5\textwidth]{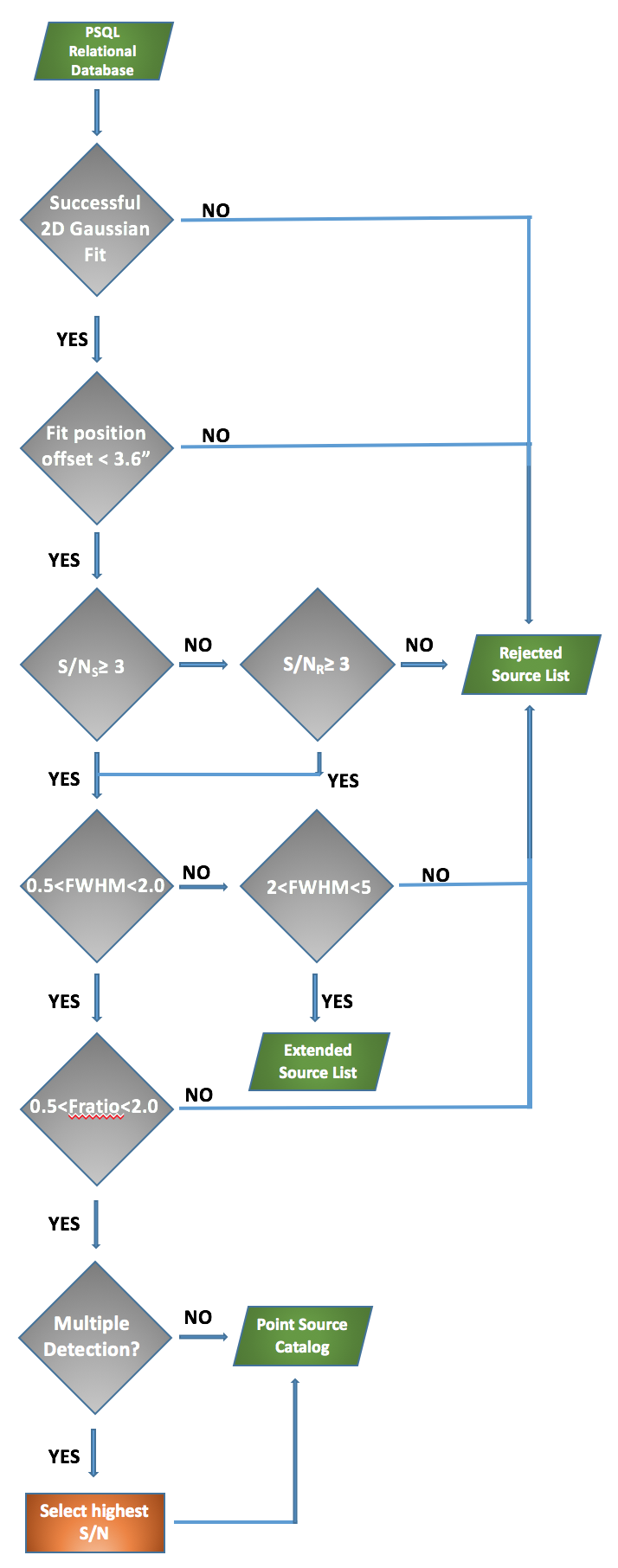}
\caption{The workflow of object consolidation. Criteria based on different properties of the sources were applied  and sources classified accordingly as point source, extended source, or rejected source.}\label{workflow3}
\end{figure}

\subsubsection{Cleaning source tables}

As stated previously, the cleaning process consists of imposing different conditions on the parameters obtained in previous steps (pipeline procedure and quality assessment) to classify all detected objects in our list as point, extended  or rejected  sources (see the workflow of object consolidation in Figure~\ref{workflow3}).\\

The first step is to confirm that a 2D Gaussian fit is achieved.  If  the fit  does not converge, the object is  automatically  rejected, and add to  the HPRSL. Of the sources  in our initial  list (8\,254\,064),  1\,041\,616 objects were discarded at this point.\\

The second condition is to examine the difference between the coordinates derived by the SUSSExtractor and the sourceFitting task. The distance between them has to be less than 3.6 $\arcsec$, i.e,  three times the mean \textit{Herschel} pointing accuracy. As shown Figure \ref{fitpos}, this limit is  supported by the positional uncertainty from simulations. Again, if  this  criterion is not  satisfied,  we do not consider this source as a good candidate  and we move it to the HPRSL. \\

As described in Section~\ref{sub:qa}, we provide two different methods for the estimation of the signal--to--noise ratio. Our condition is fulfilled if at least one of them gives S/N > 3. All sources with both S/N ratios below 3 are rejected and are included in the HPRSL. Check the columns S/N$_S$ and S/N$_R$ in the HPPSC and HPESL to identify which criteria are met.\\

As the next step, we analyse the FWHM of the 2D Gaussian fit in order to reject other possible spurious sources, although mainly to classify the source as "point" or "extended". An object is  considered a point source if the value of FWHM satisfies the criterion that:
\begin{equation}
    0.75 \times PSF < FWHM < 2\times PSF
\end{equation}

The thresholds were determined using a FWHM based on simulations (see Figure \ref{fwhmsim}).  
As explained previously, the 2D Gaussian fit provides values for the FWHMs in both the X and the Y direction. If the 2D Gaussian is elliptical, the FWHM values in both axes need to meet this criterion. The PSFs were taken from Table \ref{tab:psffwhm}. Alternatively,  if a source is not included in the HPPSC and the FWHM is between 2 and 5 times the tabulated PSF, it is  added to  the Extended Source List, all other sources are rejected. 
After this step,  the HPESL is  complete.  The final number  of extended  sources identified is 546\,587.\\

\begin{figure}[H]
\centering
\includegraphics[width=0.45\textwidth]{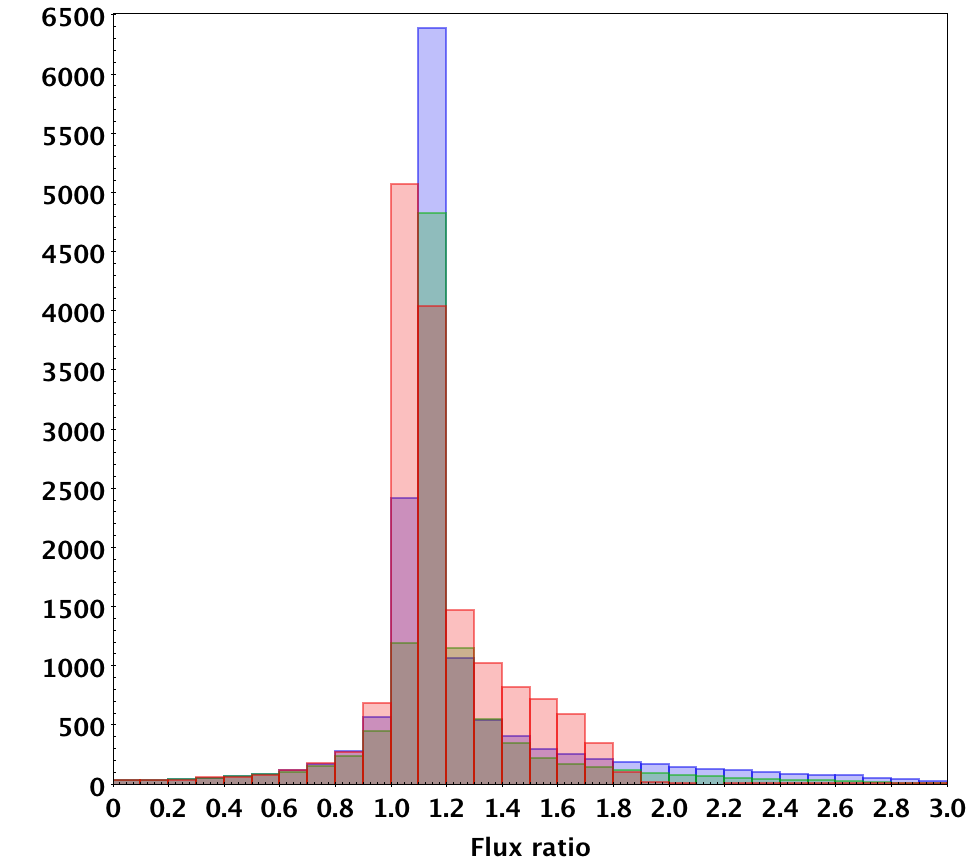}
\includegraphics[width=0.45\textwidth]{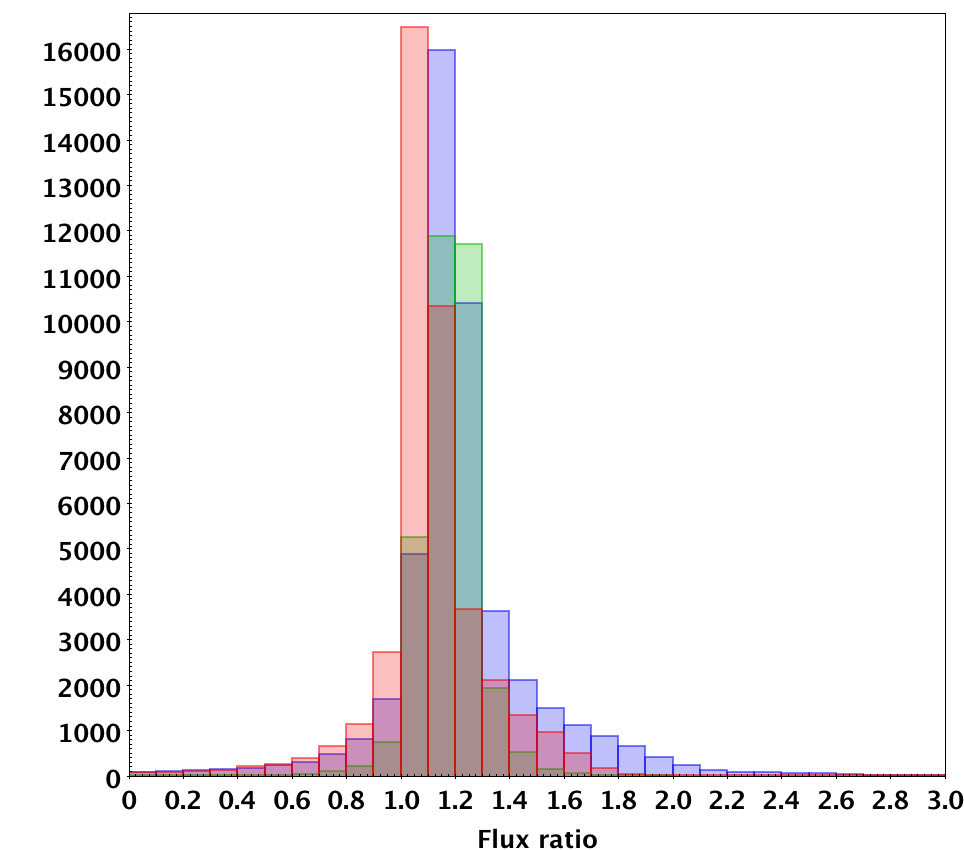}
\caption{Histograms of ratio fluxes (aperture6/ aperture1, see text) for the scan map (left) and Parallel Mode (right). The blue, green and red bands are indicated with the corresponding colors}\label{fluxratios}
\end{figure}

The final condition takes for the flux distribution of the source. To do this, we check the ratio between two fluxes measured from two different apertures, aperture 6 (9$\arcsec$ for blue and green bands, 14$\arcsec$ for the red) and aperture 1 (4$\arcsec$ for blue and green bands, 9$\arcsec$ for the red band). To include an object in our final point source catalogue, this flux ratio must be between 0.5 and 2. These limits are established using results from simulations. Figure \ref{fluxratios} shows the flux ratios obtained to simulated sources, for both Scan and Parallel mode. Note that this condition does not apply to extended sources. At  this point, the number of consolidated point sources is 522317. However, the catalogue still contains duplicated sources.  In the next section, we explain the procedure used to  group  these duplicated objects to a  single  source detection in the HPPSC.

\subsubsection{Multiple detections}
Once the Source List is consolidated using the quality criteria described above, a further step in necessary to generate the final catalogue table. The source extraction is performed on a map repository that includes, the highest processing Level available within the HSA for every sky region, and does not prevent multiple detection to the same source appearing in different sky maps. To identify single objects with multiple detections in overlapping maps, sources belonging to different observations and 
separated by a smaller distance on the sky than within the PSF FWHM are grouped. Within each group, only the source with the highest S/N (the average between RMS S/N and Structure Noise S/N is taken - see Sections~\ref{sub:snr}, ~\ref{sub:stn}) is included in the final catalogue table. The PSF values are those given in Table~\ref{tab:psffwhm}, they depend strongly on the observing mode, while the absolute pointing error for SPG13 (1.2$\arcsec$) is adopted as uncertainty.

\newpage

\section{Point Source Catalogue}
\begin{figure}[H]
\centering
\includegraphics[width=1.0\textwidth]{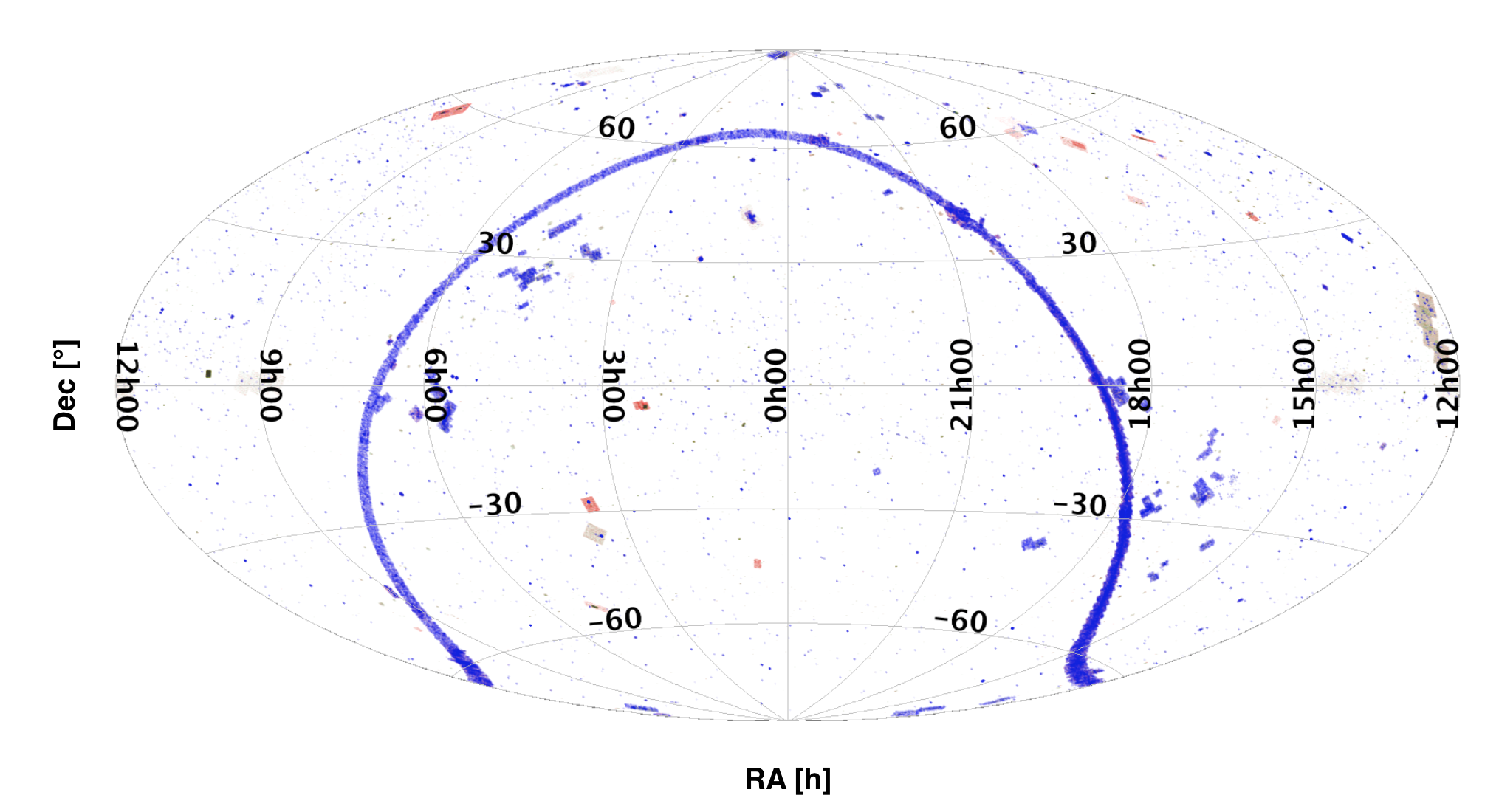}
\caption{All-sky Aitoff projection of the HPPSC sources across the three PACS bands.}\label{allsky}
\end{figure}

\subsection{Catalogue properties}
The objects of the \textit{Herschel}/PACS Point Source Catalogue has been identified  selecting carefully the parameters that best define the point sources and applying our filtering criteria to the raw source table based on positional uncertainties, FWHM, S/N, flux ratio and elongation to eliminate spurious sources and extended sources. 
The HPPSC is presented in three comma-separated value (CSV) files with headers that identify the columns. We include 251\,392, 131\,322 and 108\,319 sources in the red, green and blue bands, respectively. Detailed numbers of sources included in the different tables of the HPPSC are found in Table~\ref{hppscnumbers}. The sky coverage of the extracted sources is shown in Figure~\ref{allsky}, using the J2000 equatorial system in an Aitoff projection. The Galactic Plane is clearly visible as a blue stripe on the image, indicating area covered by the observations of the \href{http://iopscience.iop.org/article/10.1086/651314/pdf}{Hi-GAL} Key Program. 

\begin{table}[H]
\centering
\small\addtolength{\tabcolsep}{-3pt}
\begin{tabular}{|l|ccc|ccc|ccc|}
\hline
	 &\multicolumn{3}{c|}{blue} &\multicolumn{3}{c|}{green} &\multicolumn{3}{c|}{red}\\
\hline
&HPPSC&HPESL&HPRSL&HPPSC&HPESL&HPRSL&HPPSC&HPESL&HPRSL\\
\hline
Scan map 10$\arcsec s^{-1}$ &319&143&12728&391&72&9532&435&219&9810\\
Scan map 20$\arcsec s^{-1}$ &17010&10573&329657&87278&23859&1211037&46231&37597&896410\\
Scan map 60$\arcsec s^{-1}$ &177&175&5874&270&42&5682&441&578&6884\\
Parallel Mode 20$\arcsec s^{-1}$ &3343&11534&118274&7497&6043&63870&44325&48438&832131\\
Parallel Mode 60$\arcsec s^{-1}$ &87470&145634&1680660&35886&22525&479784&159960&239155&1554111\\
\hline
\end{tabular}
\caption{The number of sources at different scan speeds and in different observing modes in the products.}
\label{hppscnumbers}
\end{table}

\begin{figure}[H]
\centering
\includegraphics[width=0.46\textwidth]{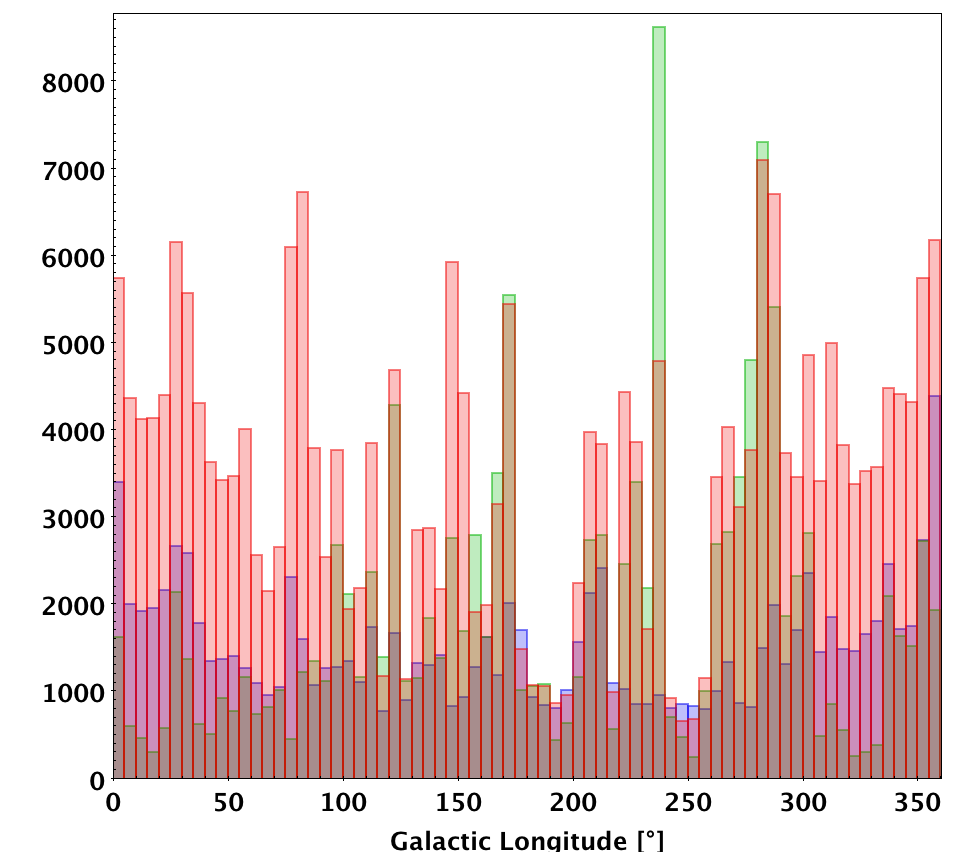}
\includegraphics[width=0.46\textwidth]{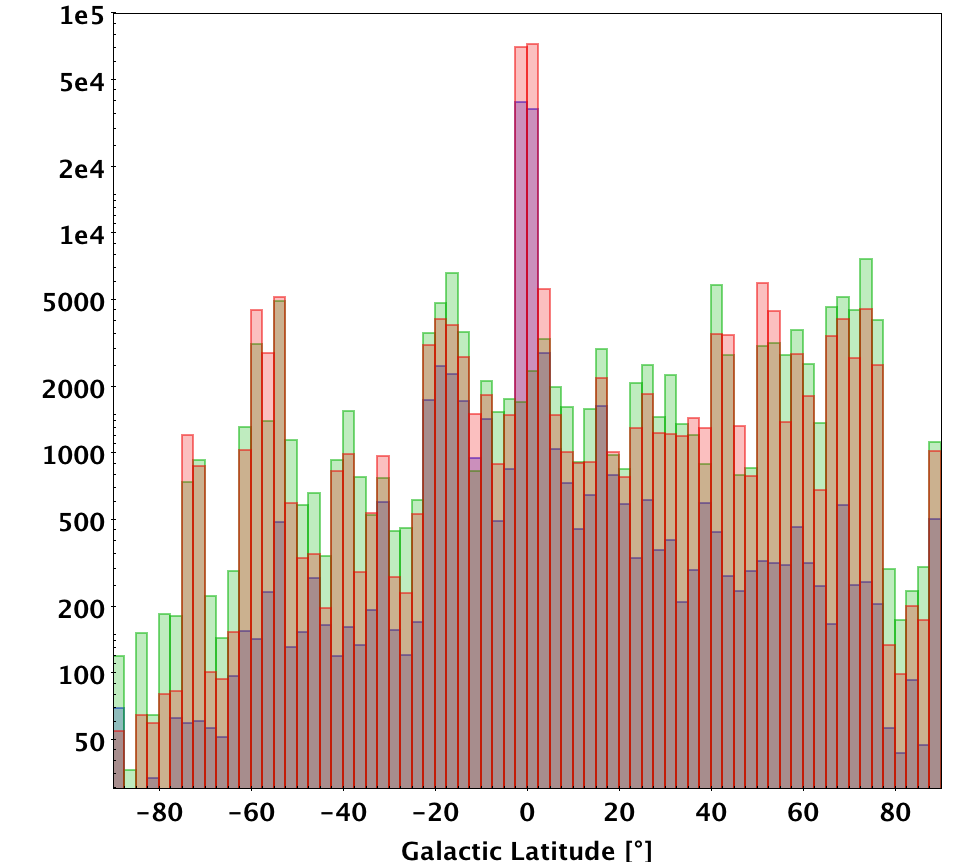}
\caption{The HPPSC source distribution in the blue, green and red bands, indicated with the corresponding bar colours, as a function of the Galactic latitude and longitude.}\label{lbhist}
\end{figure}

Figure~\ref{lbhist} shows the distribution of our HPPSC sources as a function of Galactic Longitude (left) and Latitude (right). On the Galactic latitude figure the peaks in the blue and red bands are caused by the Hi-GAL sources. An excess of source counts is visible in the green and red bands between $\sim$40$\degr$ and $\sim$60$\degr$, caused by the large extragalactic fields observed for \href{http://www.hevics.org/}{HeViCS}. Other large surveys can be also identified based on the source distribution.

\begin{figure}[H]
\centering
\includegraphics[width=0.32\textwidth]{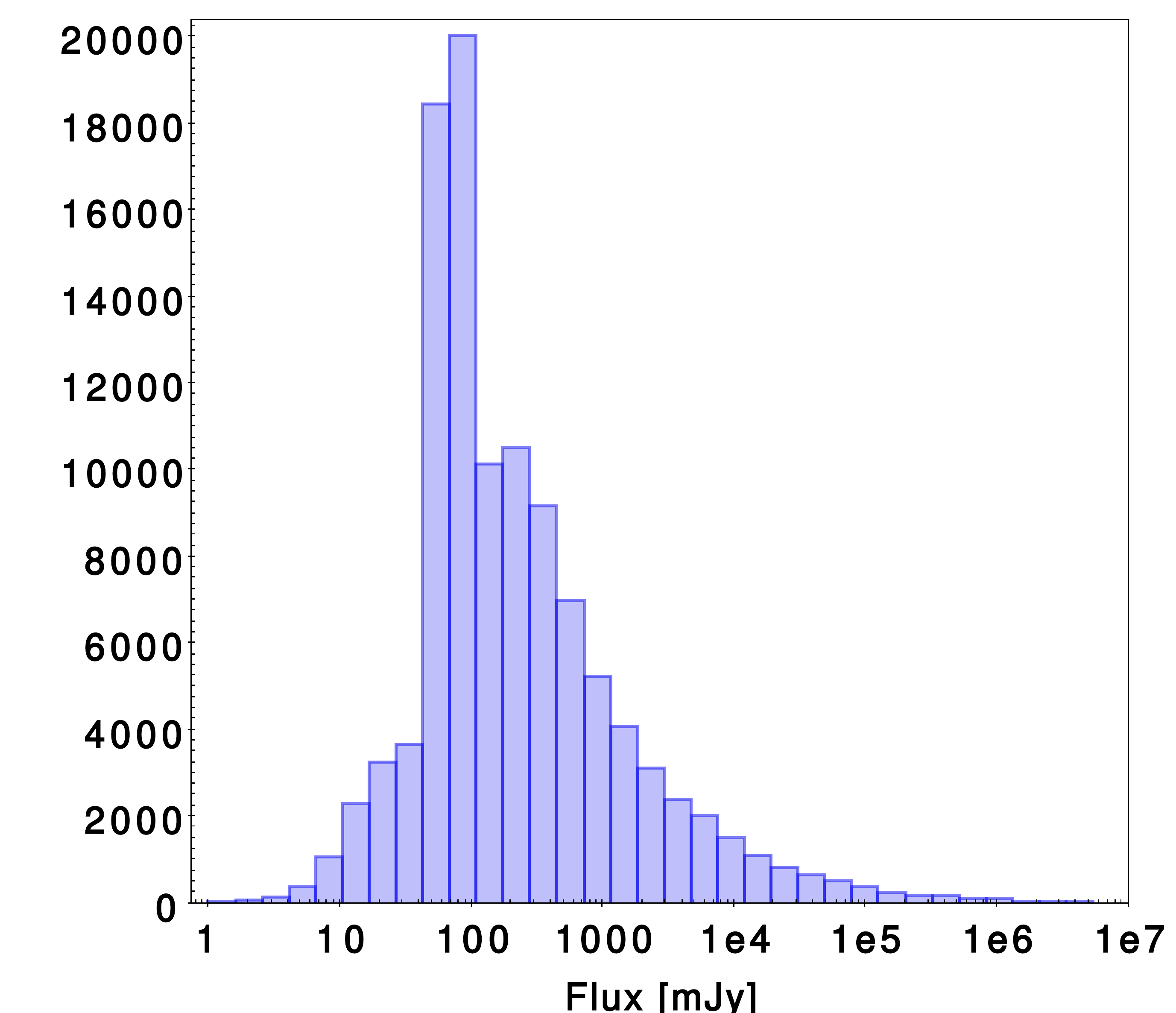}
\includegraphics[width=0.32\textwidth]{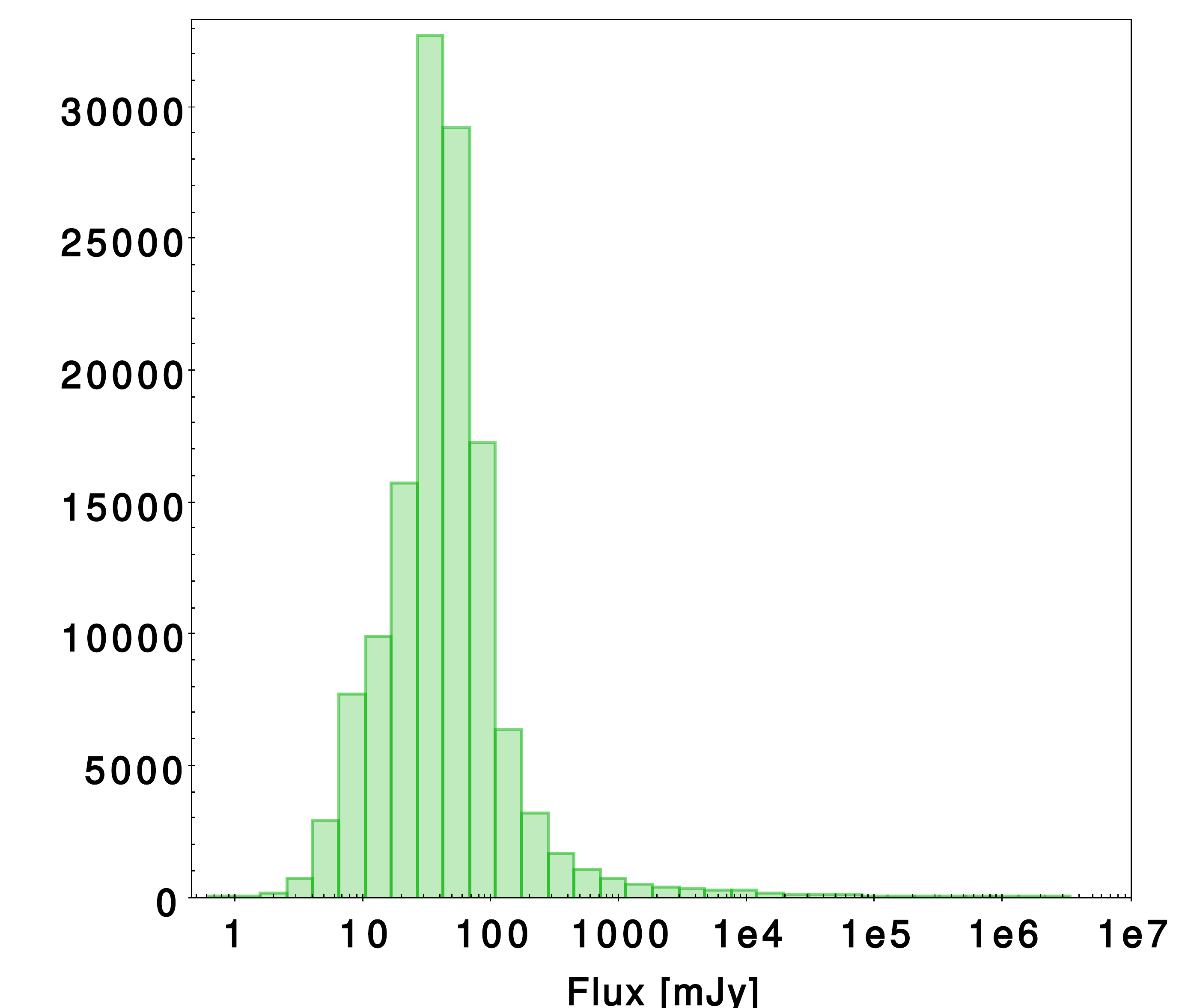}
\includegraphics[width=0.32\textwidth]{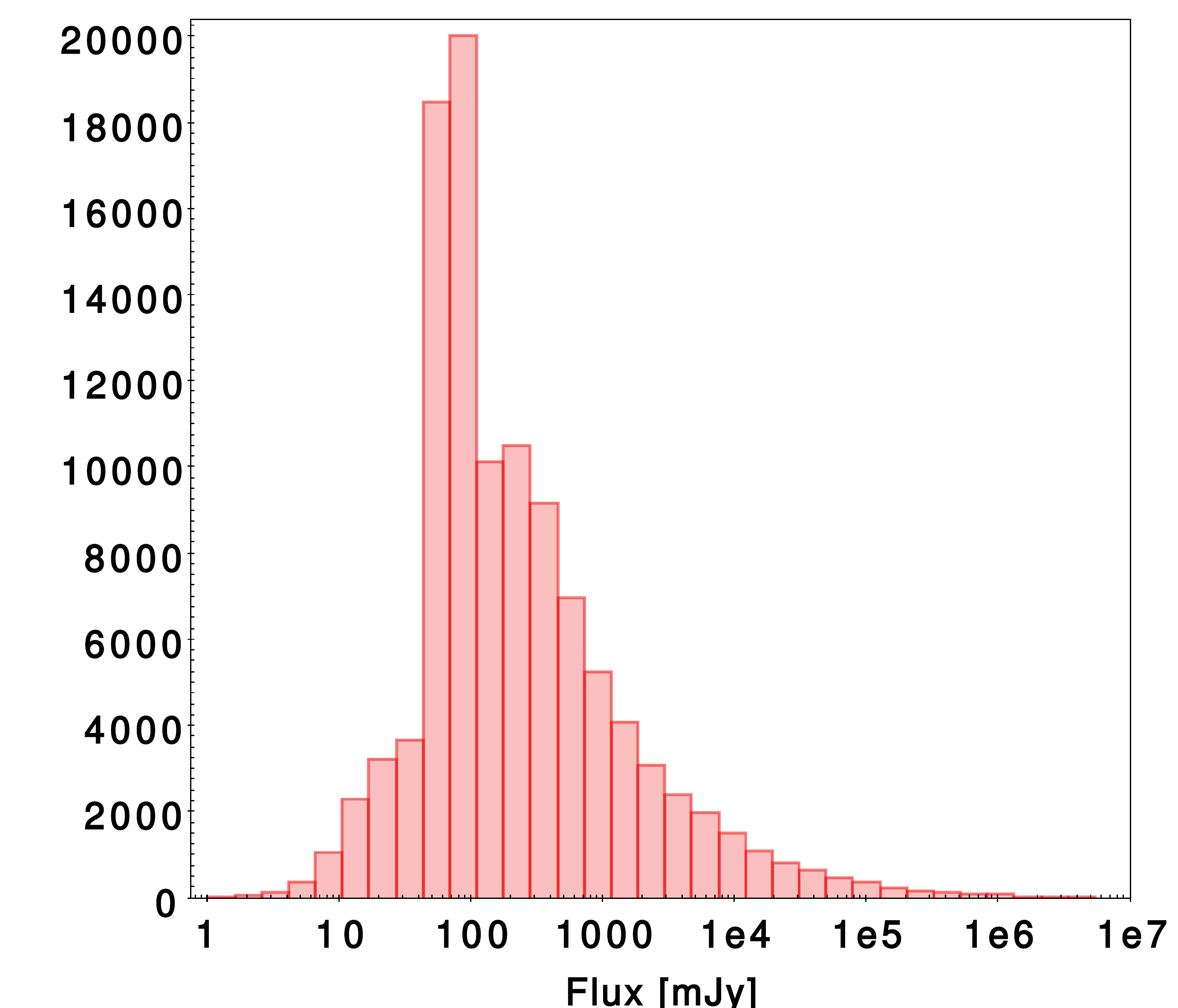}
\caption{Flux distribution in the blue, green and red bands, indicated with the corresponding colour from left to right. }\label{fluxhist}
\end{figure}

The flux distribution of our Catalogue objects is shown on Figure~\ref{fluxhist}. In all bands the flux distribution has a peak at $\sim$90 mJy. The blue and the red bands have an excess of bright sources compared to the green band. This is due to the fact that observations using the green/red filter combinations were performed in  significantly smaller numbers close to the Galactic Plane, where sources have to be bright to be detected and reliably extracted against the bright background.

\begin{figure}[H]
\centering
\includegraphics[width=0.32\textwidth]{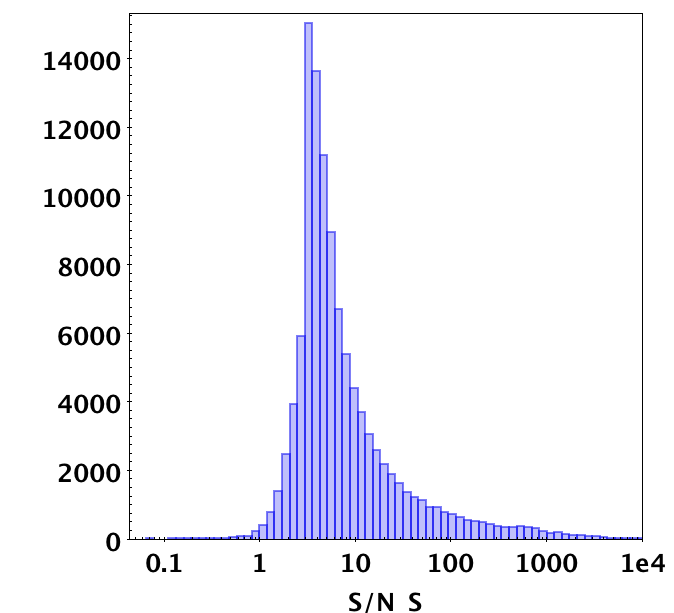}
\includegraphics[width=0.32\textwidth]{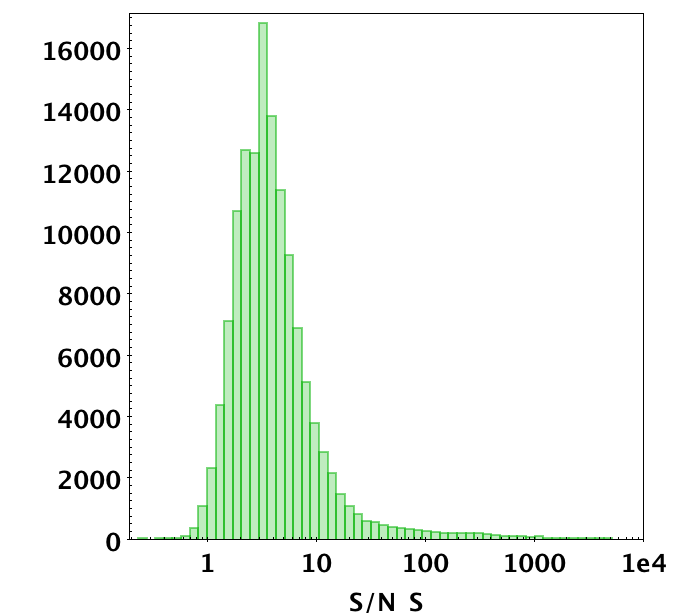}
\includegraphics[width=0.32\textwidth]{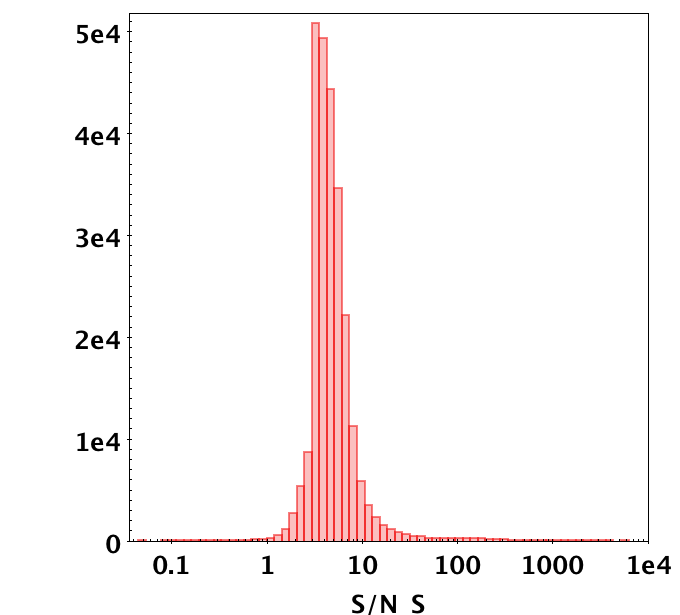}
\caption{S/N$_S$ distribution in the blue, green and red bands, from left to right.}\label{snrhist}
\end{figure}

The distribution of the S/N$_S$ values (i.e. the signal--to--noise ratios estimated from the simulations, calculated as a function of the N$_S$ and the source flux) have a peak between 3 and 5. Apparently the S/N$_S$ values in the blue band are more distributed towards higher values than in the green band.The red band S/N$_S$ values have the narrowest distribution, see Figure~\ref{snrhist}.

\begin{figure}[H]
\centering
\includegraphics[width=0.32\textwidth]{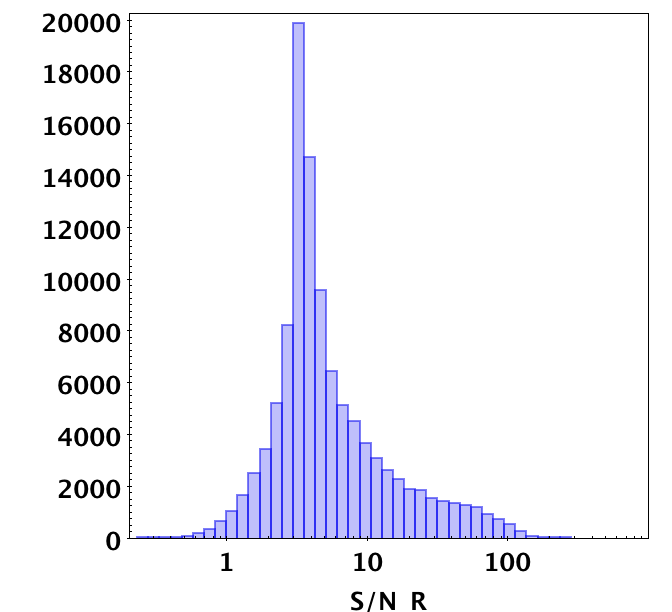}
\includegraphics[width=0.32\textwidth]{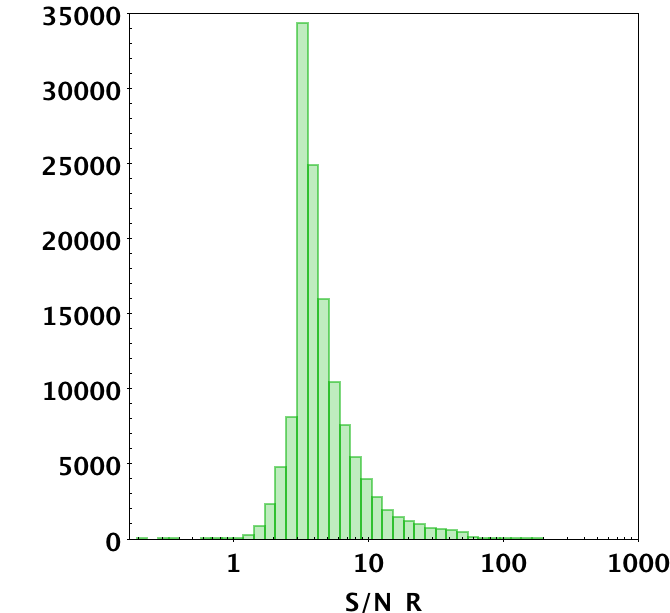}
\includegraphics[width=0.32\textwidth]{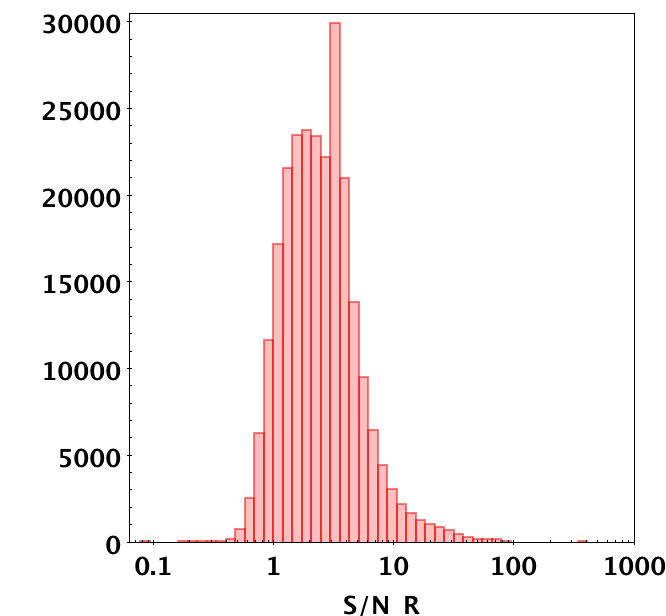}
\caption{S/N$_R$ distribution in the blue, green and red bands, from left to right.}\label{stnhist}
\end{figure}

Figure~\ref{stnhist} shows the distribution of the background RMS based S/N$_R$ values. It shows similarities with the  S/N$_S$ distribution as values in the blue band occupy a larger interval, and have an excess towards high values, while in the green and the red bands the S/N$_R$ distribution is more concentrated.

\begin{figure}[H]
\centering
\includegraphics[width=0.32\textwidth]{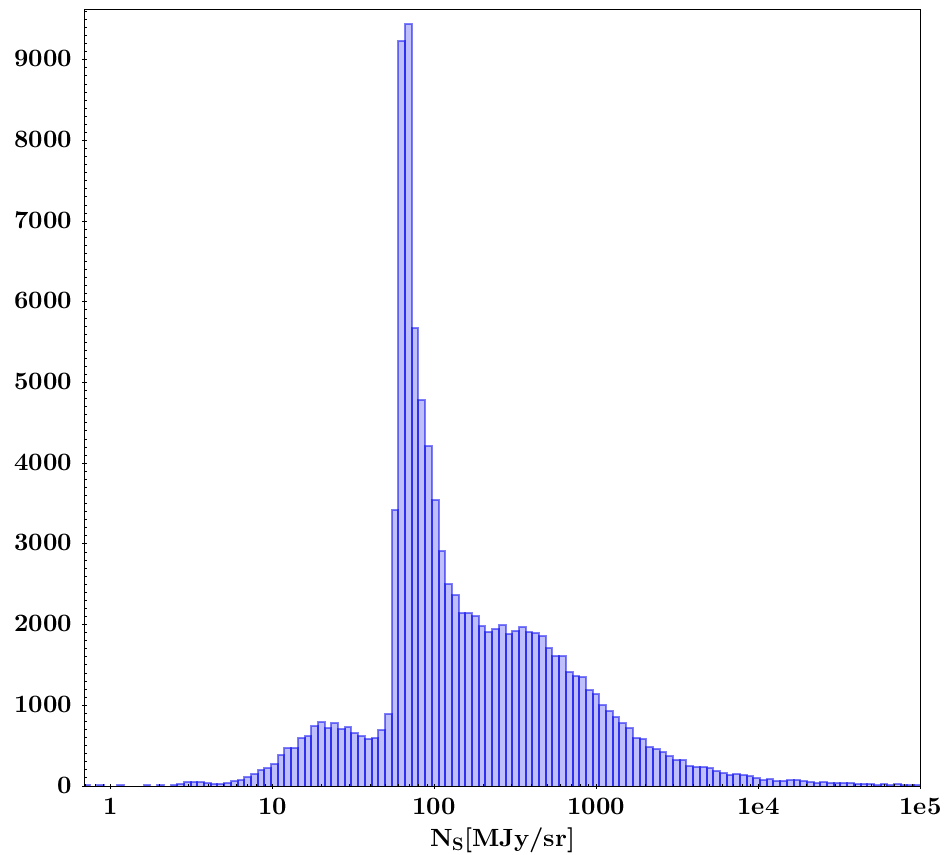}
\includegraphics[width=0.32\textwidth]{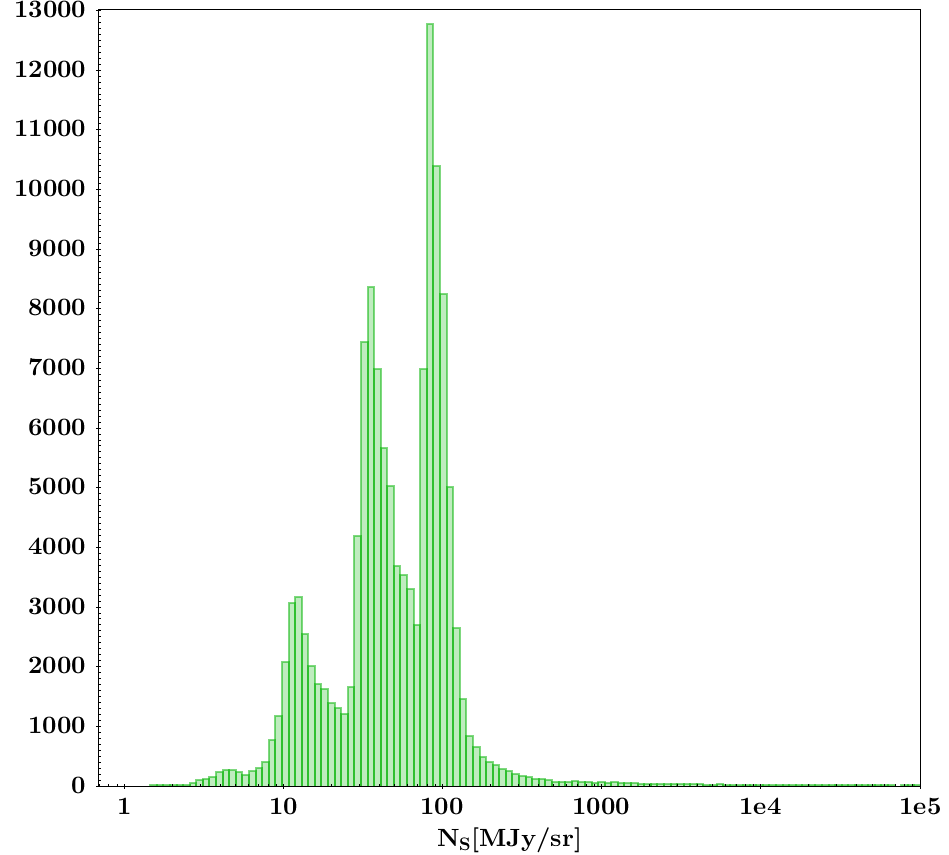}
\includegraphics[width=0.32\textwidth]{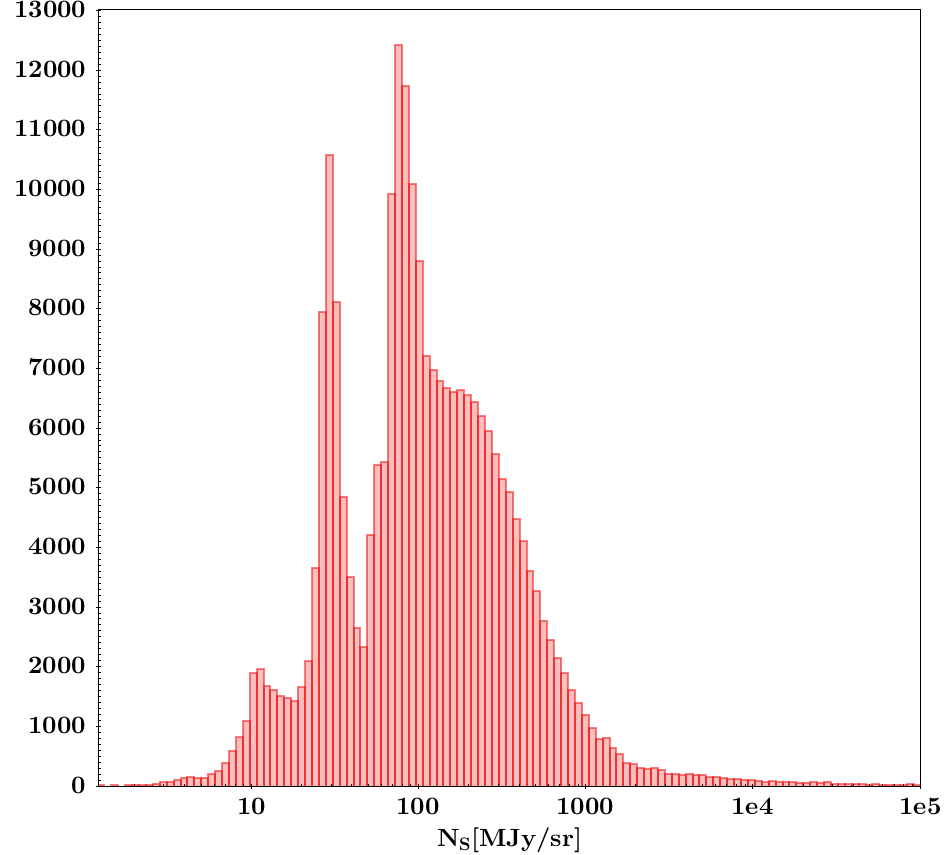}
\caption{Distribution, from left to right, of the N$_S$ values in the blue, green and red bands.}\label{strncomparison}
\end{figure}

\begin{figure}[H]
\centering
\includegraphics[width=0.9\textwidth]{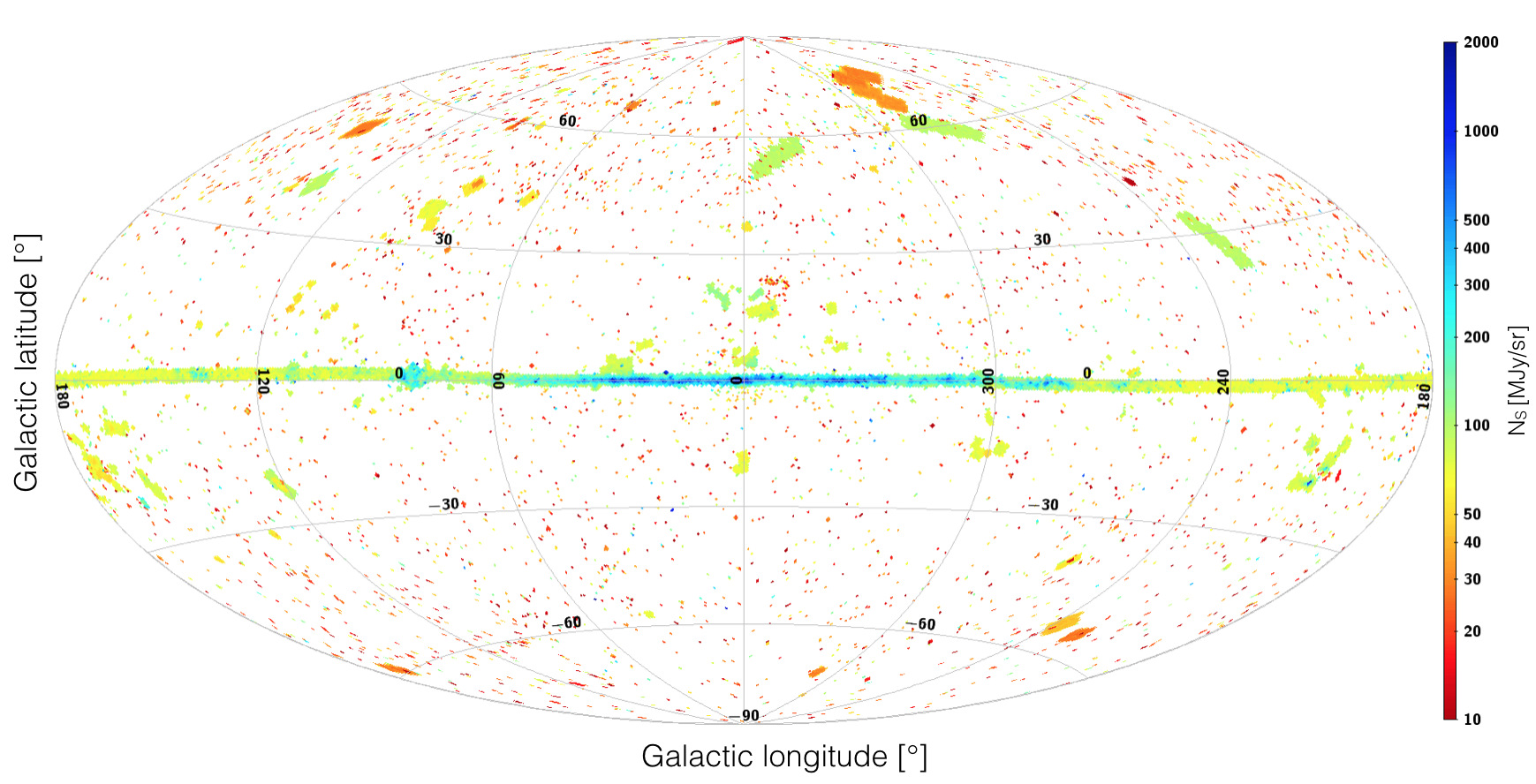}
\caption{Distribution of the red band N$_S$ values on the sky in Galactic coordinates, using an Aitoff projection. Colours from red to blue indicate the increasing structure noise.}\label{strnsky}
\end{figure}

The distribution of the N$_S$ (structure noise) values is plotted on Figure~\ref{strncomparison}. In the blue band the distribution peaks at $\sim$20, 90 and 400 MJy sr$^{-1}$. The same behaviour is present in the green and red bands. These values identify clearly regions of different complexity. 
Figure~\ref{strnsky} shows the N$_S$ values projected in the sky in Galactic coordinate system. The high values are concentrated along the Galactic plane and around the Galactic Centre.

We have summarised the distributions of the above quantities in Table~\ref{hppscmeans}. 


\begin{table}[H]
\centering
\begin{tabular}{|l|ccc|ccc|ccc|}
\hline
	 &\multicolumn{3}{c|}{blue} &\multicolumn{3}{c|}{green} &\multicolumn{3}{c|}{red}\\
\hline
&5\%&Median&95\%&5\%&Median&95\%&5\%&Median&95\%\\
\hline
Flux [mJy] &21.2&138.8&8057.1&7.9&40.7&231.7&43.4&271.5&4393.4\\
S/N$_S$  &2.1&5.1&127.7&1.4&3.5&19.1&2.6&4.4&11.1\\
S/N$_R$  &1.6&4.1&42.9&2.2&3.9&14.6&0.9&2.4&8.6\\
N$_S$ [MJy sr$^{-1}$] &19.8&110.3&1940.3&10.1&50.0&129.5&14.1&98.9&790.8\\
\hline
\end{tabular}
\caption{5\%, 50\% and 95\% quantiles of some quantities characterising our Catalogue.}
\label{hppscmeans}
\end{table}

\subsection{HPPSC columns}
\subsubsection{Name (\textit{name})}
The object name is an identifier consisting of two parts, acronym and sequence, and is formatted as 'HPPSCxxxA\_Jhhmmss.s$\pm$ddmmss'. 'HPPSC' stands for \textit{Herschel} PACS Point Source Catalogue, xxx is the central wavelength of the band in which the source was observed, identifying  the wavelength in $\mu$m, and the letter A  indicates the first version of the catalogue. The second part has the format 'Jhhmmss.s+ddmmss' derived from the source coordinates with 'J' indicating a J2000 reference system, 'hhmmss.s' is the Right Ascension in hours, minutes and seconds accuracy of one digits after the decimal, and '+ddmmss' is the Declination in signed degrees, minutes and seconds to an accuracy of one digit after the decimal.

\subsubsection{Band (\textit{band})}
This is one of the strings "blue","green","red" identifying the PACS detector array names for the filter bands centred at 70, 100, 160 $\mu$m, respectively. The wavelength is also specified in the name of the object. 

\subsubsection{R.A./Dec (\textit{ra/dec})}
Right Ascension/ Declination coordinate in a J2000.0 reference frame. These are the peak position of the SUSSEXtractor detection. The coordinates are degrees written in double precision format. 

\subsubsection{RA$_{err}$/Dec$_{err}$ (\textit{raerr/decerr})}
Positional uncertainties of each object as provided by SUSSEXtractor. Units are degrees, the data type is double precision. 

\subsubsection{Flux (\textit{flux})}
This is the flux as provided by the AnnularSkyAperturePhotometry task. For objects that were identified several times across different maps, this flux corresponds to the extraction with the best average S/N, where the N$_S$ based S/N (S/N$_S$)and the background RMS based S/N (S/N$_R$) values are taken into account. For the blue and green bands the aperture radius was 6$\arcsec$, for the red band 12$\arcsec$ (see  Section~\ref{pipeline} for more details). For all cases the raw flux density values were corrected for the aperture size, based on \href{http://herschel.esac.esa.int/twiki/pub/Public/PacsCalibrationWeb/bolopsf_22.pdf}{Lutz (2015)}. Colour corrections were not applied. The units are mJy, written as double precision.



\subsubsection{S/N$_{S}$ (\textit{snr})}
Estimated S/N based on a statistical approach, using the S/N$_S$ calibration surfaces. It is calculated as a function of the source flux and the N$_S$ value. The method is described in details in Section~\ref{sub:snr}. The data type is double precision. Units are mJy.

\subsubsection{N$_{S}$ based noise (\textit{snrnoise})}
It is calculated from the source flux and the estimated S/N$_S$, by dividing the flux by the S/N$_S$. This noise includes both the sky confusion and the instrument noise.

\subsubsection{S/N$_{R}$ (\textit{stn})}
Calculated as the ratio of the source flux and the RMS. Data type is double precision. 

\subsubsection{RMS (\textit{rms})}
RMS noise as measured in the blank sky apertures placed around the source (sse Section~\ref{sub:stn}). Double precision. Units are mJy.



\subsubsection{N$_{S}$ (\textit{strn})}
Structure noise value obtained from the structure noise maps as described in Section~\ref{sub:snr}. Double precision, in units of MJy\,$sr^{-1}$. 

\subsubsection{Flux ratio (\textit{fratio})}
The ratio of the flux density values measured in the 6th and the 1st aperture, corresponding to  4$\arcsec$ and 9$\arcsec$ in the blue and green bands, and  9$\arcsec$ and 14$\arcsec$ in the red band. 
\subsubsection{FWHM$_X$/FWHM$_Y$ \textit{fwhmxfit/fwhmyfit}}
This column contains the FWHM values measured by the 2D Gaussian model in X and Y directions, fitted by the sourceFitter() task. If the elongated Gaussian fit was possible then both FWHM$_X$ and FWHM$_Y$ are listed. If only a circular Gaussian could be fitted then FWHM$_Y$ is equal to -99.9. It is written in double precision format, in units of second of arc.
\subsubsection{ELONGATION FLAG  (\textit{flag\_elong})}

If an elliptical Gaussian model was fitted in the source, this column contains the ratio of the FWHM$_Y$ and FWHM$_X$. A value of 0 means that the source is circular, 0.5 means that FWHM$_Y$ = 0.5$\times$ FWHM$_X$. If FWHM$_Y$ = -99.9 (circular detection), FLAG\_ELONG = 0. FLAG\_ELONG is also set to 0 if the source eccentricity (FWHM$_X$ - FWHM$_Y$)/FWHM$_X$) defined in a similar way with the FWHM values of the PACS PSF reported in Table \ref{tab:psffwhm}, is smaller than the PSF eccentricity, 
Double precision format.

\subsubsection{EDGE FLAG (\textit{flag\_edge})}

The position of a source  with respect to the edges of the image affects the quality of the photometry, as the extraction procedure needs a large enough area to compute properly both the flux density and the background to be subtracted. Also, due to the scan speed inhomogeneities during the turnarounds, the map edges may have different noise properties than the rest of the image. 

The edge flag for each source is obtained using a Jython procedure that defines a subimage from the corresponding ROI map using the background external aperture (35$\arcsec$)
 centred on the source. We count the total number of pixels and the number of good values (1) in this subimage, obtaining the ratio between the two quantities. The conditions to set the edge flag are:\\
-The external radius is off the map

-The ratio is less than 0.9

The map edge flag is set when the source shows $\geq$10\% of pixels withhin the outer radius of the background annulus, with 0 value in the ROI maps. 

The data type is Boolean.

\subsubsection{BLEND FLAG (\textit{flag\_blend})}
Objects with distance within the PSF FWHM have the same value (>0) of FLAG\_BLEND and they can be considered as blended sources. FLAG\_BLEND = 0 means that the object is not blended with other objects. The PACS PSFs depend on the observing mode and the adopted values are the ones reported in Table~\ref{tab:psffwhm}. The absolute pointing error for SPG13 (1.2$\arcsec$) is adopted as uncertainty. 

\subsubsection{WARM ATTITUDE (\textit{warmat})}
Flag indicating that the observation was affected by "warm" attitude. The term refers to spacecraft attitudes in which the star-tracker support structure is subject to thermal distortions as was not completely shaded from sunlight. This has an impact on the pointing accuracy. Data type is Boolean. 

\subsubsection{OBSERVATION ID (\textit{obsid})}
The term is the observation identifier for the \textit{Herschel} observations . It is a 10 digit integer, starting always with 1342. The column lists the first OBSID from the list that built the L2.5/L3 map. It corresponds to the meta keyword \textit{obsid001}. 

\subsubsection{SOLAR SYSTEM MAP FLAG (\textit{ssomapflag})}

Maps of solar system objects have typically been observed in tracking mode, but were reprocessed in the 'rest' sky frame for the HPPSC, and may be contaminated by the moving targets. The affected OBSIDs are marked with an 'ssomap' flag. 

\begin{table}[H]
\centering
\small\addtolength{\tabcolsep}{-3pt}
\begin{tabular}{lccc}
\hline
Column & Format & Unit & Description\\
 \hline
name&String&&IAU compatible object identifier\\
&&&PACS detector array\\
band&String&&blue=70$\mu$m\\
&&&green=100$\mu$m\\
&&&red=160$\mu$m\\
RA&Double Precision&deg&Right Ascension (J2000)\\
Dec&Double Precision&deg&Declination (J2000)\\
raerr&Double Precision&deg&Positional uncertainty in RA\\
decerr&Double Precision&deg&Positional uncertainty in Dec\\
flux&Double Precision&mJy&non-color corrected flux density\\
snr&Double Precision&&Structure noise based signal--to--noise ratio\\
snrnoise&Double Precision&mJy&Noise derived from S/N$_S$\\
stn&Double Precision&&Background RMS based signal--to--noise\\
rms&Double Precision&mJy&Background RMS\\
strn&Double Precision&MJy sr$^{-1}$&Structure noise\\
fratio&Double Precision&&Flux ratio measured in apertures 6 and 1\\
fwhmxfit&Double Precision&asec&FWHM of the fitted Gaussian along the major axis\\
fwhmyfit&Double Precision&asec&FWHM of the fitted Gaussian along the minor axis\\
flag\_elong&Double Precision&&ratio of the FWHM$_Y$ and FWHM$_X$\\
flag\_edge&Boolean&&T = at least one contributing source shows ≥ 10\%\\
&&&of map pixels as NaN within the outer background annulus\\
flag\_blend&Integer&&Sources with the same numbers are blended\\
warmat&Boolean&&T=the observation was affected by the "warm" attitude\\
obsid&Integer&&Observation identifier\\
ssomapflag&Boolean&&T=SSO map reprocessed in the \textit{rest} sky frame\\
\hline
\end{tabular}
\caption{Columns in the HPPSC tables. Column names, format, unit and a short description are included.}
\label{hppsccolumns}
\end{table}


\subsection{Additional material}
\subsubsection{Extended Source List}

The \textit{Herschel}/PACS Extended Source List (HESL) contains slightly extended objects, extracted from database source table.  Quality  criteria are similar to those for point sources with two exceptions. We do not impose any condition on the ratio of flux values measured in different apertures, and also the FWHM of the extended source must satisfy:

\begin{equation}
2\times PSF  <  FWHM  <  5 \times PSF
\end{equation}

The detailed workflow of object consolidation, including the HPESL compilation is presented in Figure~\ref{workflow3}. 

The columns of the Extended Source Table are similar to the ones derived for HPPSC, with a few exceptions:
\begin{itemize}
\item The column NAME is similar to HPPSC name,  but  the acronyms for extended sources  is  HPESLxxxA. HPESL means  \textit{Herschel}/PACS Extended  Source  List,  xxx the wavelength in units of $\mu$m, and the letter A indicates the version of the catalogue - currently A, indicating the first version.
\item The column \textit{flux} contains the flux value measured in aperture 11 (18$\arcsec$ in the blue/green, 22$\arcsec$ in the red band). It is important to note, that unlike for point sources, for  extended sources the aperture correction was \textbf{not} applied.
\item The S/N$_S$ and S/N$_R$ values were calibrated for point sources, but were used  as a quality indicator, so they were kept in the HPESL. However, the noise values derived for them are not valid for extended sources, so they were removed.
\item The WARMAT and FLAG\_BLEND flags are also not included.

\end{itemize}

\textbf{We want to  emphasise that our pipeline was tuned for sources that have a shape and flux distribution similar to the instrument PSF. The flux density values listed in the Source List are only approximations. We  recommend strongly at users retrieve the observations from the HSA and extract fluxes and source parameters using their a preferred method. Our values are not suitable for use in scientific publications.}

\subsubsection{Rejected Source List}

This  list contains  all sources that are neither included in the HPPSC  nor in the  HPESL, because they did not verify our quality criteria  (see Section~\ref{sub:qa}), and were thus rejected.  The columns  of the HPRSL are  Source ID, RA, Dec, RA$_{err}$, Dec$_{err}$, OBSID and band. \textbf{Intentionally, other data are not provided. The detections listed here are completely unreliable and any further investigation of them has to be done by the user.}

\subsubsection{Observation Table}\label{ot}
The source extraction technique for the generation of the HPPSC was applied to the map repository as described in the Section~\ref{mapRep}. The \textit{Observation Table}, available for download, reports the main parameters that characterise the maps on which the source extraction was performed. Most of them correspond to Meta-keywords of the Observation Context. The \textit{level} parameter states which map level was adopted for each map and the \textit{nobsids} parameter specifies how many observations (obsID) contribute to the generation of the map. The \textit{obsid} parameter always refers to the first observation identifier in the Observation Context (obsid001). For every \textit{obsid} value, there are two entries in the table, corresponding to the two filters used.
\begin{table}[H]
\centering
\begin{tabular}{lcc}
\hline
Parameter & Meta-keyword & Description\\
 \hline
obsid & obsid001 &  Observation identifier\\
startod & odNumber &Operational day at the beginning of the first observation \\
targetname & object & target name, as given by the observer \\
proposal & proposal  & proposal name\\
obsmode & obsMode &Parallel Mode/Scan map/Point-source photometry\footnotemark[1]\\
ra & ra  &RA of the map centre\\
dec & dec &Dec of the map centre\\
starttime &  startTime & Start date/time of the first observation\footnotemark[2]\\
endtime & endTime & End date/time of the last observation\footnotemark[2]\\
aorlabel & aorLabel &AOR label as centred in HSPOT\\
nrep &  repFactor & Repetition factor\footnotemark[3]. 0 value for Parallel Mode\\
mapx & naxis1 from WCS & X map dimension in pixels\\
mapy & naxis2 from WCS & Y map dimension in pixels\\
mapscanspeed &  mapScanVelocity &10, 20 or 60 $\arcsec$s$^{-1}$\\
band & blue & green, blue or red filter\footnotemark[4]\\
level & - & SPG level\\
nobsids & - & number of obsids\\
swversion & creator & SPG version\\
\hline
\end{tabular}
\caption{Description of the columns (Parameter) of the \textit{Observation Table} and of the Observational Context meta-keywords to which they relate}
\label{observationtable}
\end{table}

{\footnotesize
\footnotemark[1] \textit{Point-source photometry} observations are in chop/nod mode and they are discarded for the catalogue generation.\\
\footnotemark[2] StartTime and endTime are in expressed TAI by using the format YYY-MM-DDTHH:MM:SS.\\
\footnotemark[3] The repFactor keyword doesn't exist for \textit{Parallel Mode} and nrep is set to 0.\\ 
\footnotemark[4] PACS is a dual-band photometer: an image is always acquired by the Red detector. The \textit{blue} meta-keyword value specifies which filter is applied for the Blue camera (blue1 = 70 $\mu$m, blue2 = 100 $\mu$m)\\
}



\pagebreak
\newpage

\section{Validation}
\subsection{Internal validation -- Simulations}\label{int_validation}

Simulations played an essential role in the validation of the HPPSC. One way to test our photometry is to measure the flux of sources that are injected artificially into the observational timelines with well controlled flux values. The completeness can also be characterised by simulations: as the number of injected sources per brightness bin is known we can compare these values with the source counts obtained through our pipeline. 

For PACS the PSF shape is significantly different from a simple 2D Gaussian function, especially in the blue band at high scan speed. Unlike for the SPIRE PSC, we had to use actual PSF images and add them to the Level1 TOD. The process of creating the simulations consisted of the following steps:
\begin{itemize}
\item Random position generation: A number of random positions were generated with the criteria of a minimum separation of 35$\arcsec$. The number of randomly generated sky positions is determined by the extent of the map. In general, for large Parallel Mode observations we were able to generate more positions. 
\item PSF back-projection: in order to simulate the real sources, we used the official Vesta PSFs that are specific to the observing mode, band and scan speed. After the correct PSF was selected, it was rotated according to the scan angle of the observation in which the PSF was injected. Then, the Level1 TOD of the observations was duplicated and emptied. The back-projection from the L25 PSF image to the Level TOD was carried out with a dedicated task in HIPE, called Map2signalCubeTask(). We used PSFs that were scaled so that source flux was exactly 1\,Jy. As a result of the process we were left with a TOD that was of the same length as the observation, but contained only 1\,Jy flux  sources at the randomly generated positions (source-only timelines). 
\item Map projection: The source-only timeline has the advantage that it can be scaled, multiplying it by 10 will result in 10 Jy sources. Similarly dividing by 10 results in sources with 100 mJy source flux. After scaling to the desired flux level, we added the PSF data to the actual observational TOD. The combined data was then processed with JScanam SPG13.0.0, so the same algorithm was used as for the vast majority of the maps in our repository. 
\end{itemize}

\begin{figure}[H]
\centering
\includegraphics[width=0.9\textwidth]{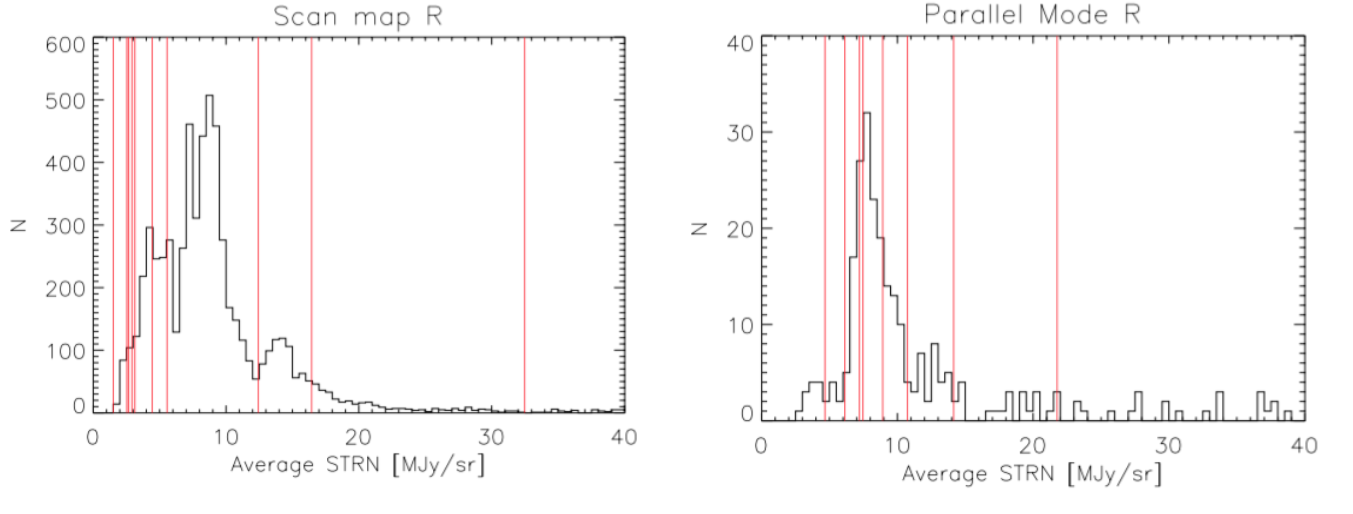}
\caption{Distribution of the average structure noise values calculated for Scan maps (left) and for Parallel Mode maps in the red band. The structure noise values of the simulated fields are marked with red lines.}\label{strnsimulations}
\end{figure}

\begin{table}[H]
\centering
\small\addtolength{\tabcolsep}{-3pt}
\begin{tabular}{llccccccc}
Field name	&OBSID		&RA, Dec [$^{\circ}$]	&	size [$^{\prime}$]&	Mode&	Speed [$\arcsec$s$^{-1}$]	&	Band	& N$_{S} (red)$ [MJy\,$sr^{-1}$] & N$_{inj}$ \\
\hline
Atlas		&1342189661	&212.46,0.55	&331$\times$345&	parallel&60	&BL+R	& 7.45 	&300\\
Field 2\_0	&1342204104	&267.87,-27.05	&208$\times$210&	parallel&60	&BS+R 	& 91.10 &200\\
Field 171\_0&1342250342 &81.10,36.15	&208$\times$210&	parallel&60	&BS+R	& 8.92	&200 \\
G159.23		&1342239263&44.37,19.751	&99$\times$103 &	parallel&20	&BL		& 7.21	&200 \\
Globule2	&1342188683	&187.54,-63.70	&80$\times$84  &	parallel&60	&BS+R	& 14.15	&200 \\
GP field-1	&1342183070 &266.61,-28.93	&125$\times$126&	parallel&60	&BS+R	& 576.69&200 \\
L1489		&1342202088	&61.00,26.28	&81$\times$81  &	parallel&60	&BS+R	& 10.74 &200\\
LDN1780		&1342224989	&235.10,-7.22	&153$\times$156&	parallel&20	&BL+R	& 6.14	&300 \\
M31			&1342211604	&10.99,41.36	&186$\times$224&	parallel&20	&BS+BL+R& 4.67 	&200\\
N6357		&1342204847	&261.71,-34.43	&136$\times$139&	parallel&20	&BS		& 172.50&300\\
Rosette		&1342186121 &98.22,4.27		&121$\times$127&	parallel&20	&BS+R	& 21.77	&200 \\
\hline	
Abell370	&1342223332	&39.97,-1.58	&20$\times$21	&	scan	&20	&BL+R	& 2.53 	& 20\\
A851		&1342271023 &145.75,46.99	&20$\times$21	&	scan	&20	&BS		& 2.65 	& 20\\
Arp 244		&1342187836 &180.47,-18.87	&22$\times$26	&	scan	&20 &BS+BL+R& 16.45 & 20\\
Cas A		&1342188205	&350.85,58.81	&66$\times$68	&	scan	&20	&BL+R	& 12.44 & 150\\
Crab		&1342204442 &83.63,22.01	&66$\times$70	&	scan	&20	&BS+BL+R& 5.56 	& 50\\
IRAS16132	&1342216487 &244.25,-50.79	&63$\times$64	&	scan	&20	&BS+R	& 82.33 & 100\\
IRDC010.70	&1342191803	&272.46,-19.73	&37$\times$40	&	scan	&20	&BL+R	& 95.68 & 50\\
M81			&1342186085	&148.90,69.07	&103$\times$103	&	scan	&20	&BS+R	& 2.91 	& 100\\
NGC1365		&1342183021 &53.40,-36.14	&75$\times$78	&	scan	&20	&BL		& 1.50 	& 100\\
NGC6946		&1342191947 &308.72,60.16	&47$\times$47	&	scan	&20	&BS+BL+R& 3.14 	& 30\\
NGC7026		&1342223919	&316.58,47.85	&24$\times$23	&	scan	&20	&BS+R	& 4.45 	& 30\\
RCW79		&1342188880	&205.06,-61.73	&88$\times$92	&	scan	&20	&BL+R	& 32.48 & 100\\
RCW120		&1342216585 &258.09,-38.47	&88$\times$91	&	scan	&20	&BS+R	& 48.93 & 150\\
\end{tabular}
\caption{Observations used to create simulations.}
\label{simulations}
\end{table}

\subsubsection{Completeness}\label{completeness}
\begin{figure}[H]
\centering
\includegraphics[width=0.9\textwidth]{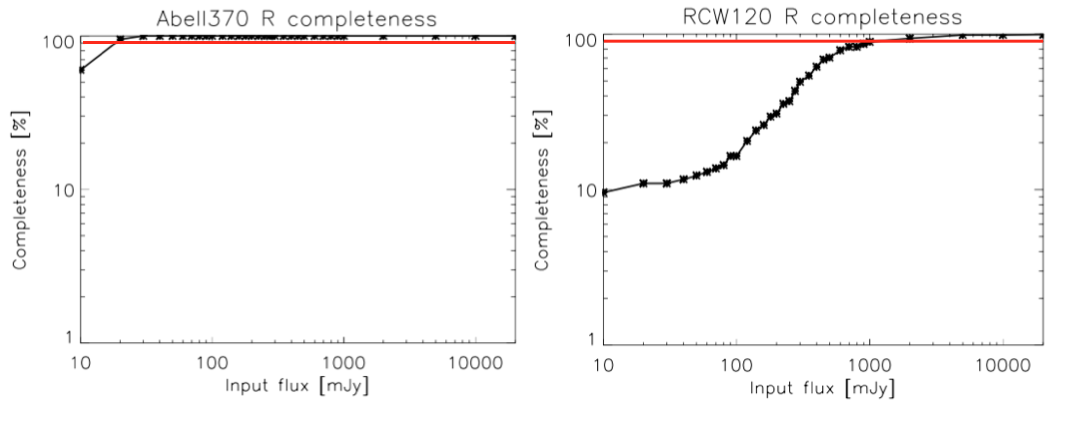}
\caption{Completeness in two of the simulated fields. Abell370 (left) is a deep cosmological observation. RCW120 (right) is a very complex HII region.}\label{completenesscomparison}
\end{figure}

The completeness is calculated as a ratio between the number of sources injected and the number of sources detected by our pipeline. We present it here for two different fields. The Abell 370 field is a deep cosmological observation, mainly free of extended emission, therefore it has a low structure noise. The RCW120 field is a HII region with ongoing star formation, close to the Galactic Plane, and therefore very structured. As one can see from the curves in Figure~\ref{completenesscomparison},  90\% completeness is reached at $\sim$20 mJy, while the same completeness in the Galactic field is reached at $\sim$1000 mJy.

As it is described in the Section~\ref{strnmaps}, the structure noise includes noise  both from the instrument and the sky confusion. Figure~\ref{completenesscomparison} clearly shows that the completeness highly depends on the complexity of the image. To characterise the relation between completeness and structure noise for each injected flux level we created N$_S$ bins, and counted the number of sources that were injected in the given N$_S$ interval and the number of sources what we detected. The same criteria that were used to clean the source tables were used here, hence the results reflect the true completeness (see Section~\ref{sub:consolidation}). A logarithmic scale was used to set size of the N$_S$ bins. The bins were centred on N$_S$ values of 10$^x$, where x varied between 1 and 3 with an interval of 0.05. 

\begin{figure}[H]
\centering
\includegraphics[width=0.9\textwidth]{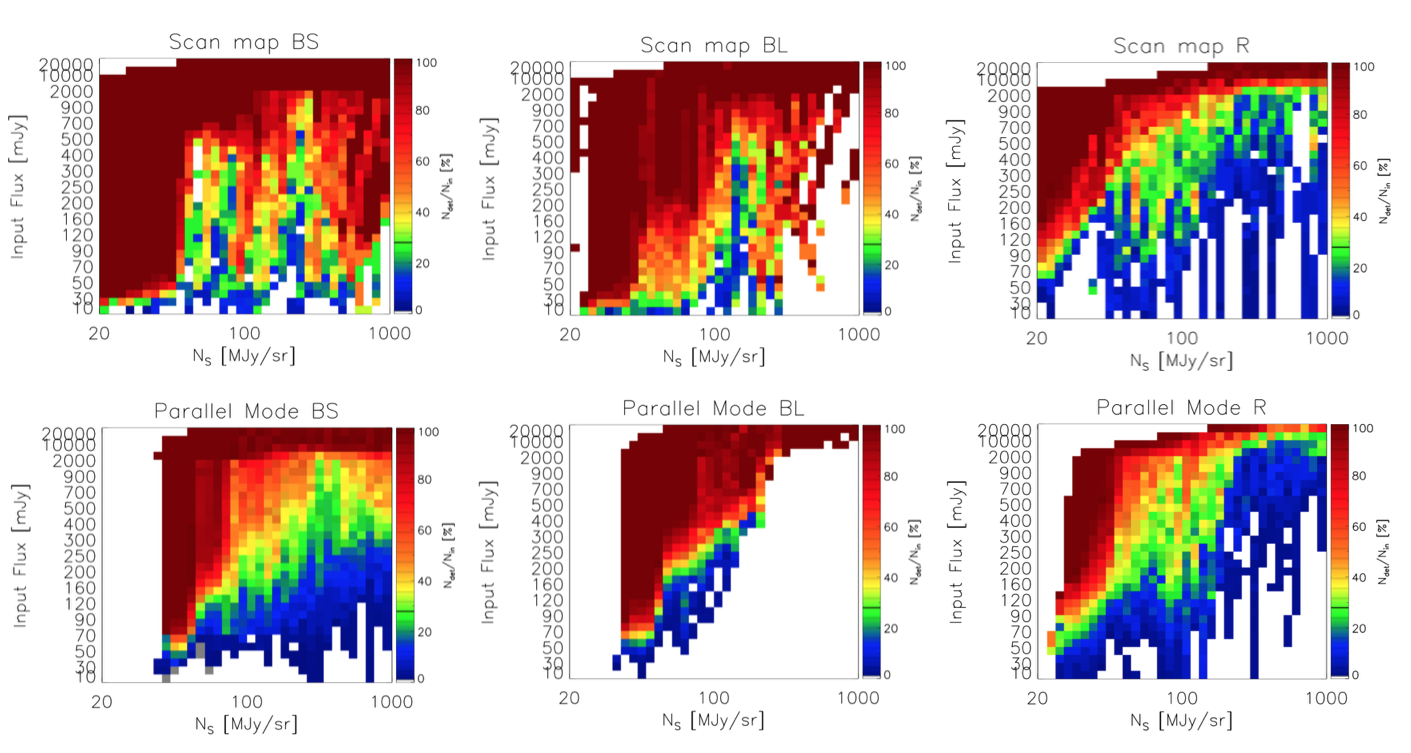}
\caption{The simulation based completeness in the blue, green and red bands (column 1,2 and 3, respectively) for the Scan map (row 1) and parallel (row 2) observing modes. The colour corresponding to from blue to red means low to high completeness.}\label{simscompplot}
\end{figure}

As expected, the completeness is greater at low N$_S$ values. Figure~\ref{simscompplot} also shows that at high flux levels we are always almost complete, while the faint sources are undetectable on the complex and bright background.

\subsubsection{Photometric accuracy}

The photometric accuracy is characterised as the ratio of the input flux to the measured flux. The photometric error is the standard deviation of the measured flux values at a given injected flux levels. For example if the injected flux is 60 mJy and the measured flux is 58.74$\pm$9.44\,mJy, the photometric accuracy is 58.74/60.0=0.979, the error is 9.44, and the S/N is 58.74/9.44=6.22. 



\begin{figure}[H]
\centering
\includegraphics[width=0.9\textwidth]{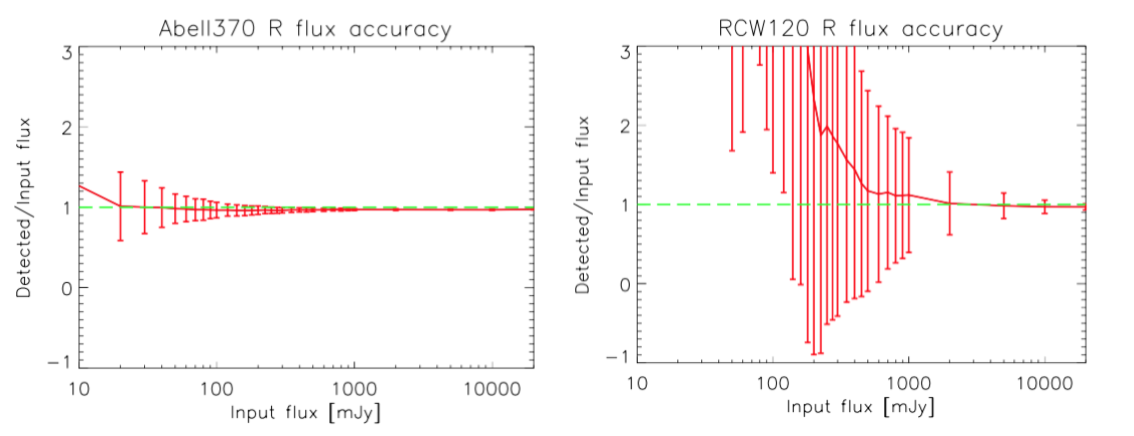}
\caption{Photometric accuracy in two of the simulated fields. RCW120 (right) is a very complex HII region. Abell370 (left) is a deep cosmological observation. }\label{fluxaccuracycomparison}
\end{figure}

As shown in Figure~\ref{fluxaccuracycomparison}, the photometric accuracy depends strongly on the environment. On the left the photometric accuracy was studied in the same Abell\,370 field used for the completeness study. On the right we used the observation of the RCW120 field. Both datasets were obtained in the red band. The dashed green lines show perfect photometry. As expected, our photometry performs better in a less complex environment. The error bars present the photometric errors, and are also lower in the cosmological observation.

\begin{figure}[H]
\centering
\includegraphics[width=0.9\textwidth]{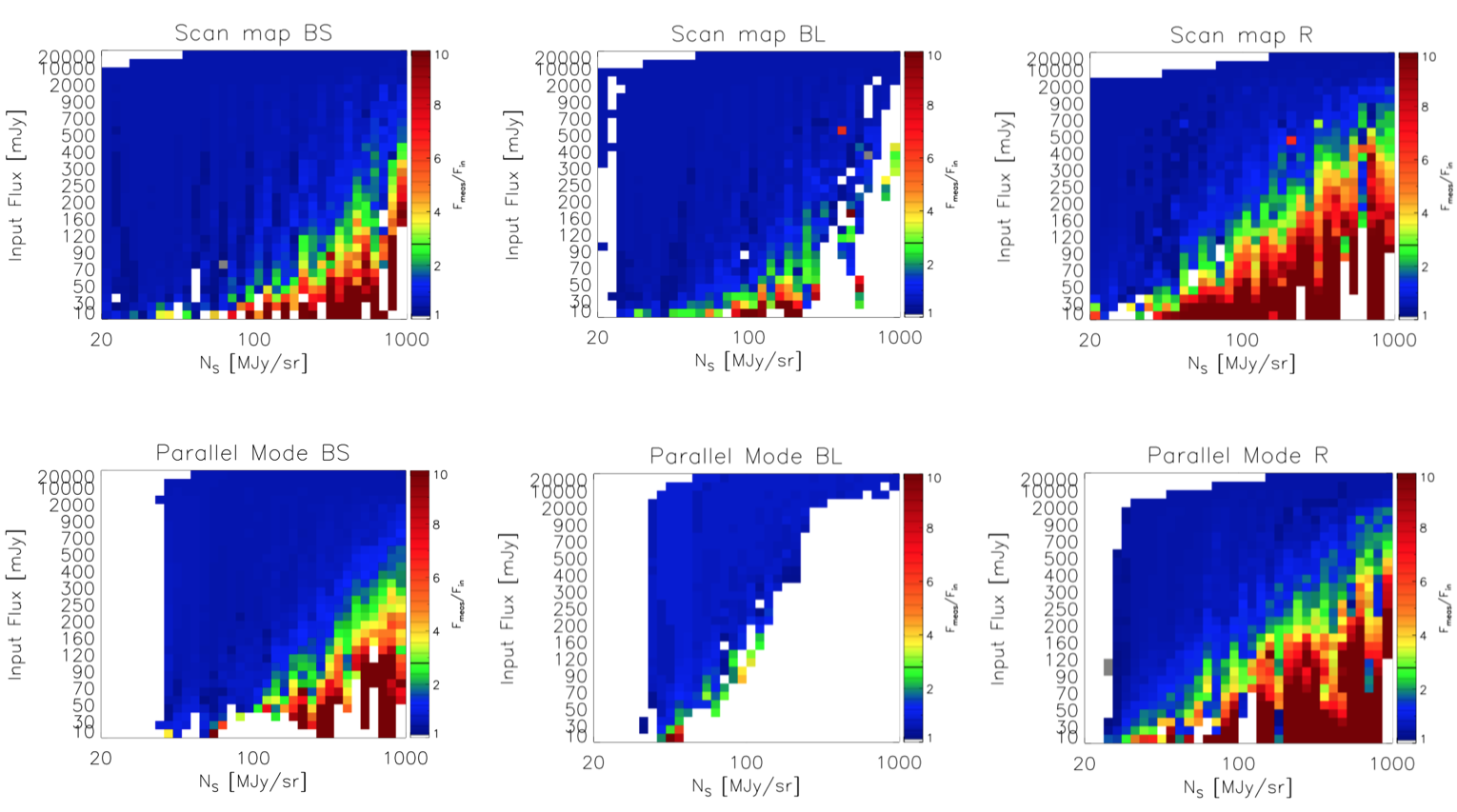}
\caption{The photometric accuracy in the blue, green and red bands (column 1,2 and 3, respectively) for Scan map (row 1) and Parallel (row 2) observing modes. The colour coding from blue to red means high to low photometric accuracy.}\label{simsaccplot}
\end{figure}

These tests also led us to the conclusion that our photometry in areas in the Galactic Plane and in star forming regions (i.e. high complexity) is less accurate. The same strategy of N$_S$ binning was applied here, so we were able to determine the flux accuracy as a function of the N$_S$ and the input flux. In Figure~\ref{simsaccplot} we show that at high flux levels our photometric performance is very good, with flux ratios close to unity. However, at low flux levels and in environments of high complexity the flux determination is quite unreliable.

\subsubsection{Positional accuracy}

We used the fitted coordinates of standard stars and compared them with their known sky positions. To characterise the positional accuracy of the catalogue. We considered the same sample for this purpose as was used for the photometric accuracy tests, i.e. objects in Balog et al. (2014) for the bright standard stars and in Klaas et al. (2017) for the faint standard stars. The median positional difference was found to be 1.5$\arcsec$ for the blue PACS detector (70 and 100\,$\mu$m filters) and 1.7$\arcsec$ for the red detector (160\,$\mu$m filter). This accuracy is comparable with the typical $\sim$2$\arcsec$ pointing accuracy of the Herschel Space Observatory (see \href{http://adsabs.harvard.edu/abs/2014ExA....37..453S}{Sanchez-Portal et al. 2014}).  

\begin{figure}[H]
\centering
\includegraphics[width=0.45\textwidth]{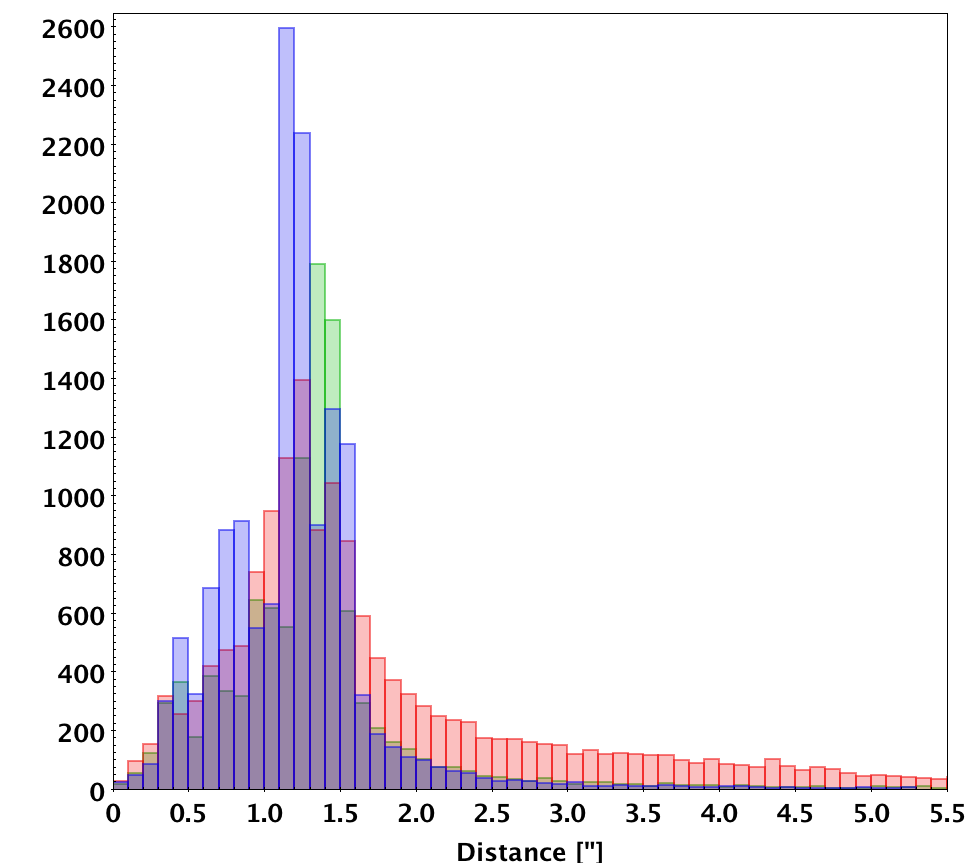}
\includegraphics[width=0.45\textwidth]{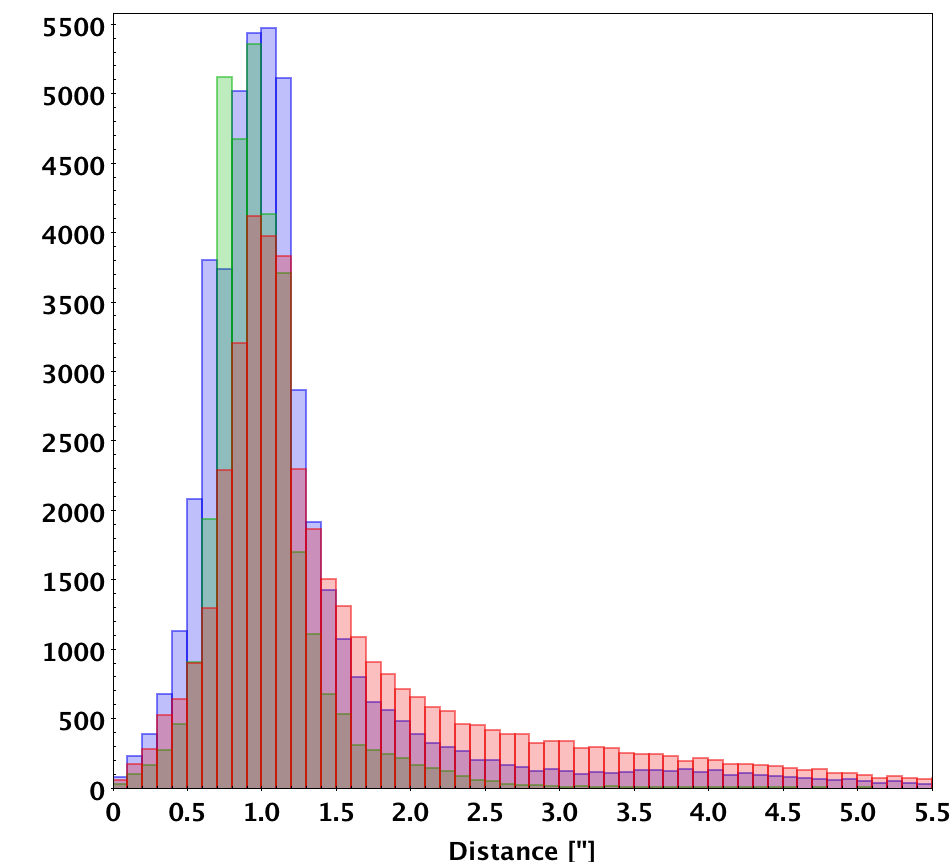}
\caption{Position uncertainty based on simulations for the Scan map (left) and Parallel Mode (right). The uncertainty is measured as the angular distance between the injected source position and the SUSSEXtractor source position. The blue, green and red bands are indicated with the corresponding colours. }\label{suspos}
\end{figure}

A second test that we performed was to compare the recovered positions of the sources that we injected into the simulations. The difference between the input coordinates and the output coordinates calculated by the SUSSEXtractor algorithm includes the errors from the source back-projection to the observational timeline, the error introduced by the map re-projection and the error introduced by SUSSEXtractor. However, as shown on Figure~\ref{suspos} the peak difference is between 1.0$\arcsec$ and 1.5$\arcsec$ for all bands and observing modes. This value is still well below the FWHM values of the instrument PSFs.

\begin{figure}[H]
\centering
\includegraphics[width=0.45\textwidth]{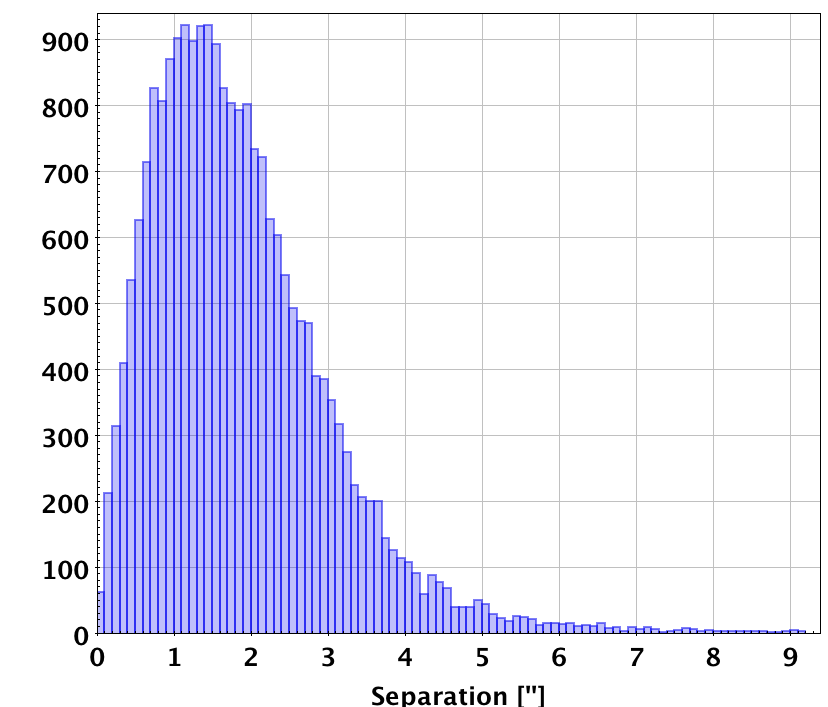}
\includegraphics[width=0.45\textwidth]{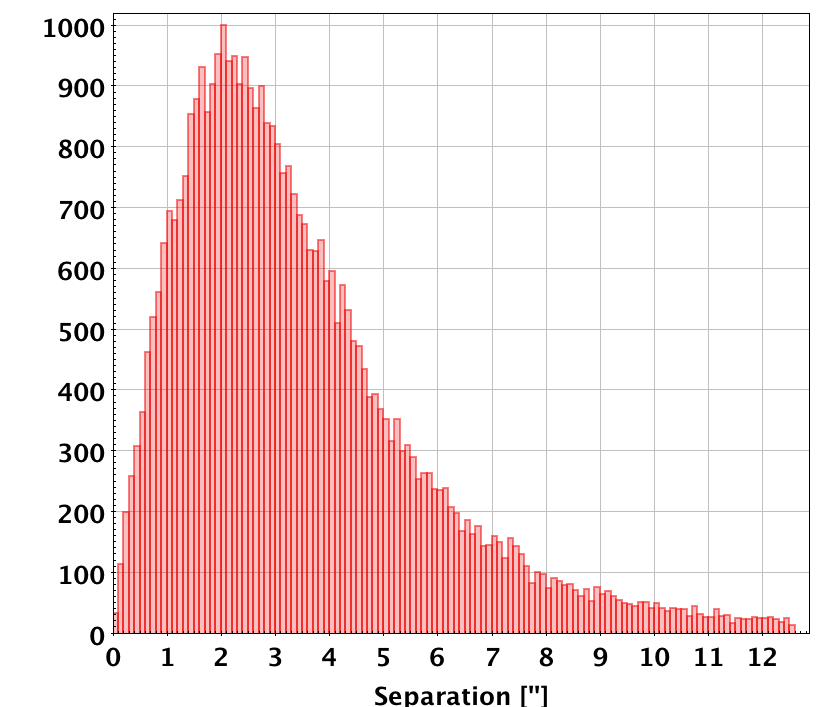}
\caption{Distribution of the distances between the HiGal and the HPPSC source positions for the blue (left) and red (right) PACS detectors. \label{fig:higalposacc}}
\end{figure}

As an additional test, we compared the source positions from the Hi-GAL project with those of the corresponding sources in the HPPSC (see Fig.~\ref{fig:higalposacc}). The distances between the Hi-GAL and the HPPSC positions show similar median values as those obtained in the previous tests.

While the positional accuracy of the HPPSC is in general comparable to the pointing accuracy of the observatory and this positional accuracy is notably better than the  spatial resolution achievable due to the finite beams in the PACS bands there are some situations in which the absolute positional accuracy of the telescope may be degraded. This may happen in the case of those measurements which were taken at 'hot' solar aspect angles ($>$\,110\degr, see \href{http://adsabs.harvard.edu/abs/2014ExA....37..453S}{Sanchez-Portal et al. 2014}, for a detailed discussion). To warn the users of the HPPSC of these conditions an additional flag 'WARMAT' was introduced for sources extracted from maps with SAA\,$>$\,110\degr. About 3\% of all maps used for HPPSC source extraction may be affected.   

\subsection{External Validation\label{sub:extval}}

While our extracted sources show internally consistent properties in the framework that we set for the PACS Point Source Catalogue, it is important to check our results against already existing catalogues of large surveys based on the same data. These calibration samples and large programmes used diverse approaches and data reduction/evaluation techniques -- typically significantly different to ours -- to reach their goals, therefore comparison with these results gives information about the robustness of our techniques, and the general quality and reliability of our data evaluation algorithms. 

\subsubsection{Comparison with point source flux calibration}\label{balog}

\begin{figure}[!h]
\includegraphics[width=0.45\textwidth]{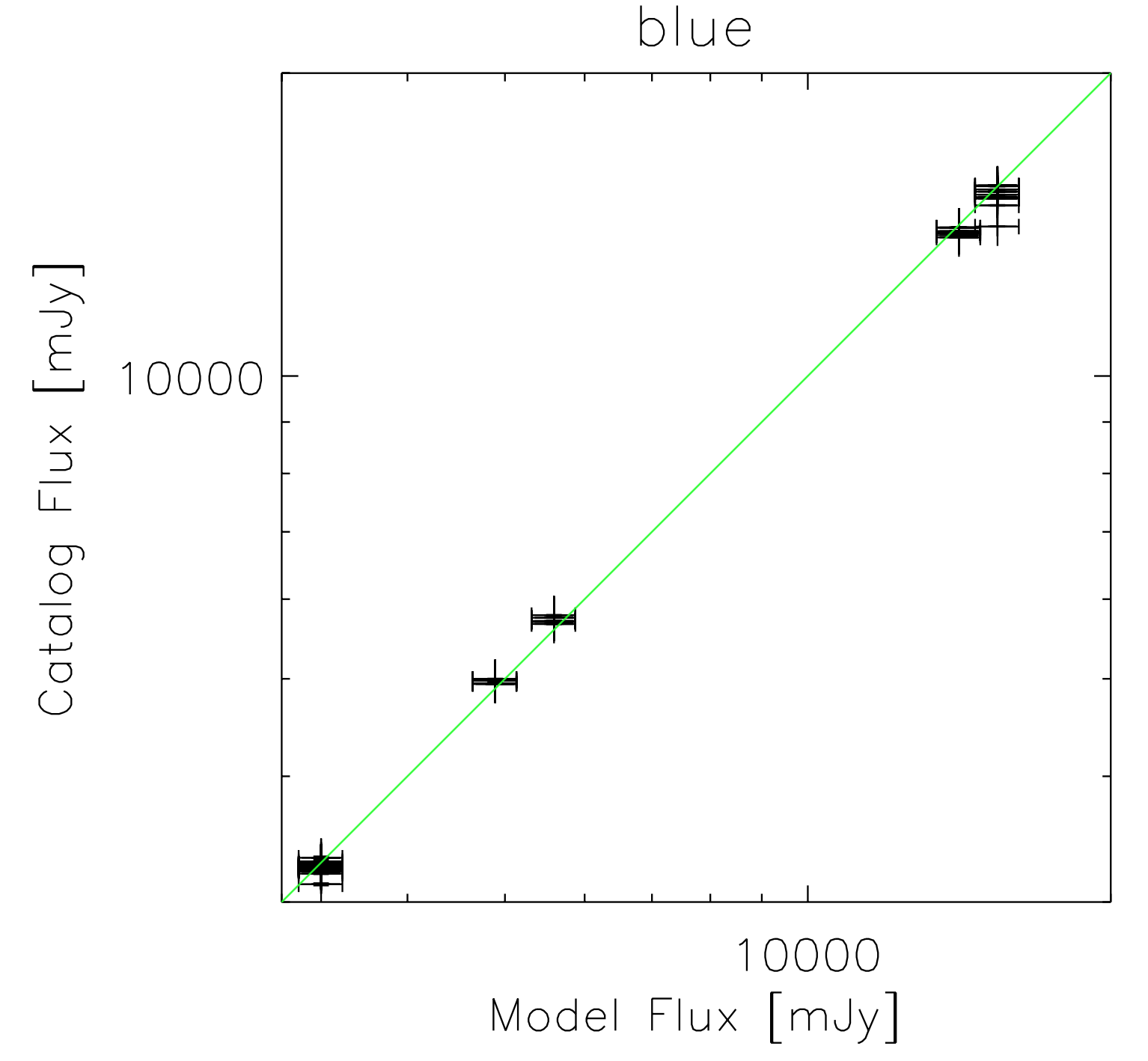}
\includegraphics[width=0.53\textwidth]{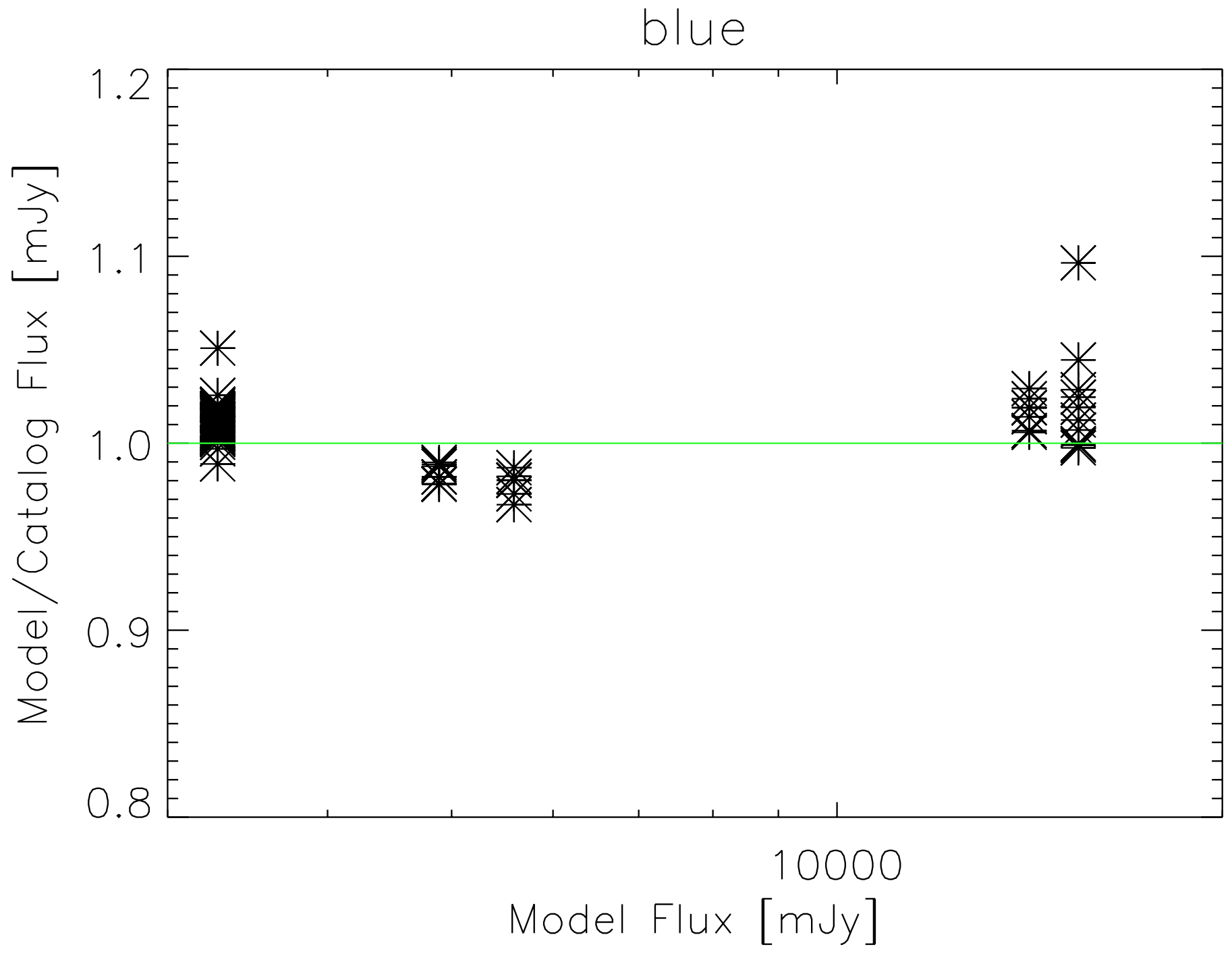}
\caption{Left: Catalogue flux density values as a function of the flux density predicted by photospheric models in the blue (70$\mu$m) band. The solid green line presents the 1:1 line. Right: Model flux to catalogue flux ratio as a function of the model flux in the blue band. The green horizontal line presents ratio of one.}\label{bluecalibrators}
\end{figure}

\begin{figure}[H]
\includegraphics[width=0.45\textwidth]{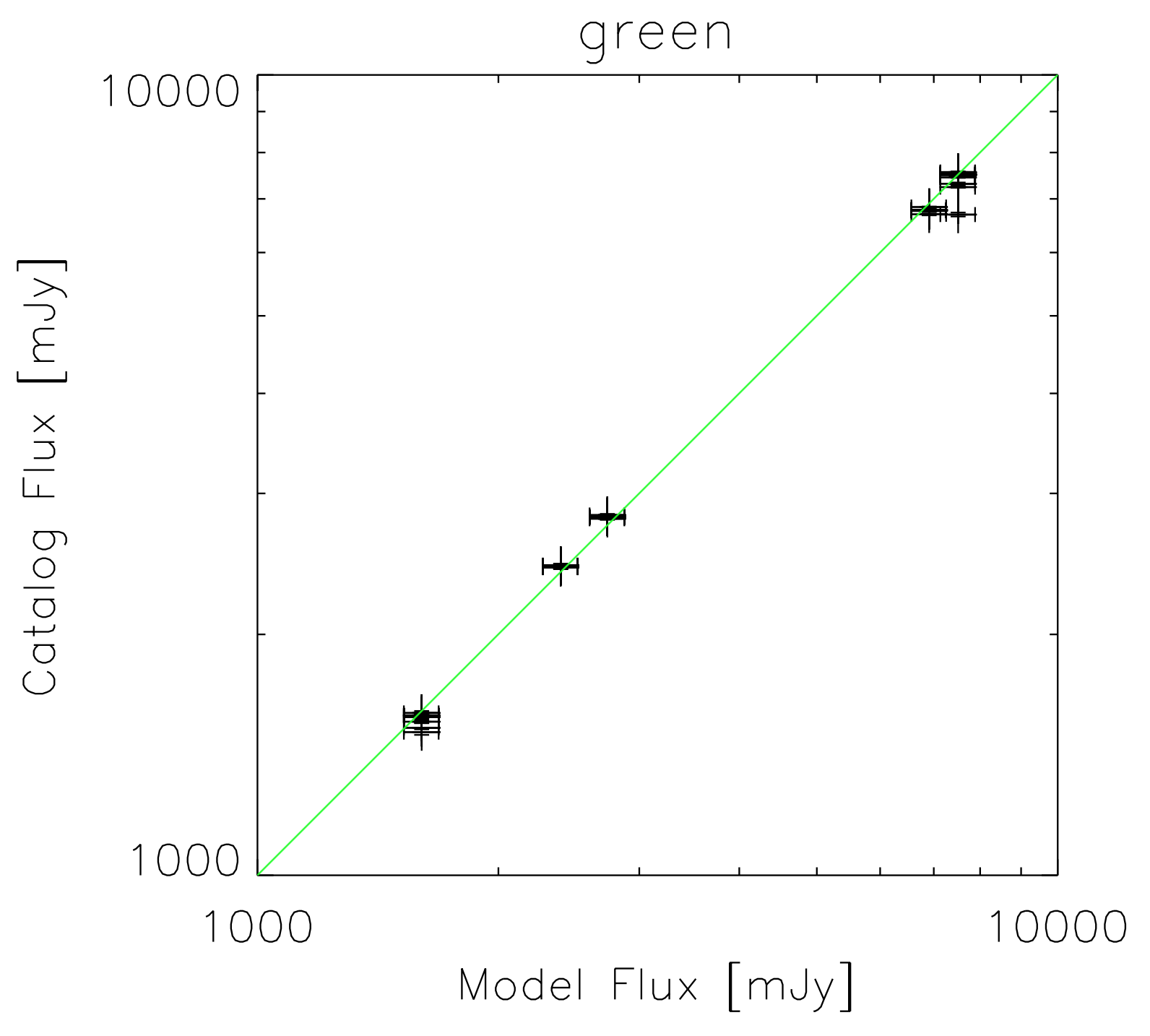}
\includegraphics[width=0.53\textwidth]{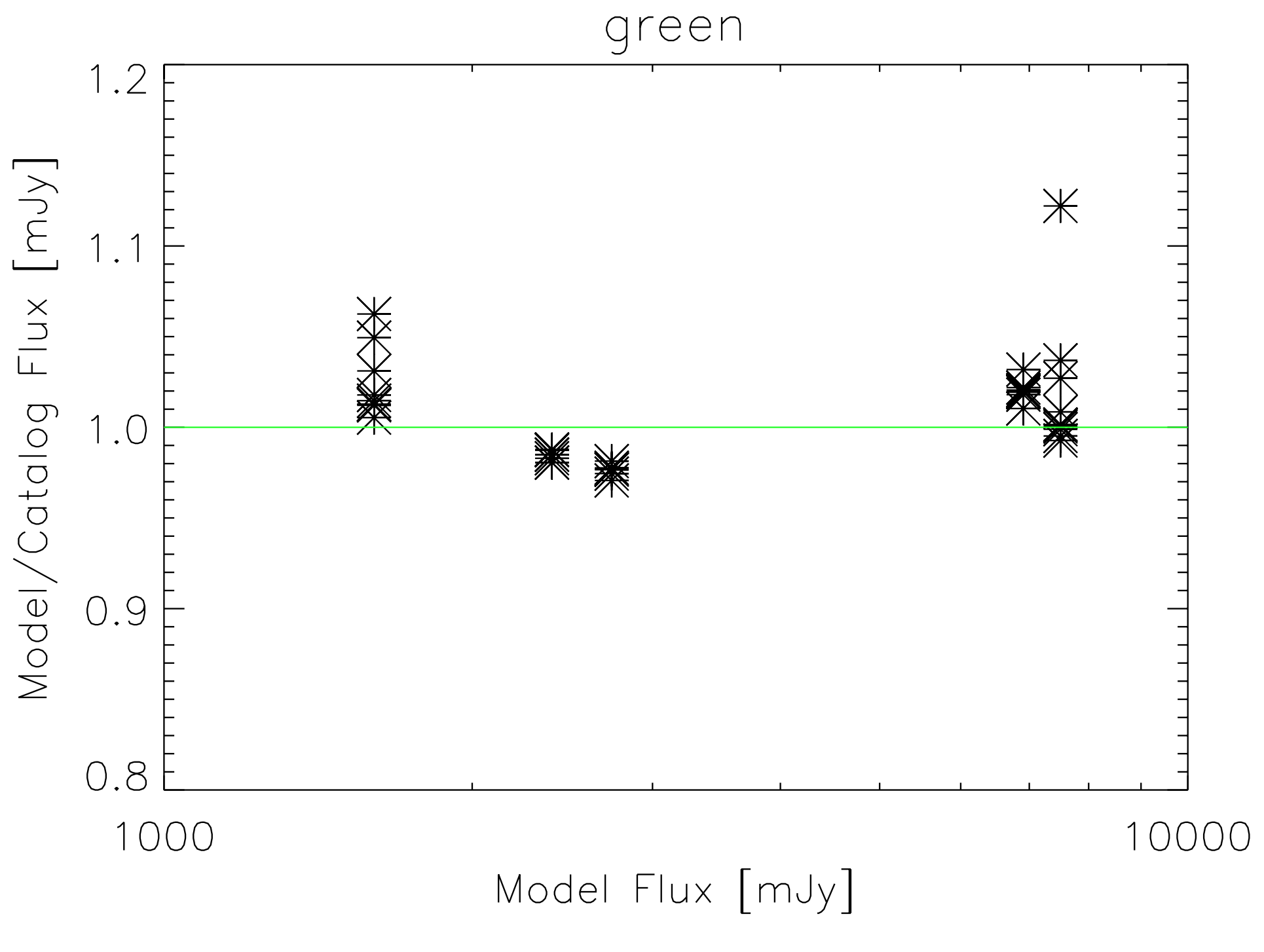}
\caption{Left: Catalogue flux density values as a function of the flux density predicted by photospheric models in the green (100$\mu$m) band. The solid green line presents the 1:1 line. Right: Model flux over the catalogue flux as a function of the model flux in the green band. The green horizontal line presents ratio of one.}\label{greencalibrators}
\end{figure}

\begin{figure}[H]
\includegraphics[width=0.43\textwidth]{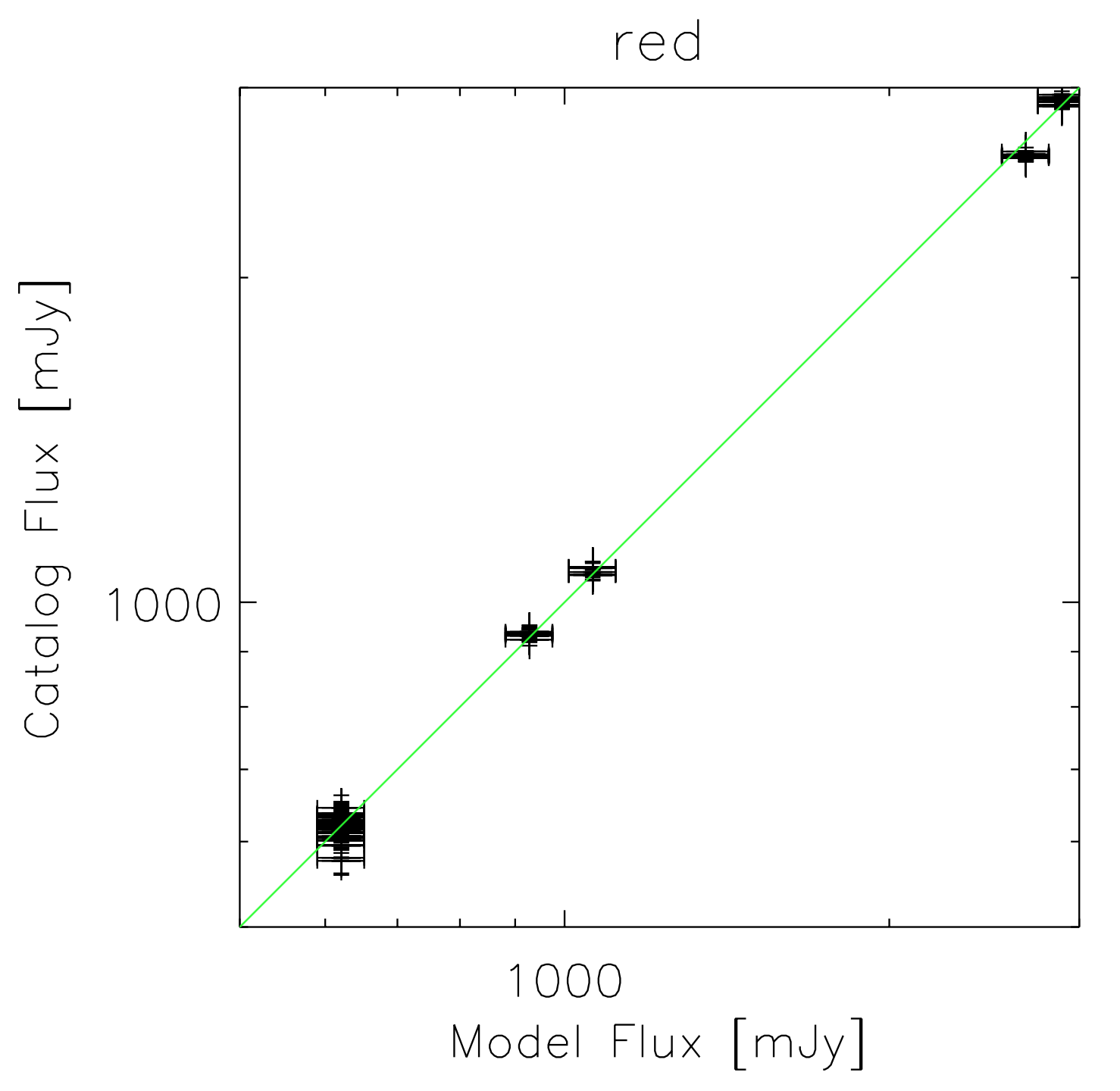}
\includegraphics[width=0.53\textwidth]{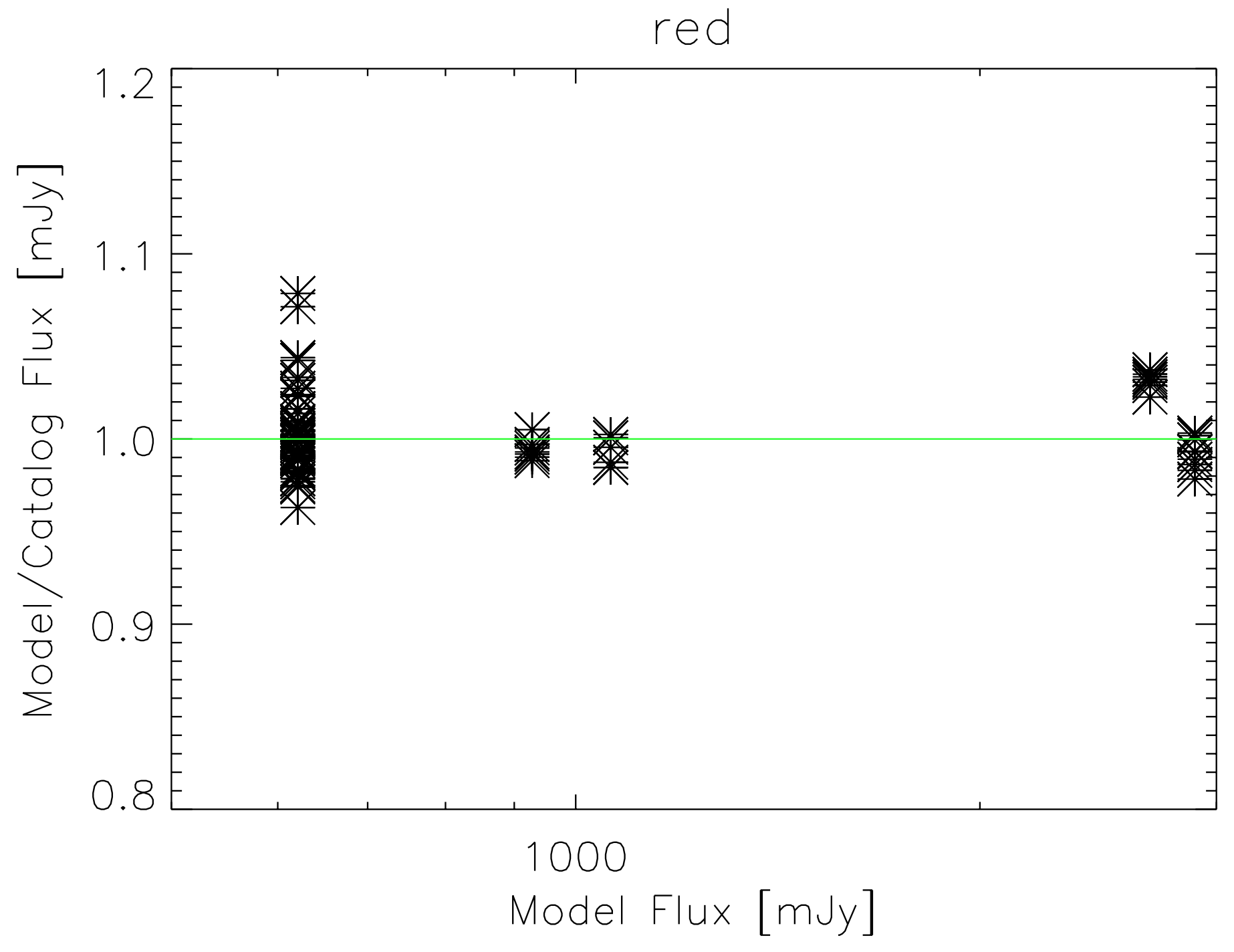}
\caption{Left: Catalogue flux density values as a function of the flux density predicted by photospheric models in the red (160$\mu$m) band. The solid green line presents the 1:1 line. Right: Model flux to catalogue flux ratio as a function of the model flux in the red band. The green horizontal line presents ratio of one.}\label{redcalibrators}
\end{figure}

\href{http://adsabs.harvard.edu/abs/2014ExA....37..129B}{Balog et al. (2014)} provided an overview of the PACS photometer flux calibration concept, in particular for the principal observation mode, the scan map. The absolute flux calibration was tied to the photospheric models of five fiducial stellar standards. We selected the same objects from our catalogue and compared our colour corrected flux values to their reported flux density values. Our photometry was carried out using 12$\arcsec$ and 24$\arcsec$ aperture for the blue and red camera and applying aperture correction. For the comparison colour correction was also applied, assuming 5000K black body temperature. The colour correction values based on the PACS report by \href{http://Herschel.esac.esa.int/twiki/pub/Public/PacsCalibrationWeb/cc_report_v1.pdf}{M\"uller et al. (2011)} was 1.016, 1.033 and 1.074 for the blue, green and red bands. Table~\ref{fluxratiotable} lists the name of the calibrators, the model flux, the average measured flux and the average ratio of the model and catalogue flux. The left panels of Figures~\ref{bluecalibrators}, ~\ref{greencalibrators} and ~\ref{redcalibrators} show the catalogue flux as a function of the model flux. The 1:1 line is shown in green. Taking into account the 5\% error of the model, our flux values agree very well with the predicted values. The right panels show the flux ratios. These figures highlight the fact that in most of the cases our flux values are within 5\% of the model flux.

\begin{table}[H]
\centering
\begin{tabular}{llcccc}
\hline
\hline
Name & band & F$_{model}$ [mJy]& F$_{catalogue}$ [mJy]& F$_{model}$/F$_{catalogue}$\\
\hline
	  &blue  &    5594 & 5718.506 $\pm$ 45.171 & 0.978$\pm$0.008\\
HD6860&green &    2737 & 2803.471 $\pm$ 11.109 & 0.976$\pm$0.004\\
      &red   &    1062 & 1070.047 $\pm$  9.263 & 0.993$\pm$0.009\\
\hline
       &blue &    4889 & 4969.804 $\pm$ 28.160 & 0.984$\pm$0.006\\
HD18884&green&    2393 & 2431.604 $\pm$  7.849 & 0.984$\pm$0.003\\
       &red  &    928  & 933.292  $\pm$ 5.058  & 0.9943$\pm$0.005\\
\hline
       &blue &   14131 & 13897.883$\pm$ 134.419& 1.017$\pm$0.010\\
HD29139&green&    6909 & 6771.814 $\pm$ 34.735 & 1.020$\pm$0.005\\
	   &red  &    2677 & 2595.375 $\pm$ 12.287 & 1.031$\pm$0.005\\
\hline
&blue&   15434 & 15099.168$\pm$ 348.331& 1.023$\pm$0.023\\
HD124897&green&   7509 & 7381.912 $\pm$105.521 & 1.018$\pm$0.014\\
&red  &   2891 & 2914.236 $\pm$ 24.985 & 0.992$\pm$0.008\\
\hline
&blue &   3283 & 3247.424 $\pm$ 22.110 & 1.011$\pm$0.007\\
HD164058&green&   1604 & 1568.252 $\pm$  9.618 & 1.023$\pm$0.006\\
&red  &    621 & 619.558  $\pm$12.230  & 1.003$\pm$0.020\\
\end{tabular}
\caption{Comparison of the flux predicted by photospheric models and the flux value measured by our pipeline. The names of the calibrators are listed in the first column. The second column gives the band. The predicted flux value is shown in the third column. The average measured flux and the corresponding 1$\sigma$ uncertainty are given in the fourth column. Column five shows the flux ratio.}
\label{fluxratiotable}
\end{table}

We conclude from the comparison that our flux density values are in very good agreement with those predicted by the photospheric models. The average flux ratios are 1.0026$\pm$0.01, 1.0042$\pm$0.006 and 1.0026$\pm$0.04 in the blue, green and red bands, respectively. These are also in almost perfect agreement with the values reported by Balog et al. (2014).

\subsubsection{Faint star comparison}\label{klaas}
Klaas et al. (in prep.) determined flux densities and photometric accuracy for a set of 17 stars that range in flux from intermediate brightness ($\sim$2.5 Jy) to faint ($\sim$5 mJy). Their data processing followed that of Balog et al. (2014).  It should be noted here that Klaas et al. performed the photometry on Level 2 high-pass filtered maps, while in the preparation of the HPPSC we used maps generated by the JScanam method. Therefore, the comparison of the faint standard star fluxes from the two different sources tests also the compatibility of the two mapping algorithms.

\begin{table}[H]
\centering
\small\addtolength{\tabcolsep}{-3pt}
\begin{tabular}{llccccc}
\hline
\hline
Name & band & F$_{model}$ [mJy]& F$_{catalogue}$ [mJy]& F$_{model}$/F$_{catalogue}$ & F$_{Klaas}$ [mJy] & F$_{catalogue}$/F$_{Klaas}$\\
\hline
	   &blue  &    2457 & 2702.094$\pm$3.835 &    0.909$\pm$0.001&2649.400&   1.020$\pm$   0.001\\
HD62509&green &    1190 & 1334.990$\pm$3.963 &    0.891$\pm$0.003&1287.100&   0.835$\pm$   0.004\\
       &red   &    456  & 526.971$\pm$6.069 &    0.865$\pm$0.010&497.400&   1.059$\pm$   0.012\\
\hline
       &blue &    1707 & 1704.437$\pm$  13.264&   1.002$\pm$   0.008&1668.300&   1.022$\pm$   0.008\\
HD12929&green&    831 & 851.174$\pm$   7.679&   0.977$\pm$   0.009&831.600&   1.024$\pm$   0.009\\
       &red  &    321& 343.872$\pm$   2.094&   0.934$\pm$   0.006&336.000&   1.023$\pm$   0.006\\
\hline
       &blue &   1182&1194.326$\pm$   6.118&   0.990$\pm$   0.005&1157.100&   1.032$\pm$   0.005\\
HD32887&green&   576& 592.220$\pm$   2.975&   0.973$\pm$   0.005&568.800&   1.041$\pm$   0.005\\
	   &red  &   222& 234.146$\pm$   4.685&   0.951$\pm$   0.019&217.500&   1.077$\pm$   0.022\\
\hline
		&blue&   857& 868.183$\pm$   3.489&   0.988$\pm$   0.004&839.000&   1.035$\pm$   0.004\\
HD198542&green&  418& 428.658$\pm$   1.955&   0.975$\pm$   0.004&414.300&   1.035$\pm$   0.005\\
		&red  &  161& 170.364$\pm$   6.439&   0.949$\pm$   0.037&168.600&   1.010$\pm$   0.038\\
\hline
		&blue &   479& 519.299$\pm$  -&   0.923$\pm$  -&517.800&   1.003$\pm$-\\
HD148387&green&   232& 256.012$\pm$  -&   0.909$\pm$  -&237.900&   1.076$\pm$-\\
		&red  &   89&  86.323$\pm$  -&   1.047$\pm$  -&116.400&   0.742$\pm$-\\
\hline
		&blue &   428& 443.526$\pm$   3.371&   0.968$\pm$   0.007&433.600&   1.023$\pm$   0.008\\
HD180711&green&   207& 214.613$\pm$   3.839&   0.972$\pm$   0.016&214.000&   1.003$\pm$   0.018\\
		&red  &   79&  81.978$\pm$   4.573&   1.007$\pm$   0.047&89.000&   0.921$\pm$   0.051\\
\hline
		&blue &   286& 291.876$\pm$   -&   0.981$\pm$   -&278.700&   1.047$\pm$\\
HD139669&green&   140& -$\pm$-&   -$\pm$   -\\
		&red  &   53&  55.507$\pm$   -&   0.971$\pm$   -&72.800&   0.762$\pm$\\
\hline
		&blue &   196& -$\pm$   -&   -$\pm$   -&\\
HD41047 &green&   95& 101.709$\pm$   -&   0.938$\pm$  -&99.400&   1.023$\pm$   0.000\\
		&red  &   36&  33.121$\pm$   -&   1.114$\pm$   -&52.800&   0.627$\pm$   0.000\\
\hline
		&blue &   153& 150.013$\pm$  -&   1.025$\pm$  -&146.500&   1.024$\pm$-\\
HD170693&green&   75&  75.222$\pm$  -&   1.001$\pm$  -75.100&   1.002$\pm$-\\
		&red  &   29&  32.628$\pm$  -&   0.902$\pm$  -31.200&   1.046$\pm$-\\
\hline
		&blue &   115& 114.789$\pm$   -&   1.010$\pm$   -&109.200&   1.051$\pm$   -\\
HD138265&green&   56&  56.997$\pm$   -&   0.997$\pm$   -&57.200&   0.996$\pm$-\\
		&red  &   22&  29.010$\pm$  -&   0.766$\pm$  -&31.200&   0.930$\pm$-\\
\hline
		&blue &   64&  64.754$\pm$   -&   0.991$\pm$   -&60.700&   1.067$\pm$\\
HD159330&green&   31&  33.194$\pm$  -&   0.951$\pm$  -32.900&   1.009$\pm$-\\
		&red  &   12&  17.104$\pm$   -&   0.719$\pm$   -&-&-$\pm$-\\
\hline
		&blue &   39&  40.308$\pm$  -&   0.977$\pm$  -&35.100&   1.148$\pm$-\\
HD152222&green&   19&  21.103$\pm$   -&   0.915$\pm$   -22.400&   0.942$\pm$-\\
		&red  &   7&  11.406$\pm$   4.415&   0.726$\pm$   0.388&6.200&   1.840$\pm$   0.712\\
\hline
		&blue &   30&  32.308$\pm$   -&   0.956$\pm$   -&29.800&   1.084$\pm$   -\\
HD39608 &green&   15&  -$\pm$   -&   -$\pm$   -&\\
		&red  &   5&  17.459$\pm$   -&   0.338$\pm$   -&18.200&   0.959$\pm$   -\\
\hline
		&blue &   22&  24.930$\pm$   -&   0.919$\pm$   -&7.700&   3.238$\pm$-\\
HD15008&green&   11&  12.118$\pm$   -&   0.924$\pm$   -&-&   -$\pm$-\\
		&red  &   4&  -$\pm$   -&   -$\pm$   -\\

\end{tabular}
\caption{Comparison of the flux predicted by photospheric models and the flux value measured by our pipeline. The names of the calibrators are listed in the first column. The second column shows the band. The predicted flux value is listed in the third column. The average measured flux and the corresponding 1$\sigma$ uncertainty are in the fourth column. Column five shows the flux ratio.}
\label{faintfluxratiotable}
\end{table}

\begin{figure}[H]
\includegraphics[width=0.33\textwidth]{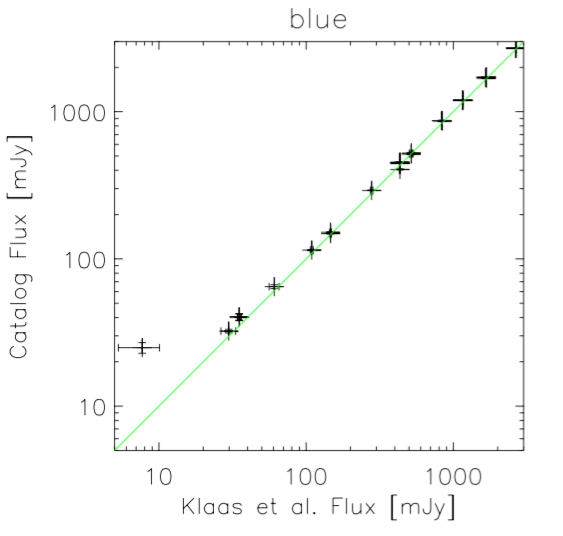}
\includegraphics[width=0.33\textwidth]{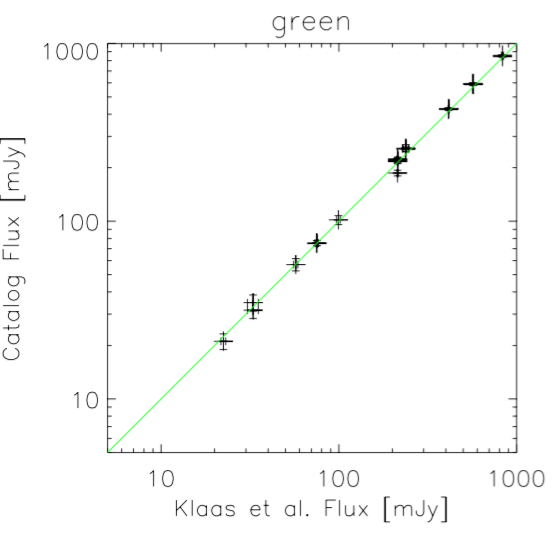}
\includegraphics[width=0.33\textwidth]{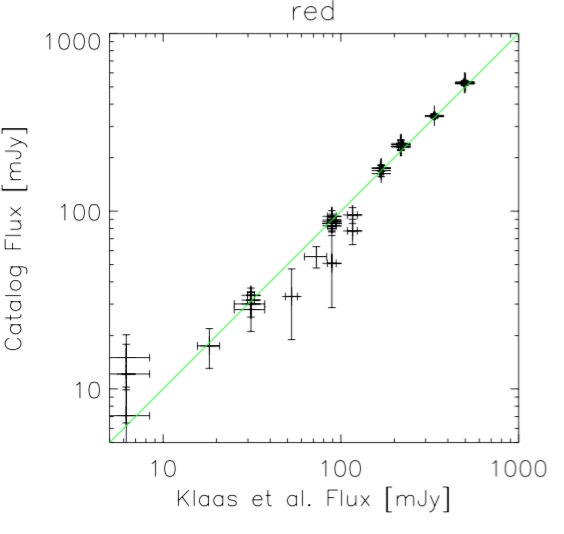}
\caption{From left to right, respectively, HPPSC flux densities for faint stars as a function of the values reported by Klaas et al. in the blue, green and red bands.}\label{faintcalibrators}
\end{figure}

\subsubsection{PEP-COSMOS field}

The COSMOS field (RA:150.11917	Dec:2.20583) has been observed by \href{http://www.aanda.org/articles/aa/full_html/2011/08/aa17107-11/aa17107-11.html}{Lutz et al (2011)}  for the Guaranteed Time Key Programme PACS Evolutionary Probe (PEP).  COSMOS is the  largest field observed  by  this programme (2 square degrees, 200h). We compared the PEP blind catalogues with  the HPPSC catalogue.  PEP sources have been  extracted via the Starfinder code \href{https://arxiv.org/abs/astro-ph/0009177}(Diolaiti et al 2000). Only sources with a S/N higher than three have been included in DR1 PEP catalogues.

We have carried out the cross-match applying TOPCAT between our Point Source Catalogue  and the PEP-COSMOS list, 
using a search radius  of  4 and  6$\arcsec$ for the 100 and 160$\mu$m catalogues, respectively (these values are near to the corresponding HWHM in each band). 

The HPPSC catalogue has 5607   and 2663  sources  in the COSMOS region. It must be noted that there are other observations, corresponding to the \href{(http://candels.ucolick.org/)}{CANDELS} proposal  (not included in PEP catalogues), that contribute to populate our catalogue. In this comparison, we do not use the slightly extended sources in the field ( i.e., the HPESL) hence the percentages of  cross-matches will be smaller than true values.  The PEP catalogue has 7443 objects at 100$\mu$m and 7047 sources at 160 $\mu$m in the COSMOS field. The number of matching sources is 3866 and 2335 respectively for the green and the red bands.  Hence  65\%-83\% of HPPSC sources match  PEP objects, while    
52\%-33\% of PEP sources found in the HPPSC catalogue.  

\begin{itemize}
\item  Separation

The position of matched sources is consistent, with most of them agreeing to a few arcsec (see Figure~\ref{rmatch_cosmos}). Mean values for the separation between objects are 1.9 and 2.3$\arcsec$ (for 100 and 160$\mu$m, respectively), values much lower that the selected search radius in each band. 

\begin{figure}[H]
\begin{center}
\includegraphics[width=0.43\textwidth]{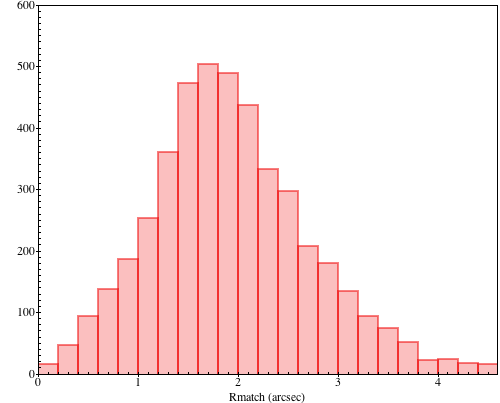}
\includegraphics[width=0.43\textwidth]{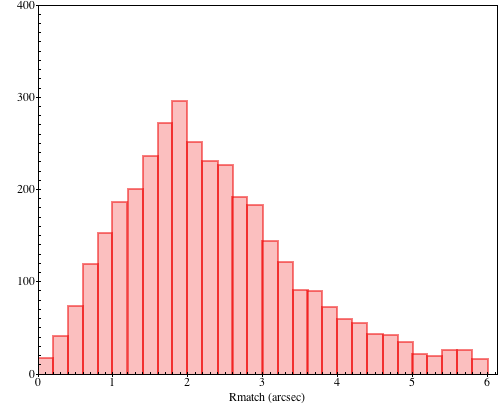}
\caption{From left to right: 100$\mu$m and 160$\mu$m  separation between HPPSC and PEP-COSMOS sources. }\label{rmatch_cosmos}
\end{center}
\end{figure}

\item  Flux ratios

We have compared  the starfinder PEP-COSMOS fluxes to our aperture fluxes in the matched catalogue.  The results are shown in Figures~\ref{flux_cosmos} and~\ref{rflux_cosmos} . The fluxes are in good agreement, with resulting mean differences  between fluxes  of around 3\% for the green band and 8\% for the red band.

\begin{figure}[H]
\begin{center}
\includegraphics[width=0.43\textwidth]{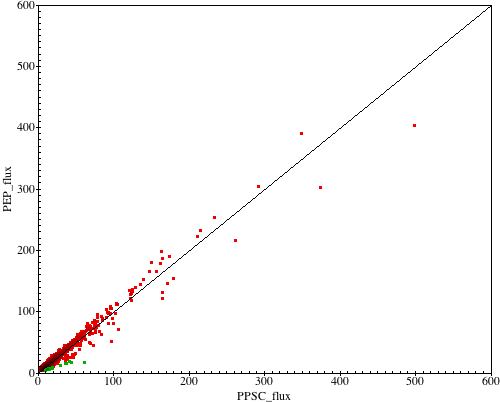}
\includegraphics[width=0.43\textwidth]{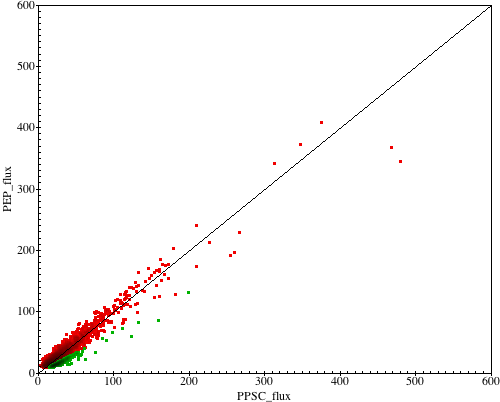}
\caption{From left to right: 100$\mu$m and 160$\mu$m HPPSC vs PEP-COSMOS fluxes. All matches (red points) and outliers (green points) are plotted. }\label{flux_cosmos}
\end{center}
\end{figure}

\begin{figure}[H]
\begin{center}
\includegraphics[width=0.43\textwidth]{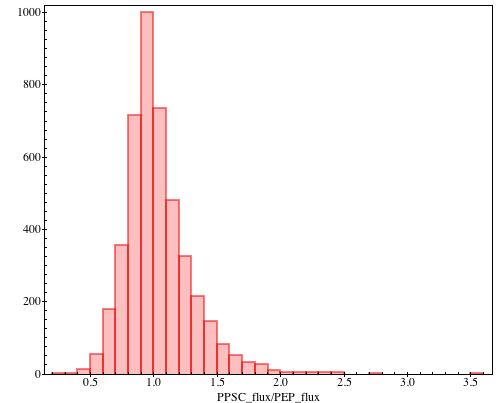}
\includegraphics[width=0.43\textwidth]{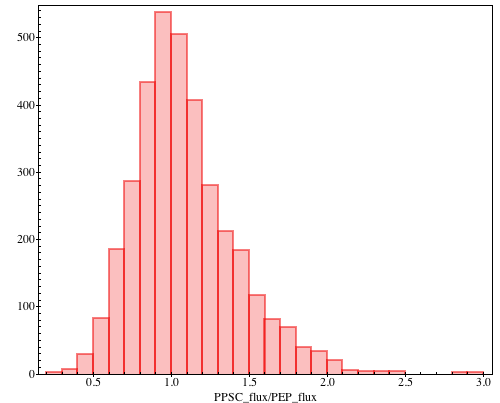}
\caption{From left to right: 100$\mu$m and 160$\mu$m HPPSC/PEP-COSMOS ratio fluxes. }\label{rflux_cosmos}
\end{center}
\end{figure}

\item Outliers

As shown Figure~\ref{flux_cosmos}, there are objects that can be considered as outliers in both bands. There are  17 objects in the green band (see Figure~\ref{outliers_cosmos100}) for which  PEP fluxes are systematically lower than  HPPSC fluxes.  Of  these, 6 HPPSC sources have a double match  in the PEP-COSMOS catalogues, and  the sum of the values from the two PEP sources in the match radius is similar to the HPPSC flux .  The other  12 sources have  a unique match. For these objects, the differences between fluxes can be explained because they are weak sources (PEP fluxes are $\sim$5mJy), the distance between peaks is larger that the median value of rmatch and by the use of two distinct procedures (aperture and  starfinder) to  obtain  the fluxes.

\begin{figure}[H]
\begin{center}
\includegraphics[width=0.63\textwidth]{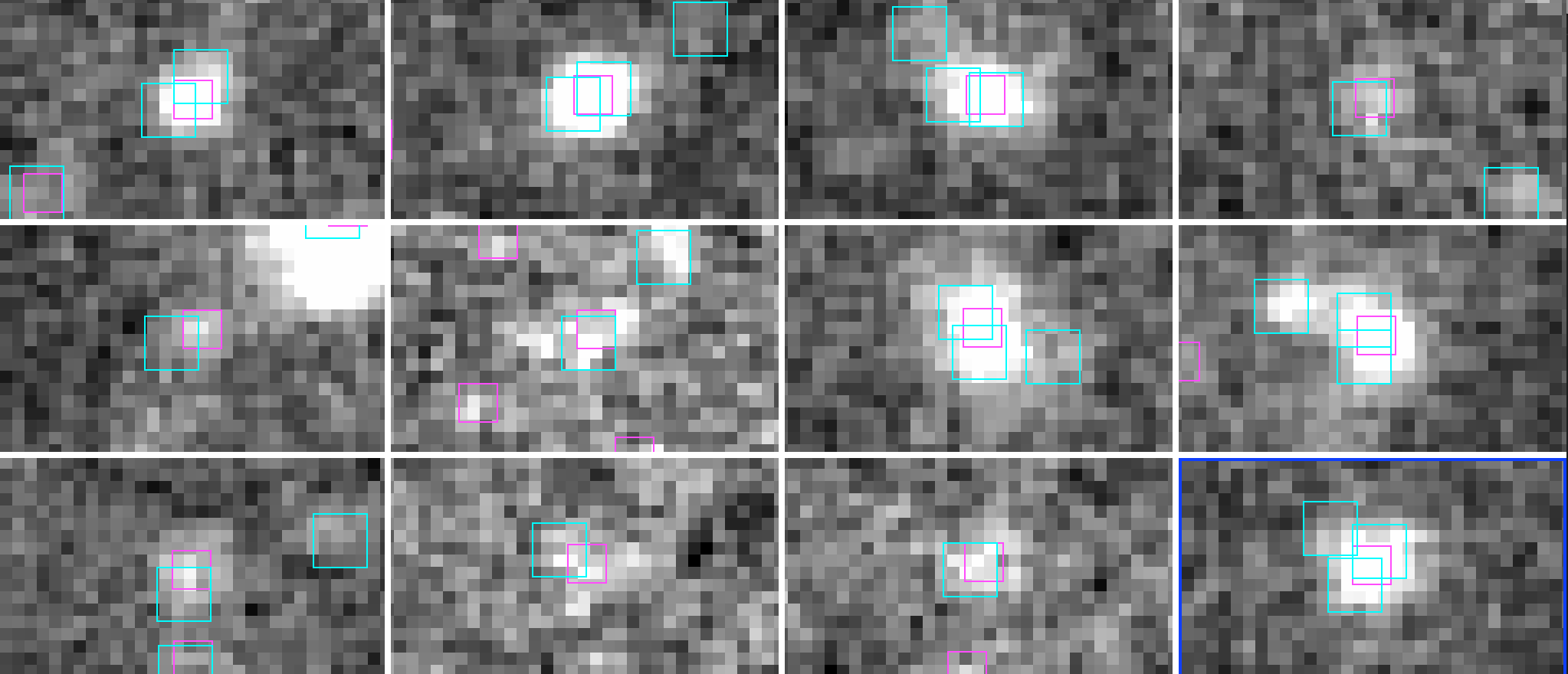}
\caption{Thumbnails of some outliers found in the HPPSC/PEP-COSMOS match at 100$\mu$m (19). Magenta squares: HPPSC sources. Cyan squares: PEP sources. }\label{outliers_cosmos100}
\end{center}
\end{figure}

In the red band, there are 39 outliers.  As well as the green band,  9 are not unique detections,  and  the flux of  HPPSC source is similar  to  the sum of  PEP fluxes of the contributing sources.  For 

\begin{figure}[H]
\begin{center}
\includegraphics[width=0.83\textwidth]{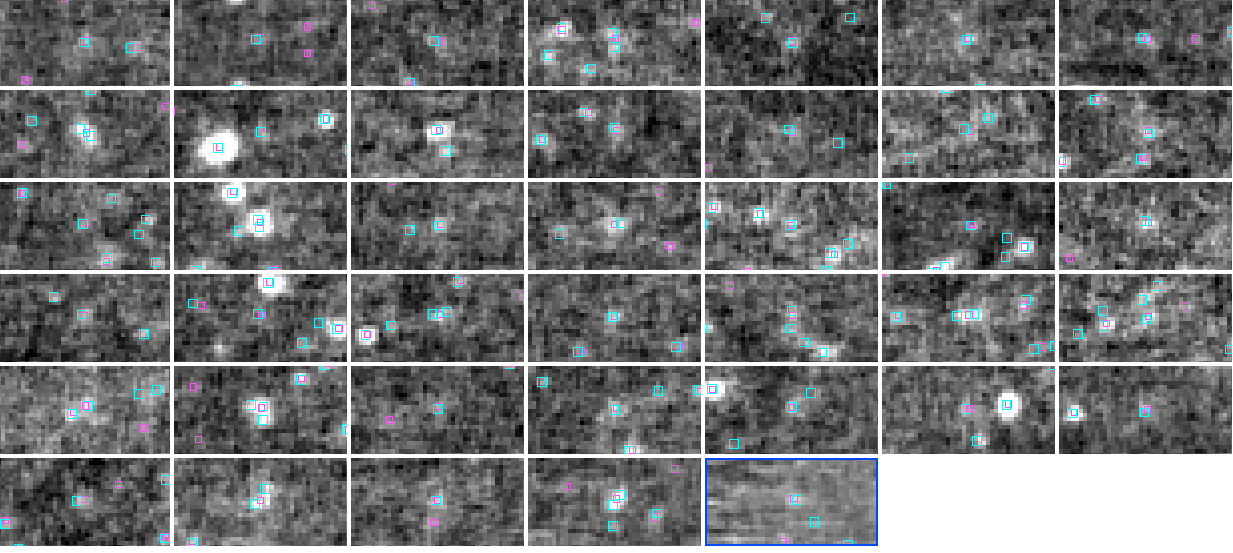}
\caption{Thumbnails of some outliers found in HPPSC/PEP-COSMOS match  at 160$\mu$m. (40). Magenta squares: HPPSC sources. Cyan squares: PEP sources. }\label{outliers_cosmos160}
\end{center}
\end{figure}

\end{itemize}

\subsubsection{PEP-GOODS field}

GOODS-South  (RA: 53.12654	Dec:–27.80467) is the only field observed by PEP  at 70$\mu$m, covering  17x11 arcmin$^2$. The number of sources in this field of the PPS catalogue is 199, while the PEP catalogue has 486 sources. We have matched both catalogues with a search radius of 3$\arcsec$. About  83\% of HPPSC sources are seen by PEP catalogue. The positions of matches  agree well (mean 2.4$\arcsec$).
Fluxes are consistent with a scatter of 10\% (see Figure~\ref{goods_70_pep}.

\begin{figure}[H]
\begin{center}
\includegraphics[width=0.30\textwidth]{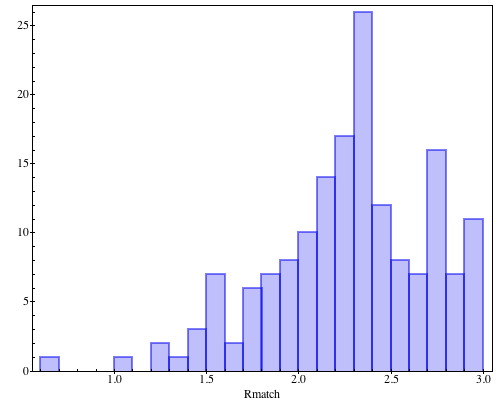}
\includegraphics[width=0.30\textwidth]{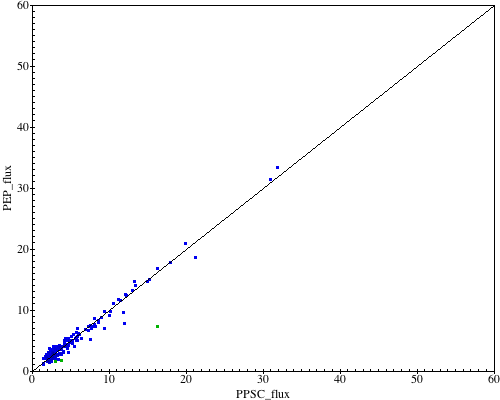}
\includegraphics[width=0.30\textwidth]{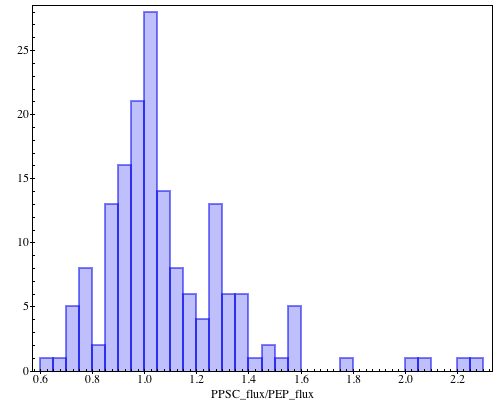}
\caption{From left to right: 70$\mu$m  separation between HPPSC and PEP-GOODS-South sources, HPPSC vs. PEP-GOODS-South fluxes and HPPSC/PEP-GOODS-South flux ratio.\label{goods70_pep}}
\end{center}
\end{figure}

There is a subgroup of matches that have  a flux that is considerably smaller in the PEP catalogue (5 sources). One of them has a double match in PEP catalogue , with  a sum of individual fluxes  very close to the HPPSC flux.  The other  four objects are weak  (see Figure~\ref{goods70_pep}) so the differences can be explained by different method to extract  fluxes.

\begin{figure}[H]
\begin{center}
\includegraphics[width=0.83\textwidth]{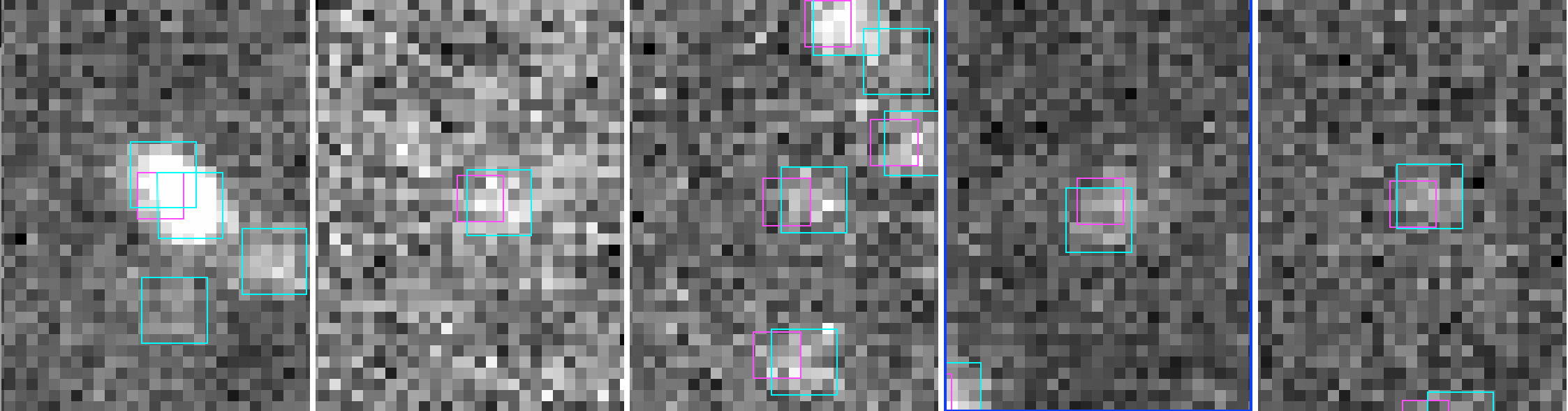}
\caption{Thumbnails of outliers found in the HPPSC/PEP-GOODS match  at 70$\mu$m. Magenta squares: HPPSC sources. Cyan squares: PEP-GOODS sources. }\label{outliers_goods70}
\end{center}
\end{figure}

\textbf{Conclusions}\\

The positions of matches generally agree well and the scatter is consistent with the absolute pointing uncertainty of \textit{Herschel}. 

The fluxes of matching sources are generally consistent with a scatter of about 10\% or better. However, there are some sources where PEP  fluxes are considerably smaller than the SPSC fluxes. These are cases in which the source is   multiple and the constituents are closer together than the FWHM at the  wavelength and thus are not distinguished by the HSPSC extraction algorithm or weak sources for which the difference between the peaks are larger than the mean difference. We conclude that the agreement between the two catalogues is quite good.

\subsubsection{Goods-{\it Herschel} survey}
GOODS-{\it Herschel} \href{http://www.aanda.org/articles/aa/full_html/2011/09/aa17239-11/aa17239-11.html}{Elbaz et al, 2011} is an ESA Open Time Key Project consisting of the deepest \textit{Herschel} observations of the two Great Observatories Origins Deep Survey (GOODS) fields in the Northern and Southern hemispheres. Flux densities and their associated uncertainties were obtained from point source fitting using  prior 24$\mu$m positions. 

\begin{figure}[!h]
\begin{center}
\includegraphics[width=0.43\textwidth]{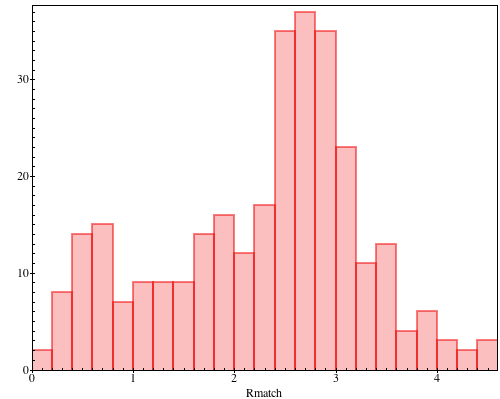}
\includegraphics[width=0.42\textwidth]{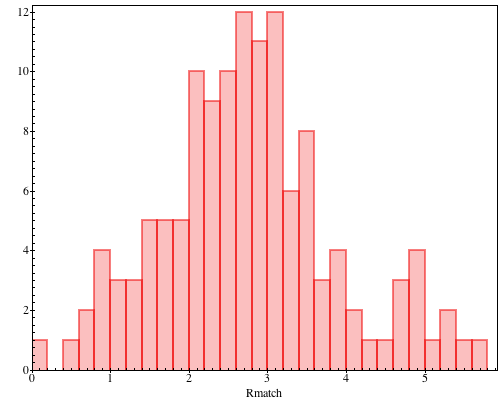}
\caption{From left to right: 100$\mu$m and 160$\mu$m separation between HPPSC and GH-GOODS-South sources.}\label{goods_gh}
\end{center}
\end{figure}

For PACS-100 $\mu$m, they have used MIPS-24 $\mu$m priors down to the 3$\sigma$ limit, imposing a minimum flux density of 20 $\mu$Jy. For PACS-160 $\mu$m and SPIRE-250 $\mu$m, they restricted the 24 $\mu$m priors to the 5$\sigma$ (30 $\mu$Jy) limit (reducing the number of priors by about 35\%).
GH-GOODS sources are flagged  with a "clean\_index" . This flag  marks contamination by nearby sources. Only sources with a clean\_index = 0 have a reliable flux in the GH catalogue.

\begin{figure}[H]
\begin{center}
\includegraphics[width=0.43\textwidth]{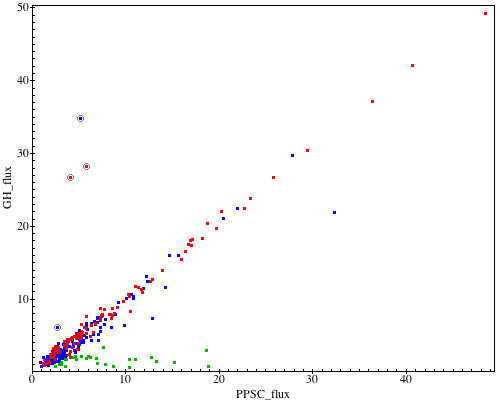}
\includegraphics[width=0.42\textwidth]{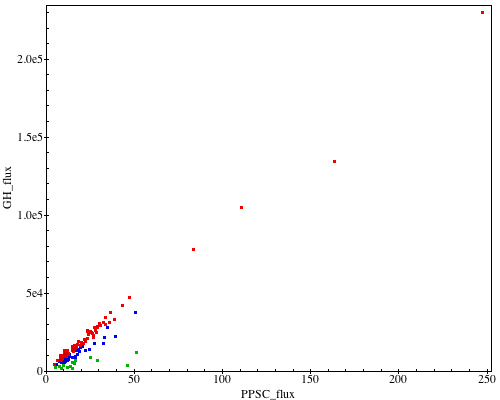}
\includegraphics[width=0.43\textwidth]{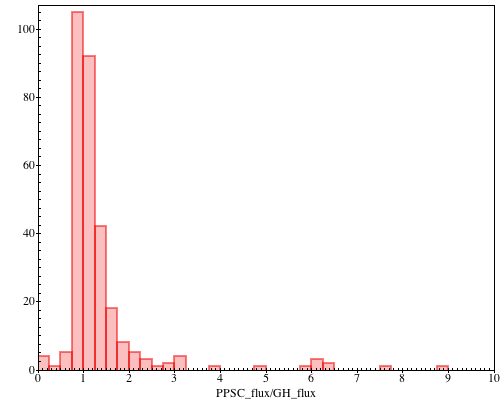}
\includegraphics[width=0.42\textwidth]{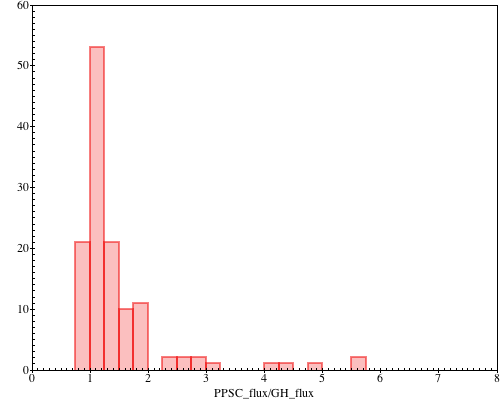}
\caption{From left to right and top to bottom: 100$\mu$m   and 160$\mu$m  HPPSC vs GH-GOODS-South fluxes and HPPSC/GH-GOODS-South fluxes. Red points: all matches, Green Points: outliers with HPPSC/GH fluxes > 2, Blue points: clean sources, Grey circles: outliers with GH/HPPSC fluxes > 2}\label{goods_gh_flux}
\end{center}
\end{figure}

\begin{figure}[H]
\begin{center}
\includegraphics[width=0.83\textwidth]{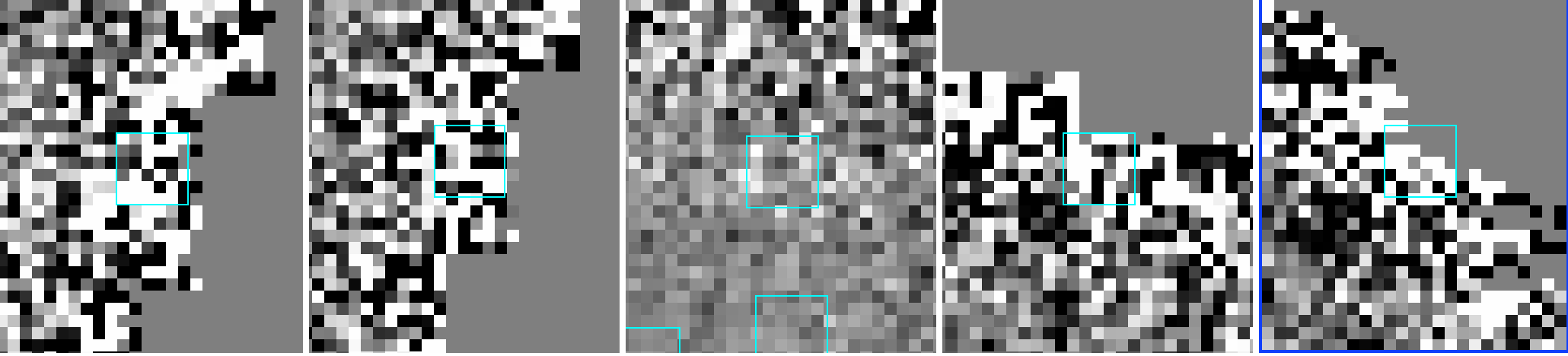}
\caption{Thumbnails of 5 objects found in HPPSC/GH-GOODS-South match  with GH-GOODS-S/HPPSC fluxes ratio larger than 2. Cyan squares: GH-GOODS-South sources. }\label{outliers2_goods100}
\end{center}
\end{figure}

The number of matches is 304 in the green band, and  130  at the red band.  Match radii have a mean below 2$\arcsec$ (green band) and 3$\arcsec$ (red band).  If we consider all matches, the ratio between the HPPSC and the GH-GOODS-S fluxes is 1.6 in two bands. We have found 29 (green) and 14 (red) outliers (with flux ratio bigger than 2), but if we constrain our comparison to clean sources, the flux ratios are 1.01 and 1.03, for the green and the red bands, respectively. Furthermore, none of the outliers are clean sources.
At 100 $\mu$m, we detect also  5 sources which have  considerably larger fluxes  in  the GH-GOODS catalogue.  We show these objects in Figure~\ref{outliers2_goods100}. All of them are at the edge of the field , such that the  GH-GOODS fluxes are not  accurate.



In conclusion,  the agreement (both in position and in flux) between the two catalogues  is really excellent, if we consider only 'clean sources ' (sources without contamination by other objects) in the GH-catalogue, despite the use of different methods to detect and extract  the flux of sources. The flux discrepancies  between clean sources and our sources are explained due to them being  at the edge  of the GH images.

\subsubsection{The Galactic Cold Cores OTKP}
The "Galactic Cold Cores: A \textit{Herschel} survey of the source populations revealed by Planck" was an Open Time Key Programme (Proposal ID: KPOT$\_$mjuvela$\_$1, PI:Mika Juvela). The goal of the project was to study starless cores and the initial conditions of star formation using the 	\textit{Herschel} satellite and the combined power of its PACS and SPIRE instruments. Their catalogue was published in \href{http://adsabs.harvard.edu/abs/2015A\%26A...584A..92M}{Montillaud et al. (2015)}. The \href{http://www.herschel.fr/Phocea/file.php?class=page&file=6/aa18797-12_Men_shchikov_etal2012.pdf}{\textit{getsources}} multiwavelength source extraction algorithm was employed in their work to build a catalogue of several thousand cold sources, extracted from 116 fields. {\textit{getsources}} is a powerful multi-scale, multi-wavelength source extraction algorithm, designed primarily for use in FIR surveys of star-forming regions observed with \textit{Herschel}. 

\begin{figure}[H]
\begin{center}
\includegraphics[width=0.90\textwidth]{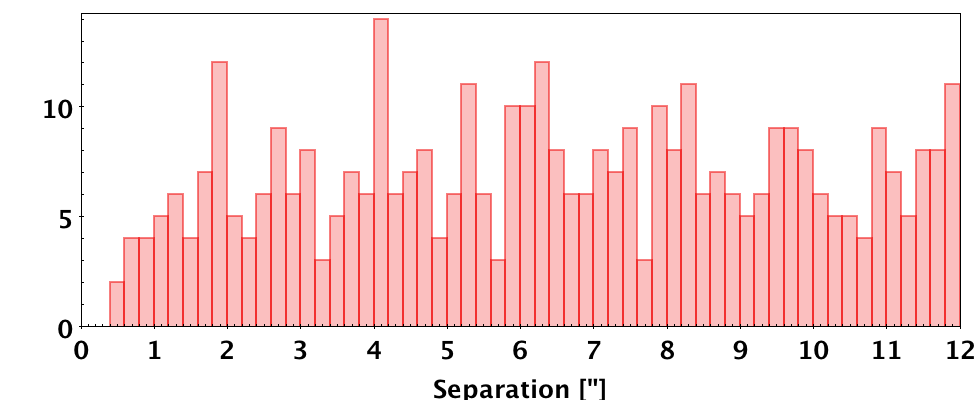}
\caption{Angular separation between the overlapping catalogue sources.}\label{gccseparation}
\end{center}
\end{figure}

The GCC catalogue contains 4466 rows, listing band-merged data. Most of their targets were very cold ($<$10K) sources, therefore their PACS data is not very reliable, and only 160$\mu$m flux densities are reported. To match their findings with our catalogue we used the source matching tool of TOPCAT. The maximum separation allowed was 12$\arcsec$. 400 cases were found in which a GCC object was found within a 12$\arcsec$ radius of our HPPSC sources. The distribution of the angular separations between the matching sources is shown on Figure~\ref{gccseparation}. The distribution is rather uniform, suggesting that the matches are more coincidental than systematic.

\begin{figure}[H]
\begin{center}
\includegraphics[width=0.44\textwidth]{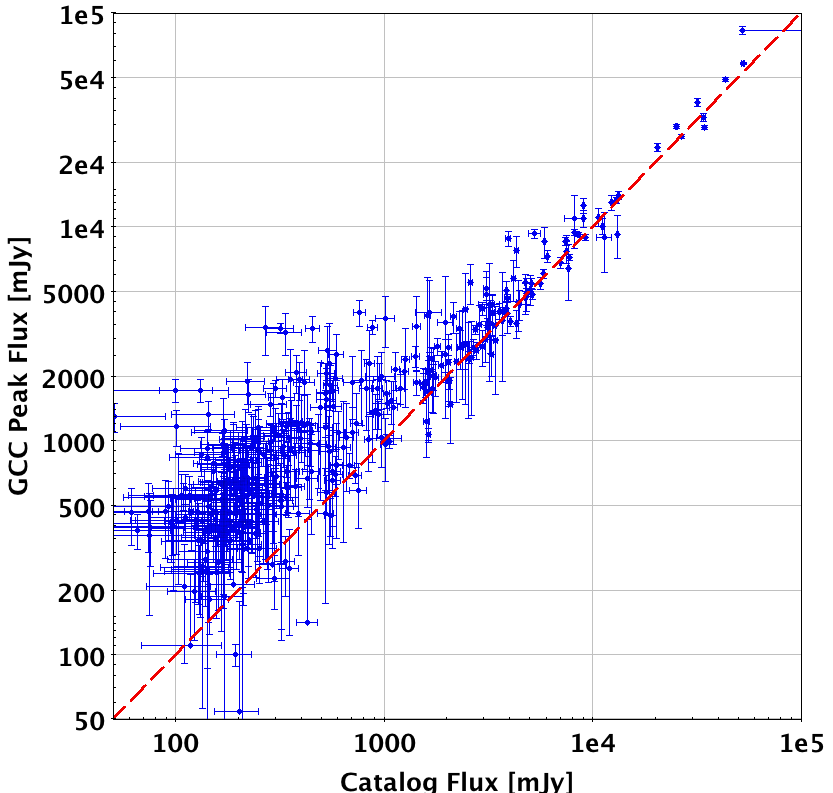}
\includegraphics[width=0.44\textwidth]{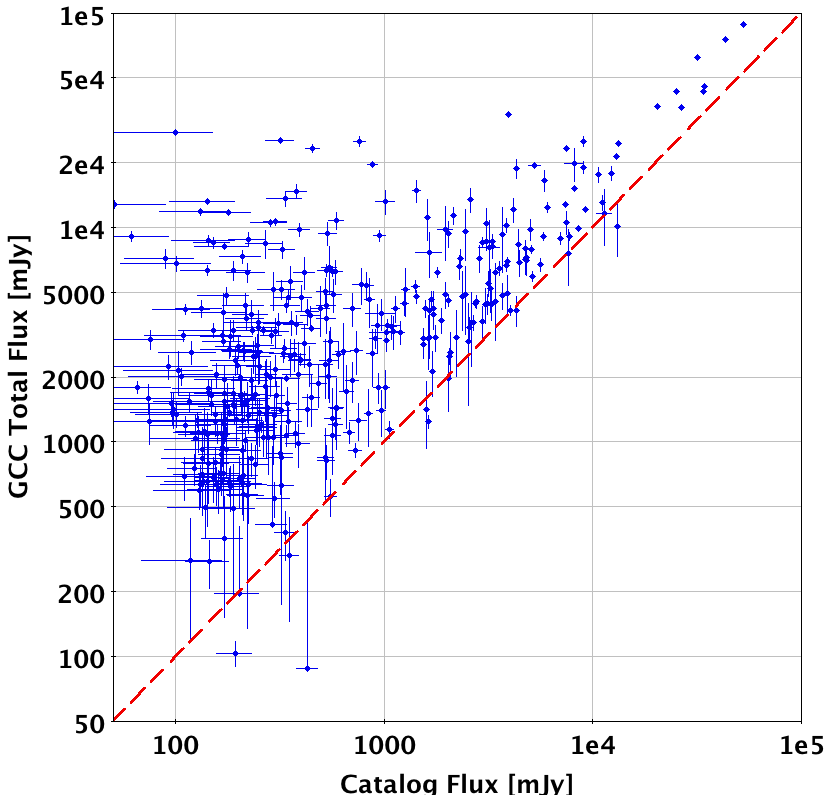}
\caption{Left: The GCC peak flux as measured by \textit{getsources} as a function of our HPPSC flux. Right The total GCC flux as integrated by getsources as a function of the HPPSC flux. Error bars represent the corresponding flux uncertainty reported by the GCC catalogue and the structure noise based statistical error for our measurements.}\label{gccflux}
\end{center}
\end{figure}

\textit{getsources} calculates two different flux values. One of them is the peak flux, while the other is the integrated flux measured inside the footprint of the source. We compared both with our HPPSC flux values. Figure~\ref{gccflux} shows the GCC peak (left) and the GCC total (right) flux as a function of our values. A perfect match is shown by the dashed red line. The corresponding errors for our values are the ones calculated from the structure noise based S/N$_S$ value. The figures show that their peak flux is in better agreement with our fluxes, especially at the bright end of the diagram. The integrated GCC fluxes are systematically higher than the ones we calculate by aperture photometry. 

\begin{figure}[H]
\begin{center}
\includegraphics[width=0.44\textwidth]{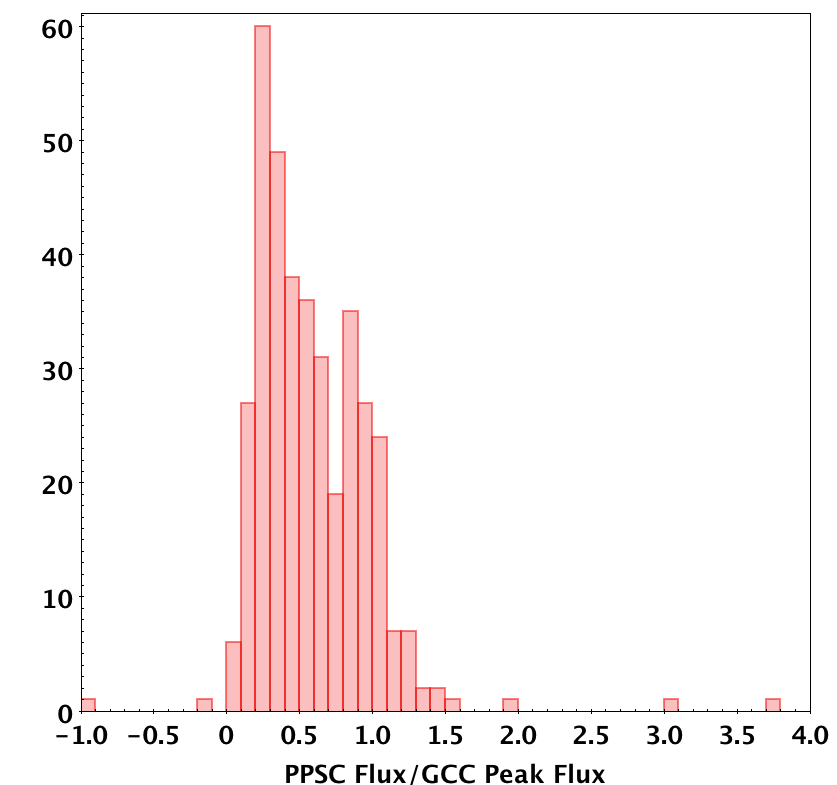}
\includegraphics[width=0.44\textwidth]{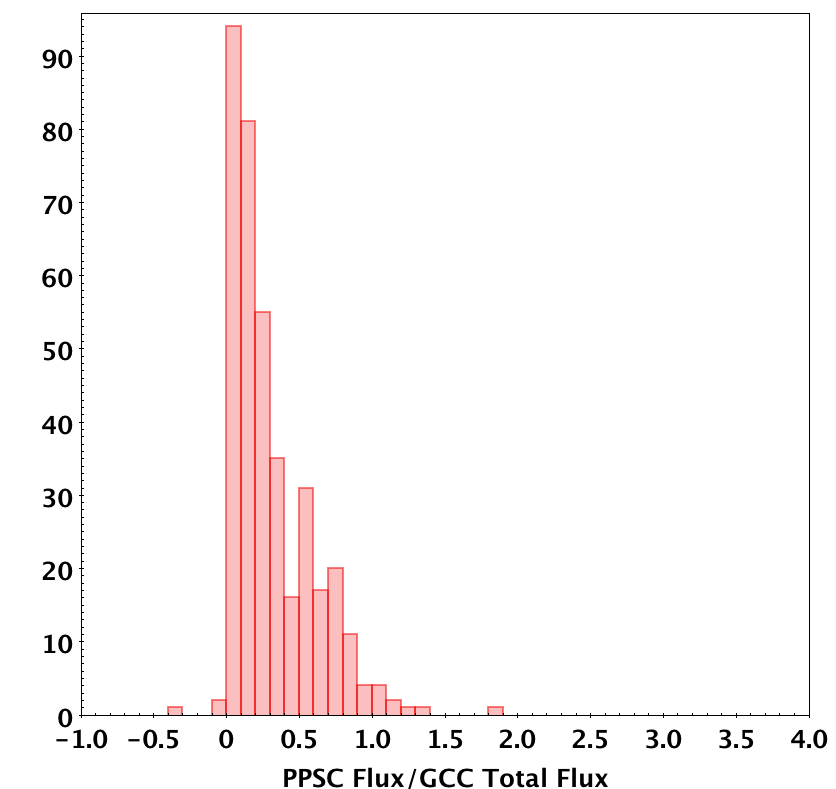}
\caption{Left: The HPPSC flux over the GCC peak flux. Right: Flux ratios obtained by using the integrated GCC flux.}\label{gccfluxratio}
\end{center}
\end{figure}

The size of the difference can be seen if we plot the flux ratios for both cases. Figure~\ref{gccfluxratio} shows the HPPSC flux over the GCC peak flux (right) and the HPPSC flux divided by the GCC integrated flux (right). A secondary maximum is seen in the left image at $\sim$1, meaning that there are sources for which flux the agreement is good. In most cases, the ratio is below 1, meaning that the GCC peak flux is larger than the correspondong HPPSC value. Using the integrated GCC flux we found that in almost all cases the flux reported in the GCC catalogue is higher than our values.

The large difference between the GCC integrated flux and the HPPSC flux can easily be understood by looking at the size of the GCC source footprints. While our definition of point sources does not allow the fitted FWHM to be larger than 2.0 times the corresponding PSF FWHM ($<25\arcsec$), the GCC FWHM is 68$\arcsec$.28$\pm$20$\arcsec$.78. 

\subsubsection{Hi-GAL DR1}
The \href{http://iopscience.iop.org/article/10.1086/651314/pdf}{Hi-GAL} (\textit{Herschel}-infrared Galactic Plane Survey) project carried out a full survey of the Galactic Plane in 5 bands, including the blue and red PACS band in Parallel Mode. The \href{https://arxiv.org/pdf/1604.05911.pdf}{first data release (DR1)} from the Hi-GAL team covered the inner Milky Way in the longitude range $68\degr\geq l\geq-70\degr$
in a $|b| \leq 1\degr$ latitude strip. The Hi-GAL project used different data processing pipelines to the HPPSC in every aspect. The data reduction was carried out using the ROMAGAL data processing software described in details in \href{http://adsabs.harvard.edu/abs/2011MNRAS.416.2932T}{Traficante et al. (2011)}. Source extraction was carried out with the \href{http://www.aanda.org/articles/aa/pdf/2011/06/aa14752-10.pdf}{CuTEx} algorithm, standing for standing for Curvature Thresholding Extractor. The algorithm looks for the pixels in the map with the highest curvature by computing the second derivative of the map. All the “clumps” of pixels above a defined threshold are analysed and those larger than a certain area are kept as candidate detections. The output fluxes and sizes are determined by fitting simultaneously elliptical Gaussian functions plus a 2$^{nd}$-order 2D surface for the background.

\begin{figure}[H]
\begin{center}
\includegraphics[width=0.9\textwidth]{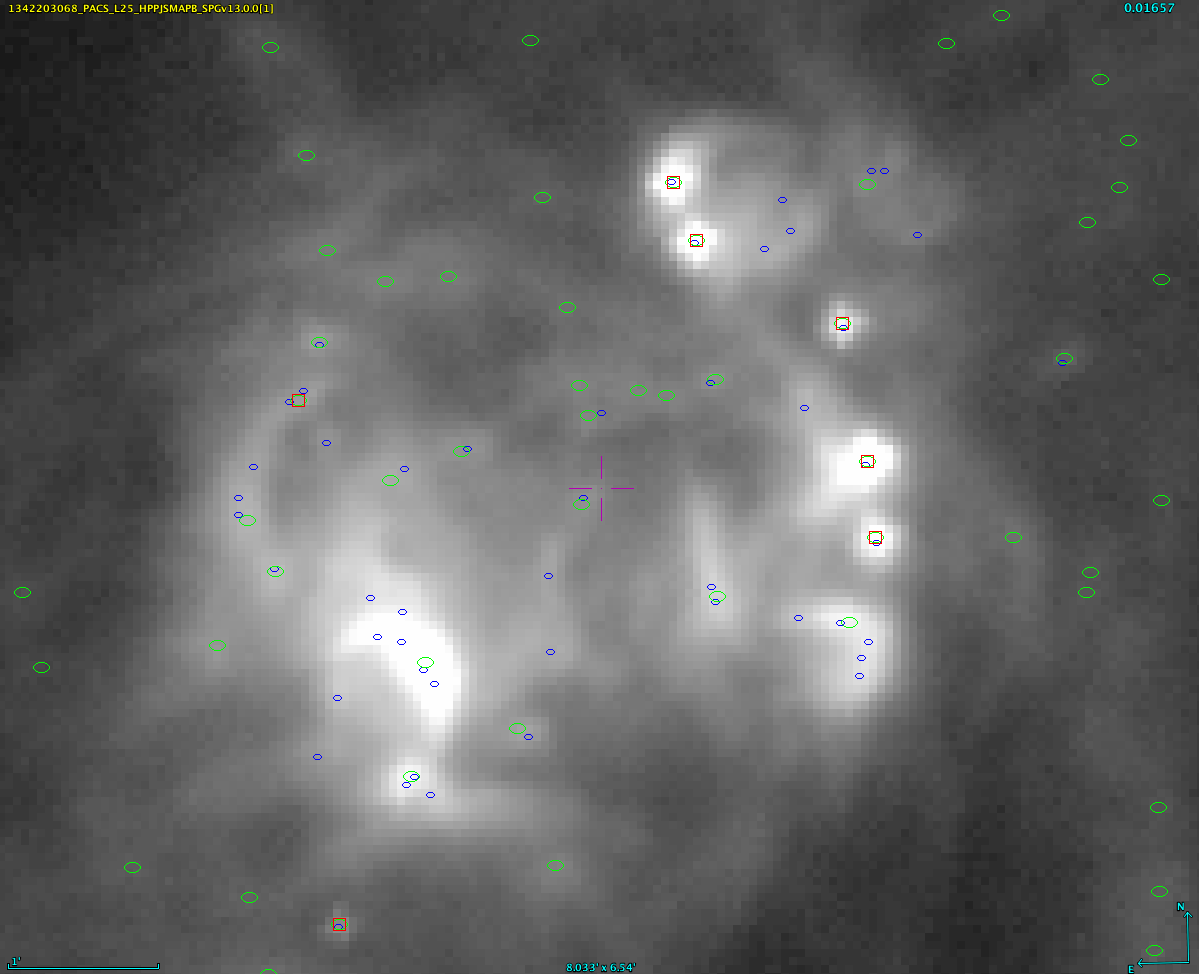}
\caption{Part of the Hi-GAL tile Field 297\_0 in the blue band. Blue circles indicate the sources from the Hi-GAL DR1 catalogue. The green ovals show all of our extractions from the region. The red squares show the objects in the final HPPSC.}\label{higalmap}
\end{center}
\end{figure}

The released catalogues contain information on 120\,581 and 291\,858 sources in the blue and red bands, respectively. We found 22\,684 and 40\,373 matching sources in the two bands. Figure~\ref{higalmap} shows a part of Hi-GAL image Field 297\_0 in the blue band. The picture demonstrates well how our criteria for source shape classification work. The HPPSC sources are well defined point sources on the image. We can also see clearly, how differently the CuTEx algorithm works. Many of the larger, extended objects are divided into individual peaks by the Hi-GAL DR1 pipeline. Similarly, there are many detections by SUSSEXtractor that are not recognised by CuTEx as sources. Our quick comparison with the HPESL showed that many objects from the Hi-GAL DR1 can be also found in our product. The number of matches is $\sim2.4\times 10^4$ and $\sim6.8\times 10^4$ in the blue and red bands, respectively, dependent on the radius used for matching the sources.

\begin{figure}[H]
\begin{center}
\includegraphics[width=0.44\textwidth]{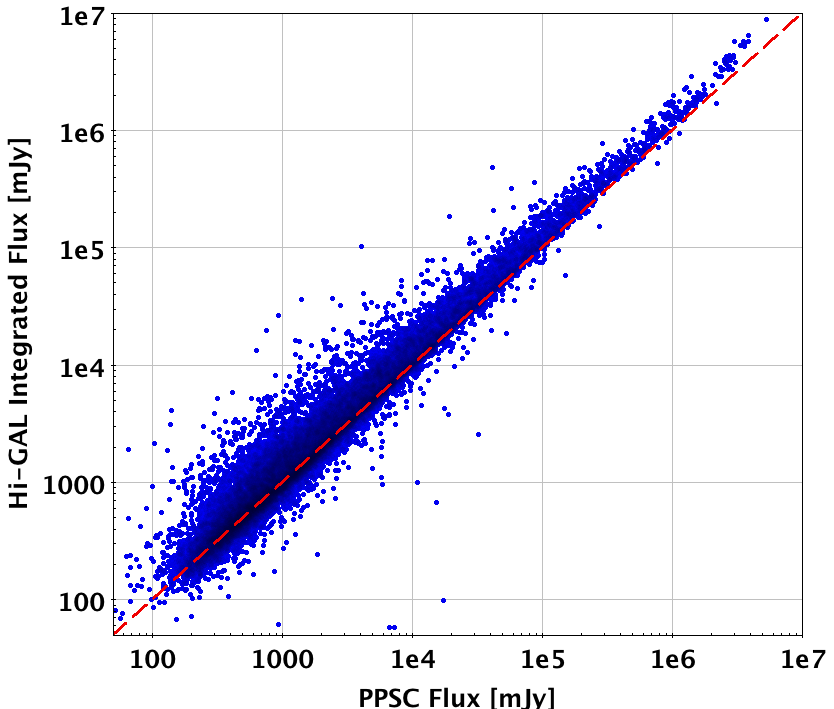}
\includegraphics[width=0.44\textwidth]{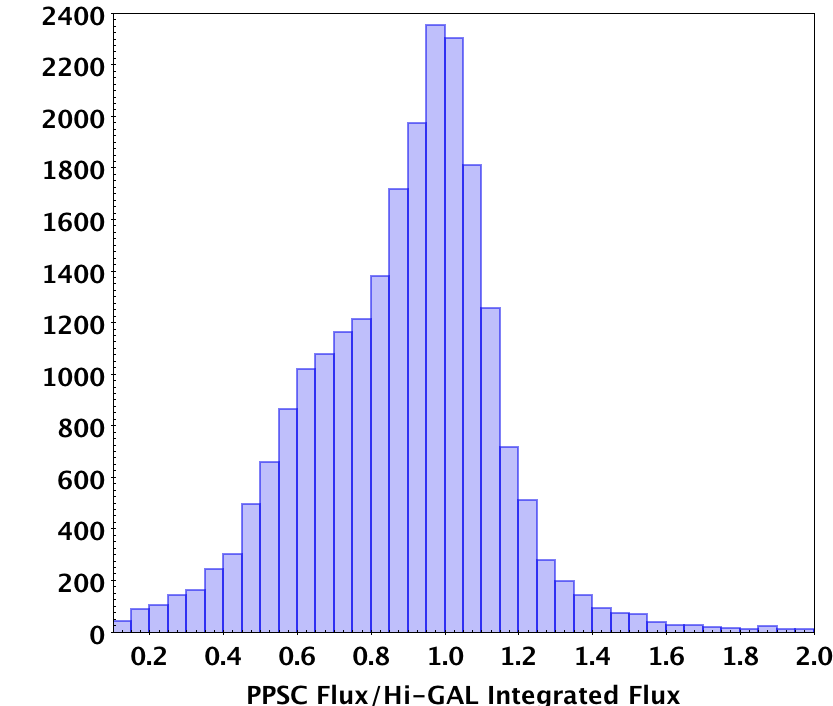}
\caption{Left: The HPPSC flux in the blue band over the Hi-GAL integrated flux. Right: Flux ratios by using the same flux values.}\label{higalblue}
\end{center}
\end{figure}

Our flux comparison shows a good agreement with the Hi-GAL extractions, especially for the blue band, see the left panel in Figure~\ref{higalblue}. The average flux ratio was found to be $0.91\pm0.24$. The distribution of the flux ratios is plotted on the right panel of Figure~\ref{higalblue}

\begin{figure}[H]
\begin{center}
\includegraphics[width=0.44\textwidth]{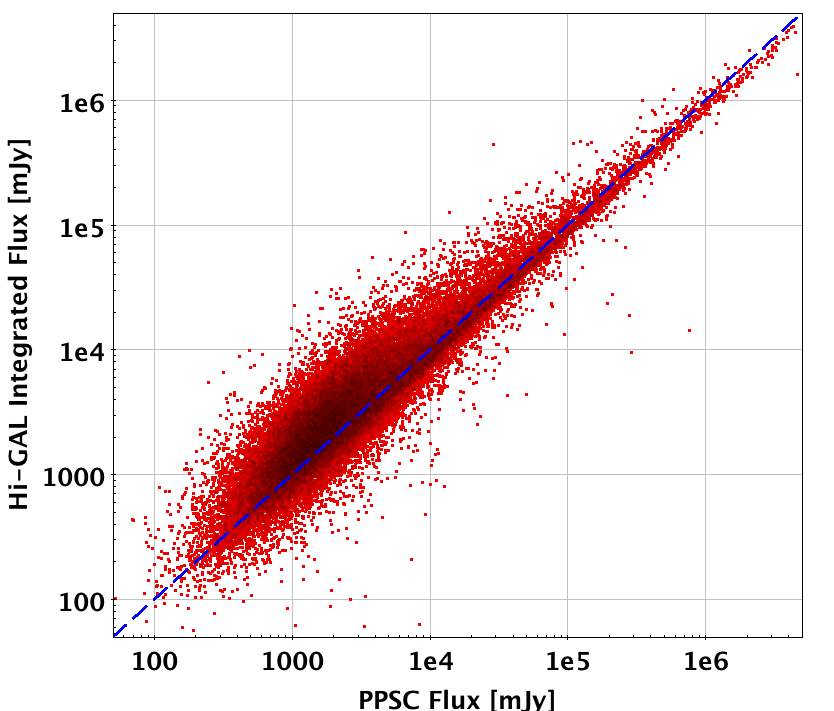}
\includegraphics[width=0.44\textwidth]{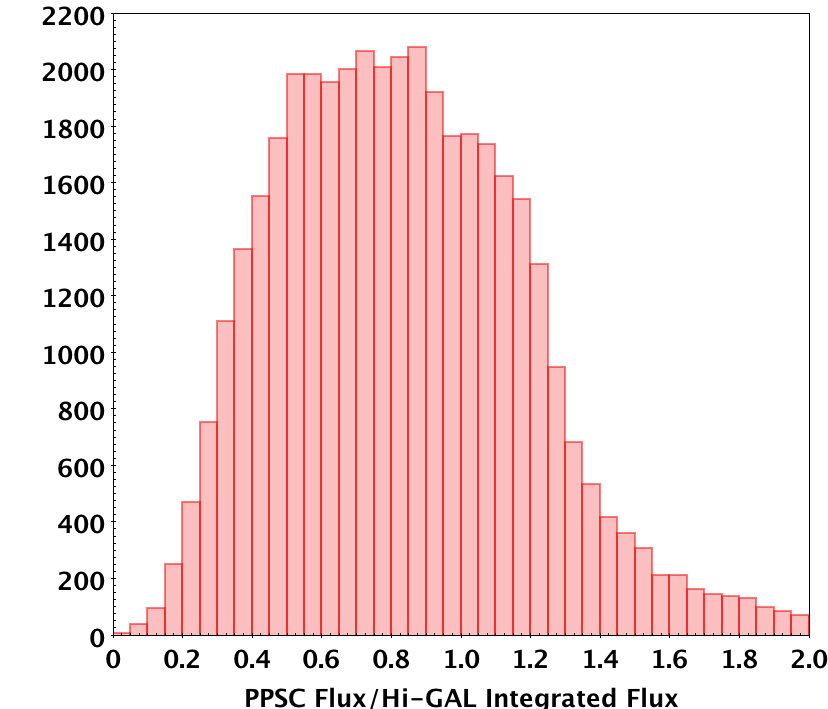}
\caption{Left: The HPPSC flux in the red band over the Hi-GAL integrated flux. Right: Flux ratios found by using the same flux values.}\label{higalred}
\end{center}
\end{figure}

The correlation is still very good, although the difference between the fluxes is slightly larger (see left panel on Figure~\ref{higalred}). The average flux ratio was found to be $0.88\pm0.36$ (right panel). 

\begin{figure}[H]
\begin{center}
\includegraphics[width=0.44\textwidth]{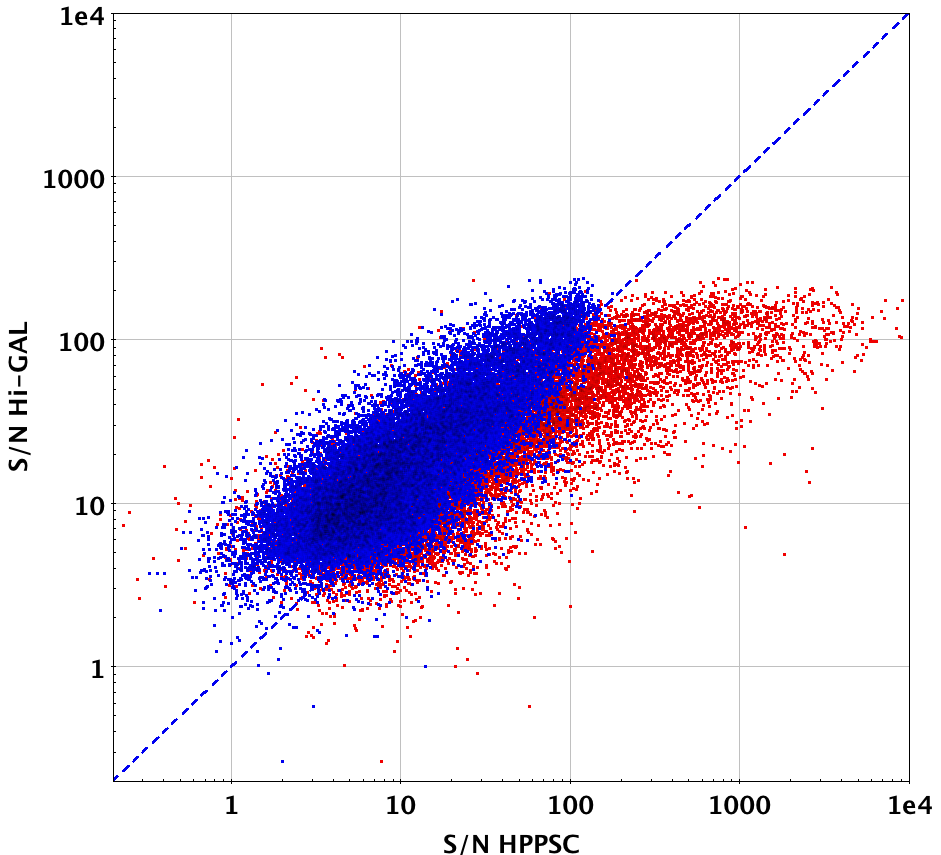}
\includegraphics[width=0.44\textwidth]{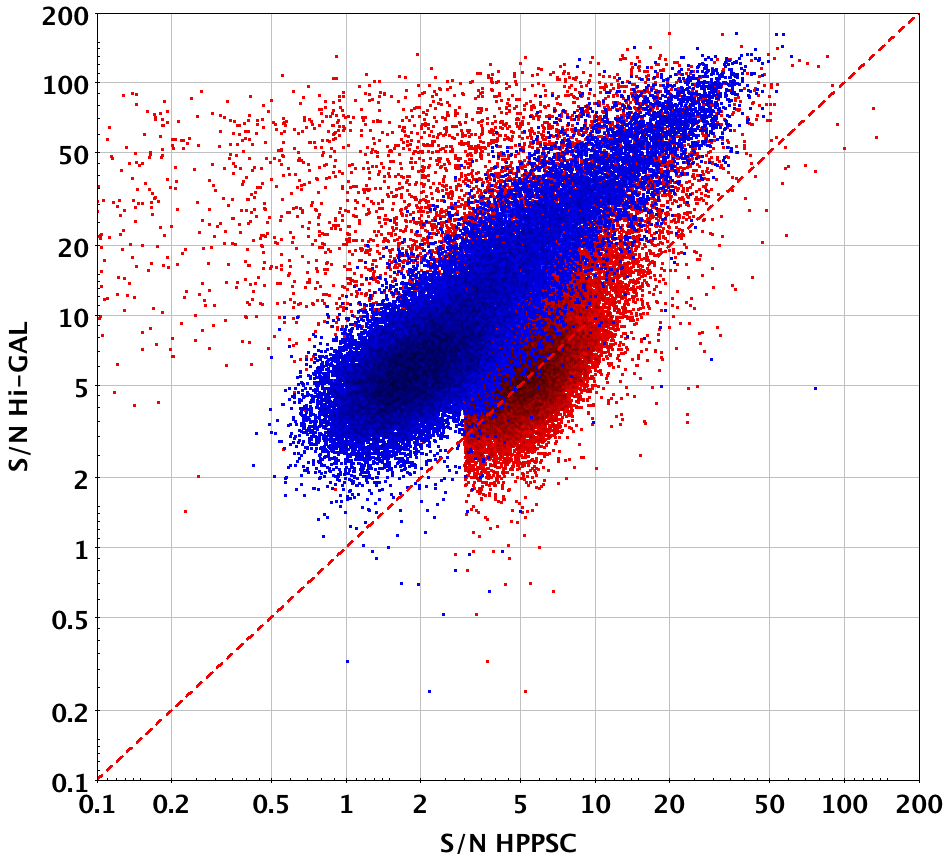}
\caption{Comparison of the HPPSC S/N$_S$ (red dots) and S/N$_R$ (blue dots) with the S/N values provided by Hi-GAL in the blue band (left panel) and in the red band (right panel).}\label{higalsnr}
\end{center}
\end{figure}

We have also compared our estimated S/N values to the S/N values calculated by the Hi-GAL team. As seen in Figure~\ref{higalsnr} for the blue band (left panel) the S/N$_R$ values (blue dots) occupy an interval of values that is similar to Hi-GAL. The Hi-GAL noise is also a background RMS but, instead of calculating it around the source, it is calculated in the position of the source, after source subtraction from the image. Our S/N$_S$ values cover a wider interval in the blue band, but they are similar to the Hi-GAL ones in the red band (right panel). The Hi-GAL S/N values are systematically higher than our S/N$_R$ values. 

\subsection{Conclusion}

Our simulations show that the photometry is most reliable in case of environments with low complexity and for sources brighter than $\sim$100 mJy. External validation by calibraton stars show that our photometry is very accurate and deviates from previous results by less than 1\%, especially in the case of the bright targets. Comparison with cosmological catalogues suggest that fluxes of matching sources are generally consistent with a scatter of about 10\% or better.  In the case of catalogues compiled from observations of star forming regions (therefore containing more complex extended emission) the agreement  is less good. The deviation of our fluxes from the Hi-GAL values is $\sim$10\%, with a scatter of $\sim$25-30\%.

Our completeness is good in fields with low N$_S$ values, but  in complex regions the bright extended emission does not allow the detection of faint sources. In deep extragalactic observations 90\% completeness can be reached at a source flux $\sim$30 mJy. In complex regions this level of completeness can be reached only at around 1 Jy based on our simulated data. Comparison with existing extragalactic catalogues shows that we are able to recover $\sim$50\% of their sources, and we have detections that are not listed in their work. The overlap with the Galactic Plane catalogues is  20-30\%, which may be caused by our criteria for point source classification. Most of the  sources catalogued in other works are slightly extended.

Recovered positions are in good agreement with the input positions in our simulations. Comparison with other works suggests that the positional accuracy is within the nominal pointing error, and definitely within the PSF FWHM.

\newpage
\subsection*{Acknowledgement}

A project of this magnitude has always many more contributors than can be found on the list of authors and small things can make a big difference. For that reason we would like to specifically acknowledge:
Bernhard Schulz and the SPIRE Point Source Catalogue team for experimenting the right path through the HPSC project and helping us on our way along. 

We would like to acknowledge Erika Varga-Vereb\'elyi and R\'obert Szak\'ats for their contribution in tests and SSO related  work. David Shupe has helped us in the exploration of our dataset. John Rector has helped us to create our database schema. G\"oran Pilbratt, Anthony Marston, David Teyssier, and Pedro Garcia Lario helped the project from the beginning. 

\textit{Herschel} is an ESA space observatory with science instruments provided by European-led Principal Investigator consortia and with important participation from NASA. 

PACS has been developed by a consortium of institutes led by MPE (Germany) and including UVIE (Austria); KU Leuven, CSL, IMEC (Belgium); CEA, LAM (France); MPIA (Germany); INAF-IFSI/OAA/OAP/OAT, LENS, SISSA (Italy); IAC (Spain). This development has been supported by the funding agencies BMVIT (Austria), ESA-PRODEX (Belgium), CEA/CNES (France), DLR (Germany), ASI/INAF (Italy), and CICYT/MCYT (Spain).

HIPE is a joint development by the 	\textit{Herschel} Science Ground Segment Consortium, consisting of ESA, the NASA \textit{Herschel} Science Centre, and the HIFI, PACS and SPIRE consortia.

This research has made use of data from the PEP (a guaranteed time key programme of 	\textit{Herschel}) and the GOODS-	\textit{Herschel} Survey  (an Open Time Key Programme of \textit{Herschel}). PEP data  was downloaded from the NASA/IPAC Science Archive (irsa.ipac.caltech.edu), and  GOODS-\textit{Herschel} data was accessed through the \textit{Herschel} Database in Marseille (http://hedam.lam.fr/). 

\smallskip
The production of the PACS Point Source Catalogue was supported by the European Space Agency and the Hungarian Space Office through grant PECS 4000109997/13/NL/KML.

\newpage

\bibliographystyle{alpha}
\bibliography{sample}

De Graauw,  Th., et al.,  "The Herschel-Heterodyne Instrument for the Far-Infrared (HIFI)",  2010,  A\&A, 518, L6

Diolaiti, E. et al. "Analysis of isoplanatic high resolution stellar fields by the StarFinder code", 2000, A\&AS, 147, 335

Elbaz D., et al, "GOODS–\textit{Herschel}: an infrared main sequence for star-forming galaxies", 2011, A\&A, 533, 119

Griffin,  M.J., et al., "he Herschel-SPIRE instrument and its in-flight performance",  2010,  A\&A, 518, L3

Kiss, Cs., et al,  " Determination of confusion noise for far-infrared measurements",  2005,  A\&A,  430, 343

Lutz, D. et al, "PACS Evolutionary Probe (PEP) - A 	\textit{Herschel} key program", 2011, A\&A, 532, 90 

Lutz, D.  "PACS photometer point spread function", 2012

Men’shchikov, A., et al., "A multi-scale, multi-wavelength source extraction method:
getsources", 2012, A\&A, 542, 81

Molinari,  S. et al.,  "Hi-GAL: The Herschel Infrared Galactic Plane Survey",  2010,  PASP,  122, 314

Molinari,  S. et al.,  "Source extraction and photometry for the far-infrared and sub-millimeter continuum in the presence of complex backgrounds", 2011,  A\&A,  530,  133

Molinari,  S. et al.,  "Hi-GAL, the Herschel Infrared Galactic Plane Survey: photometric maps and compact source catalogues.
First data release for Inner Milky Way: +68 $\geq$ l ≥ $\geq$−70", 2016,  A\&A, 591, 33 

Montillaud, J. et al. ,"Galactic cold cores. IV. Cold submillimetre sources: catalogue and statistical analysis", 2015 A\%A, 584, 92

M\"uller, T. et al, " PACS Photometer Passbands and Colour Correction Factors for Various Source SEDs", 2011

Neugebauer, B.  \&  Beichman, C. A., "INFRARED ASTRONOMICAL SATELLITE (IRAS) CATALOGS AND ATLASES Explanatory Supplement",  1987

Nielbock, M., et al.,  "The Herschel PACS photometer calibration. A time dependent flux calibration for the PACS chopped point-source photometry AOT mode",  2013, ExA,  36, 631

Ott, S.: “The 	\textit{Herschel} data processing system HIPE and pipelines”. ASP Conference Series 434, 139 (2010)

Pilbratt, G., et al., “	\textit{Herschel} space observatory. An ESA facility for far-infrared and submillimetre astronomy”. A\&A, 518, L1 (2010)

Poglistsch,  A. et al.,  "The Photodetector Array Camera and Spectrometer (PACS)
on the Herschel Space Observatory",  2010, A\&A,  518, L2

Savage, R.S. \& Oliver, S., “ Bayesian Methods of Astronomical Source Extraction” , ApJ 661, 1339 (2007)

Stetson, P.B., “DAOPHOT - a computer program for crowded-field stellar photometry”. PASP 99, 191 (1987)

Schultz, B. et al,  "SPIRE Point Source Catalog Explanatory Supplement", 2017

Taylor, M.B., "TOPCAT \& STILT: Starlink Table/VOTable Processing Software", in Astronomical Data Analysis Software and Systems XIV, eds. P Shopbell , M. Britton, and R. Ebert, ASP Conf. Ser. 347, p. 29 (2005)

Traficante, A. et al.,  "Data reduction pipeline for the Hi-GAL survey",  2011,  MNRAS,  416, 2932


\end{document}